\newif\ifmypreprint

\mypreprinttrue

\ifmypreprint

\documentclass{article}

\usepackage{authblk}

\def\myrelwidth{0.46\linewidth}

\else

\documentclass[review,a4paper]{elsarticle}

\usepackage{lineno}

\def\myrelwidth{0.48\linewidth}

\fi

\usepackage{a4wide}
\usepackage{amsmath}
\usepackage{booktabs}
\usepackage{graphicx}

\usepackage[utf8]{inputenc}
\usepackage{fancyvrb}
\usepackage{hyperref}
\usepackage{bm}
\usepackage{algpseudocode}
\usepackage{algorithmicx}
\usepackage{algorithm}

\newcommand{\smatrix}[2]{\left#2\begin{array}{#1}}
\newcommand{\ematrix}[1]{\end{array}\right#1}
\def\pdiff#1#2{\frac{\partial {#1}}{\partial {#2}}}
\def\ul#1{\underline{#1}}
\def\dM{\mathsf{M}}
\def\du{\mathsf{u}}
\def\dU{\mathsf{U}}
\def\dv{\mathsf{v}}
\def\dx{\mathsf{x}}
\def\bu{\bm{\du}}
\def\bv{\bm{\dv}}
\def\bU{\bm{\dU}}
\def\bM{\bm{\dM}}
\def\bphis{\bm{\phi}}
\def\bphiv{\bm{\Phi}}
\def\bzero{\bm{0}}

\def\dOmega{{\partial\Omega}}

\def\tmean{\bar T}
\def\twwmean{\bar T^{\rm ww}}
\def\twwmeanr{\bar T^{\rm ww}_r}
\def\twwmeand{\bar T^{\rm ww}_d}
\def\rtwwmean{\twwmean / \twwmeanr}
\def\dtwwmean{\twwmean / \twwmeand}

\def\mmax{M^{\rm max}}
\def\mmaxr{M^{\rm max}_r}
\def\rmmax{\mmax / \mmaxr}

\def\putmyabstract{

  In finite element calculations, the integral forms are usually evaluated
  using nested loops over elements, and over quadrature points. Many such forms
  (e.g. linear or multi-linear) can be expressed in a compact way, without the
  explicit loops, using a single tensor contraction expression by employing the
  Einstein summation convention. To automate this process and leverage existing
  high performance codes, we first introduce a notation allowing trivial
  differentiation of multi-linear finite element forms. Based on that we
  propose and describe a new transpiler from Einstein summation based
  expressions, augmented to allow defining multi-linear finite element weak
  forms, to regular tensor contraction expressions. The resulting expressions
  are compatible with a number of Python scientific computing packages, that
  implement, optimize and in some cases parallelize the general tensor
  contractions. We assess the performance of those packages, as well as the
  influence of operand memory layouts and tensor contraction paths
  optimizations on the elapsed time and memory requirements of the finite
  element form evaluations. We also compare the efficiency of the transpiled
  weak form implementations to the C-based functions available in the finite
  element package SfePy.

}

\begin{document}

\title{Fast Evaluation of Finite Element Weak Forms Using Python Tensor
  Contraction Packages}

\author[1]{Robert Cimrman}

\ifmypreprint

\affil[1]{New Technologies Research Centre, University of West Bohemia,
  Univerzitní 8, 306 14 Plzeň, Czech Republic}

\fi

\maketitle

\ifmypreprint

\bibliographystyle{plain}

\begin{abstract}
  \putmyabstract
\end{abstract}

\fi

\section{Introduction}

The Einstein summation convention, abbreviated as \textit{einsum} in the
following text according to the eponymous function of NumPy~\cite{numpy2020},
is a powerful and concise way of writing linear algebra or tensor contraction
expressions and as such it has received attention from computer science related
communities. Large einsum-related research is going on in tensor networks
applicable in machine learning or quantum physics calculations, see
e.g.~\cite{Ma_Ye_Solomonik_2020, Schindler_Jermyn_2020,
  Ran_Tirrito_Peng_Chen_Tagliacozzo_Su_Lewenstein_2020, Gray_Kourtis_2021,
  Liang_Xu_Deng_Yan_Hu_Zhang_Li_Xie_2021}.

In the finite element method (FEM), a mathematical model defined using partial
differential equations (PDEs) is transformed to a weak (integral) formulation
and then discretized, see e.g.~\cite{Zienkiewicz_Taylor_2000}. In the
discretization, the unknown continuous functions from infinite-dimensional
function spaces are approximated by finite-dimensional vectors of coefficients
--- the degrees of freedom (DOFs) --- that are used as linear combination
coefficients of basis functions with small support over the individual finite
elements that cover the solution domain. This leads to the necessity of
evaluating the weak form integrals over each element using a numerical
quadrature, usually using nested loops over elements, and over quadrature
points. Fast evaluation of the weak form integrals is crucial for efficient
calculations, especially when using higher order polynomials/curves for the
basis and hence the approximated variables. Many such integral forms (e.g. all
linear or multi-linear) can be written using a single tensor contraction, or
einsum, expression.

In recent years machine learning techniques and related contractions of large
tensor networks started to be involved in the solution of PDEs using various
approaches such as collocation or energy minimization approaches, see e.g.
\cite{Samaniego_Anitescu_Goswami_Nguyen-Thanh_Guo_Hamdia_Zhuang_Rabczuk_2020}.
In contrast with that we wish to use the tensor contraction expressions
directly in the classical FEM to evaluate the weak forms of PDEs. However,
unlike the huge tensor networks in machine learning or quantum calculations
\cite{Gray_Kourtis_2021}, the einsum expressions in the FE context consist
usually of only a few (e.g. three or four) tensors and the contractions have
low numerical intensity. This complicates use of modern parallel architectures
such as GPUs and is subject to an ongoing research
\cite{Swirydowicz_Chalmers_Karakus_Warburton_2019}. Nevertheless, einsums are
successfully used, for example, in the pure Python finite element package
scikit-fem \cite{Gustafsson_McBain_2020}, where vectorized FE assembling
(cf.~\cite{Rahman_Valdman_2013}) is employed for fast integral form
evaluations.

In this paper we assess several Python scientific computing packages,
that implement, optimize and in some cases parallelize the general tensor
contractions using the einsum semantics. The tested packages are:
\begin{itemize}
\item NumPy \cite{numpy2020}, with the basic implementation and some
  contraction optimizations using a version of opt\_einsum;
\item opt\_einsum \cite{opt_einsum2018}, with state-of-the-art contraction
  optimization strategies;
\item Dask \cite{dask2021}, which can use opt\_einsum as a backend, can
  parallelize the calculation and allows out-of-core calculations with very
  large data.
\item JAX \cite{jax2021github}, with JIT compilation\footnote{JIT compilation =
    Just in time compilation.} and possible parallel execution or automatic GPU
  transfer and einsum optimization for GPUS in spirit of
  \cite{Swirydowicz_Chalmers_Karakus_Warburton_2019}.
\end{itemize}
The used tensor contraction packages require rectangular (non-ragged) arrays,
which limits a single einsum call to a set of finite elements of the same shape
and approximation order. Applying the einsum semantics in the context of p- or
hp-adaptivity \cite{Babuska_Guo_1992} or XFEM
\cite{Moes_Dolbow_Belytschko_1999,
  Bordas_Nguyen_Dunant_Guidoum_Nguyen-Dang_2007} would require an additional
sorting/grouping of the same element kinds together. Such situation is not
considered in this paper.

The calculation speed and memory requirements of the various strategies are
evaluated using several integral forms common in solving multi-physical
problems and compared with the C-based implementation available in the finite
element package SfePy \cite{Cimrman_Lukes_Rohan_2019}. SfePy is used by our
team for solving relatively small ($\sim 10^6$ DOFs), but still physically
relevant, multiphysical and multiscale problems on a single computer and so far
targets only CPU computations.

We present mostly the single-thread performance of the above packages because
that is directly influenced by the contraction path optimality in terms of the
number of floating point operations (flops) and use of BLAS~\cite{BLAS_2002},
nevertheless results with enables multi-threading are presented in
Appendix~\ref{sec:pswfct}.

While we compare the performance with a particular FE solver (SfePy), the
obtained results do not depend on that and are relevant to general audience for
the following reasons. The local element tensors and resulting vectors/matrices
are tied to the mathematical problem and thus are essentially the same in every
FE code. We analyze effects of the memory layout of the tensors on the time and
memory requirements of the tensor contractions. The tensor contractions in
einsum function syntax are generated by a simple ``weak form to einsum''
transpiler that we introduce in Section~\ref{sec:wfet}. This
description/implementation of multi-linear forms does not abstract out the
details (basis functions evaluated in quadrature points etc. are numpy arrays)
and can be reused. The calculations are pure-Python from user's perspective,
and while our approach is smaller in scope than e.g. libCEED, it might offer a
simple yet efficient way of implementing the local matrix action
operator~\cite{Barra_Brown_Thompson_Dudouit_2020}.

The paper is structured as follows: first, in Section~\ref{sec:ebc} several
notions that are used throughout the text are explained. In
Section~\ref{sec:witcs} we discuss factors that influence the speed of tensor
contractions from the perspective of local FE evaluations. Then in
Section~\ref{sec:fee} we recall some basic FE concepts necessary for the
subsequent text and establish the relation to einsum expressions, and in
Section~\ref{sec:wfet} we introduce the transpiler from generalized einsum-like
expressions, suitable for describing multi-linear finite element weak forms, to
regular einsum/tensor contraction expressions. Finally, we present the results
of many numerical simulations in Section~\ref{sec:res} allowing us to assess
the performance of the several einsum implementations mentioned above and
conclude the paper in Section~\ref{sec:con}.

\section{Explanation of Basic Concepts}
\label{sec:ebc}

The presented topic combines aspects of computer science, numerical
approximation of PDEs and software engineering. This section purports to define
the common ground necessary for readers of different backgrounds to understand
the following parts.

A \emph{transpiler} (or source-to-source compiler) translates input in a
programming/domain specific language to another language that works at
approximately the same level of abstraction, unlike a traditional compiler that
translates from a higher level programming language to a lower level
programming language. In Section~\ref{sec:wfet} we use this term to denote our
translator from generalized einsum-like notation to regular einsum function
notation.

By \emph{einsum expressions} we mean both the mathematical way of expressing
implicit summation by repeated indices (the Einstein summation convention),
such as
\begin{displaymath}
  A_{ij} = B_{ik} C_{kj} \equiv \sum_{k=1}^n B_{ik} C_{kj} \;,
\end{displaymath}
where $A$, $B$ and $C$ are second order tensors (matrices), $B$ has $n$ columns
and $C$ has $n$ rows, and, more specifically, the syntax and arguments used to
call einsum functions of Python packages such as NumPy~\cite{numpy2020}. The
above matrix multiplication in this syntax is
\begin{center}
  \Verb|A = einsum('ik,kj->ij', B, C)|.
\end{center}
This operation is called \emph{tensor contraction} and the index $k$ is the
contraction index or axis. Many common linear algebra operations can be
expressed using the einsum syntax, for example ($A$ is a matrix, $u$, $v$
vectors with compatible dimensions):
\begin{center}
  \begin{tabular}{r@{ $\dots$ }l}
    trace of $A$ & \Verb|einsum('ii', A)| \\
    diagonal of $A$ & \Verb|einsum('ii->i', A)| \\
    transposition of $A$ & \Verb|einsum('ij->ji', A)| \\
    outer product of $u$, $v$ & \Verb|einsum('i,j->ij', u, v)| \\
    bilinear form $A$ applied to $u$, $v$ & \Verb|einsum('i,ij,j->', u, A, v)|
  \end{tabular}
\end{center}
All multi-linear (linear, bilinear, etc.) forms that are defined using such
multiplicative expressions can be expressed in this syntax. More examples
specific to FE evaluations are given in Section~\ref{sec:wfete}.

\section{What Influences Tensor Contraction Speed}
\label{sec:witcs}

Evaluating tensor expressions brings various trade-offs between the numerical
intensity/number of flops, the memory consumption and access patterns,
intermediate/temporary array sizes etc. In this Section we consider several
factors that influence the evaluation speed of the weak form einsum expressions
and address some of the trade-offs. The effects of memory layouts are presented
in Section~\ref{sec:mlo}. Tensor contraction paths and related optimizations
are introduced in Section~\ref{sec:tcpo}.

\subsection{Memory Layout of Operands}
\label{sec:mlo}

The einsum operands are initially, irrespective of the package used for
performing the calculations, always stored in NumPy's $n$-dimensional array
objects (\Verb|ndarray| class). A NumPy array is a continuous one-dimensional
segment of memory that can be accessed using an indexing scheme that maps $n$
integer indices into the location of a fixed-size item (e.g. a 64 bit floating
point number) in the block. The ranges in which the indices can vary is
specified by the shape of the array $(d_0, \dots, d_{n-1})$~\cite{numpy2020}.

Among many schemes of arranging items of an $n$-dimensional array in a
one-dimensional block, NumPy uses a flexible \textit{strided indexing scheme}
where an offset of an item $(i_0, \dots, i_{n-1})$ from the beginning of the
memory block is given by
\begin{displaymath}
  i_{\rm offset} = \sum_{k=1}^n s_k i_k \,,
\end{displaymath}
where the integers $s_k$ specify the array strides. The most common strides are
given by the \textit{row-major order} (as in C), where the leftmost index
varies the fastest, i.e. $s_k = \prod_{j=k+1}^{n-1} d_j$,  or the
\textit{column major order} (as in Fortran or Matlab), where the rightmost
index varies the fastest, i.e. $s_k = \prod_{j=0}^{k-1} d_j$.

The term memory layout can refer both to the layout in terms of the strides
(row-major order, column major order, \dots) and to the order of the array
dimensions. In this work we assume all operands to be in a contiguous block of
memory in the row-major order, and by layout we denote the various permutations
of array dimensions.\footnote{Both terms are related: an array of shape
  $(m, n)$ in the row-major order has exactly the same arrangement of items in
  memory as an array of shape $(n,m)$ in the column-major order.}

The operand layout significantly influences the speed of the tensor
contractions, as will be shown in Section~\ref{sec:itml}. For illustration,
let us consider the multiplication of two matrices: for all $i$, $k$,
$C_{ik} = \sum_j A_{ij} B_{jk}$. Having the innermost loop index $j$ vary
along a contiguous block of memory is usually much faster than when, for fixed
$i, k$, the $j$-th and $(j+1)$-th items of $A$ and $B$ are not next to each other
in memory for all $j$, as data access patterns with high locality help avoiding
memory cache misses.

\subsection{Tensor Contraction Path Optimization}
\label{sec:tcpo}

\begin{figure}[htp!]
  \centering
  \includegraphics[width=0.48\linewidth]{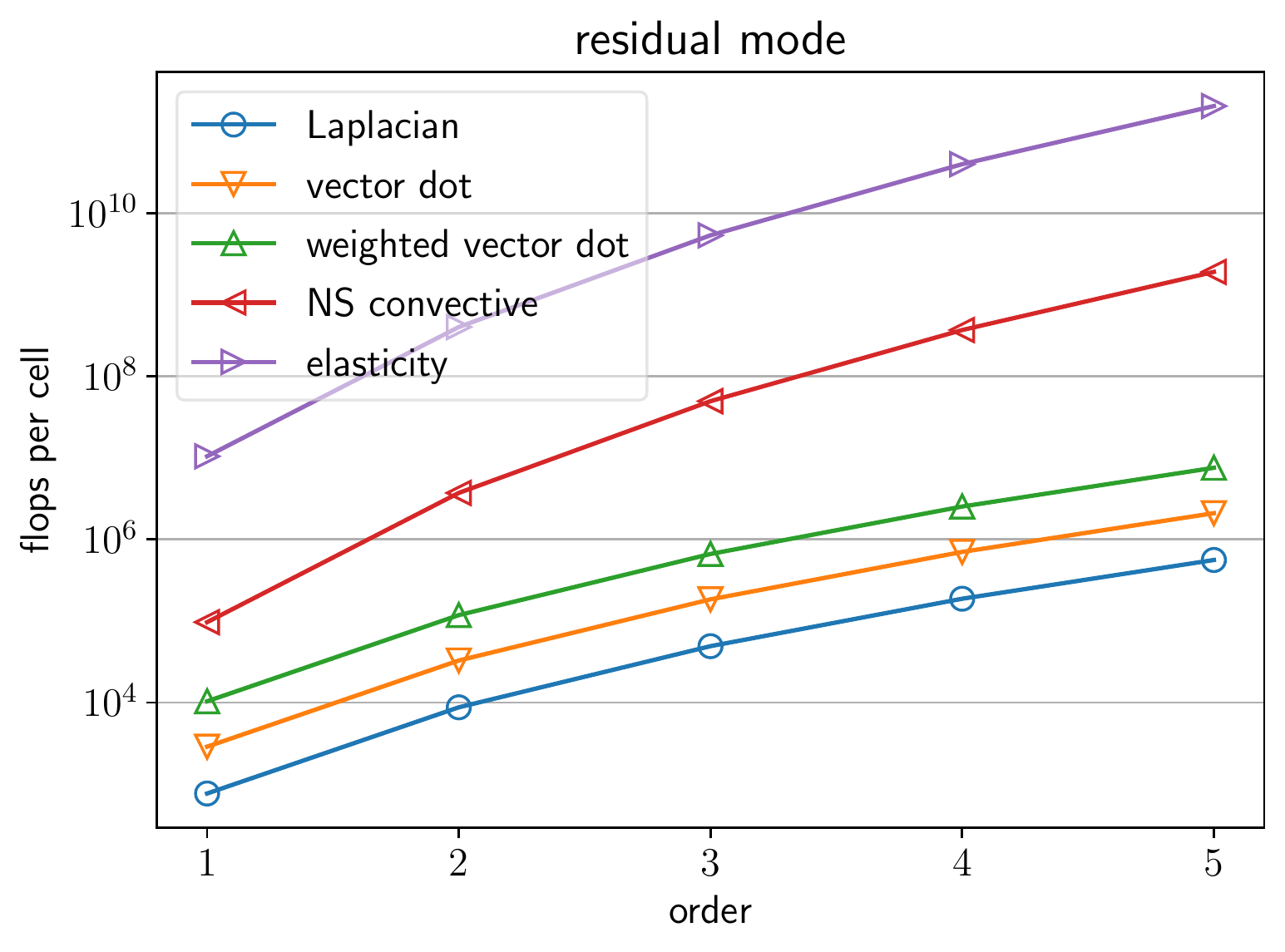}
  \includegraphics[width=0.48\linewidth]{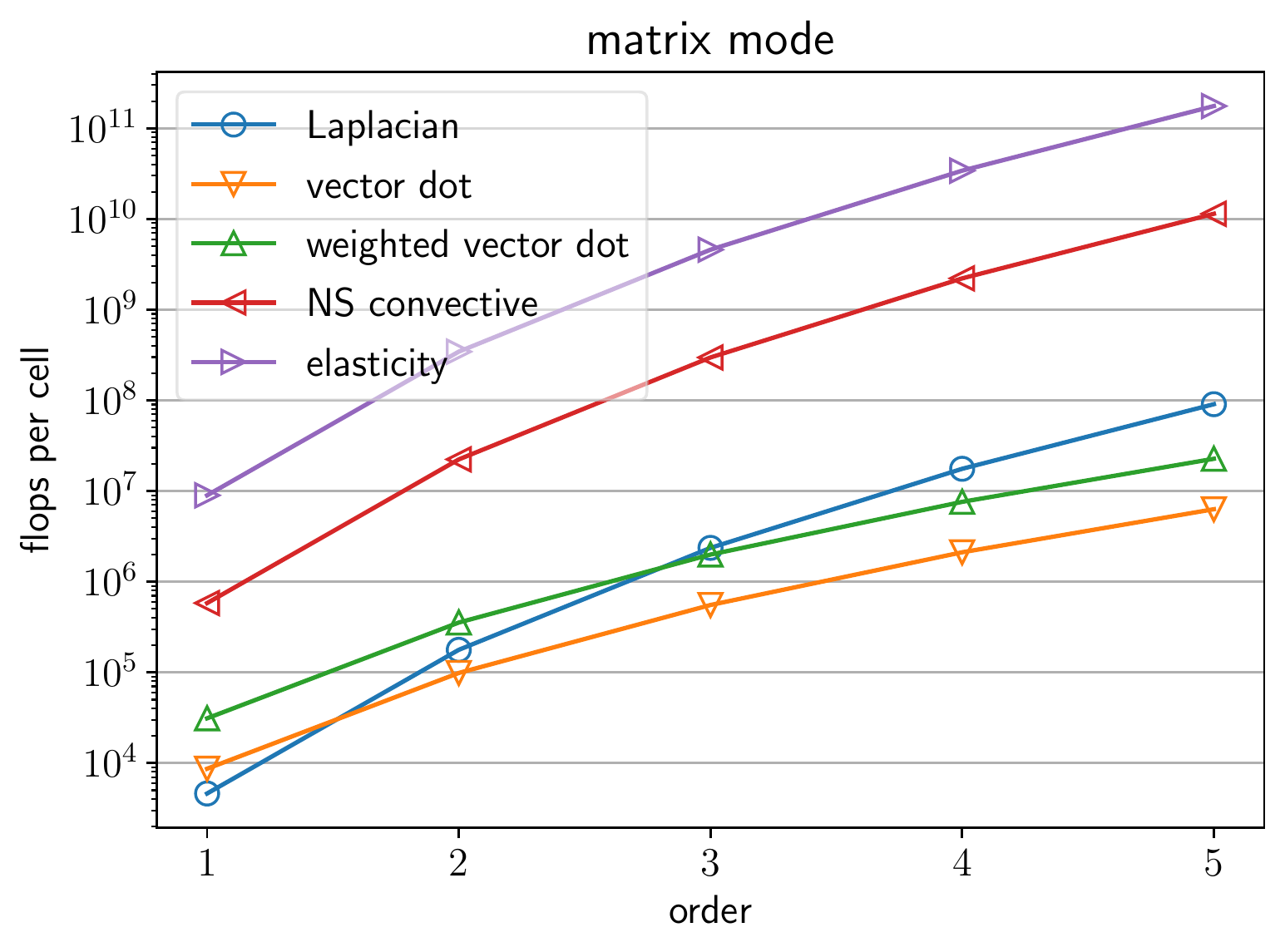}
  \caption{Dependence of the naive flops counts per FE cell on the FE
    approximation order for several weak forms considered in
    Section~\ref{sec:res}. Left: residual mode, i.e. evaluation of the weak
    form action on a vector. Right: matrix mode, i.e. evaluation of the
    local element matrices.}
  \label{fig:naive-flops}
\end{figure}

For einsum expressions with more than two operands, a number of possibilities
exist of how, or in what order, to perform the tensor contractions. The naive
flops cost, corresponding to a for loop for each index in the expression, is
proportional to the product of all operand shapes, and quickly grows
prohibitive. Even though in the context of this work the multi-linear weak form
expressions have less than 10 operands (unlike the machine learning networks),
the naive calculation time and memory requirements are not feasible for all but
the simplest forms, see Fig.~\ref{fig:naive-flops}. It is much faster to
contract a subset of operands at a time, which massively reduces the flops and
intermediate array sizes and also allows BLAS use, see \cite{opt_einsum2018}.
An ordered list of operands to contract at a time is called a contraction path.

The considered packages support various methods of contraction path
optimizations. The resulting path is a sequence of index tuples (usually
pairs), denoting the positions of operands to contract in a list of operands,
see below. While the contraction path optimization algorithms themselves
introduce an intrinsic overhead, the resulting contraction paths strongly
influence the elapsed time and memory consumption, especially as the problem
size increases and the optimization overhead becomes amortized.

For illustration let us compare the naive and optimized contraction paths of a
chained dot product $A_{ij} B_{jk} C_{kl}$, i.e. the einsum expression
\Verb|einsum('ij,jk,kl->il', A, B, C)|, where $A$ is $2 \times 2$, $B$ is
$2 \times 5$, $C$ is $5 \times 2$. In this case, opt\_einsum
\cite{opt_einsum2018} contraction path optimization results in the following:
{\small
\begin{verbatim}
  Complete contraction:  ij,jk,kl->il
         Naive scaling:  4
     Optimized scaling:  3
      Naive FLOP count:  1.200e+2
  Optimized FLOP count:  5.600e+1
   Theoretical speedup:  2.143e+0
  Largest intermediate:  4.000e+0 elements
--------------------------------------------------------------------------------
scaling        BLAS                current                             remaining
--------------------------------------------------------------------------------
   3           GEMM              kl,jk->lj                             ij,lj->il
   3           GEMM              lj,ij->il                                il->il
\end{verbatim}
} That is, instead of the naive four nested loops, it is faster to first
contract the $B$ and $C$: \Verb|D = einsum('kl,jk->jl', B, C)|, three nested
loops, and then to contract the result with $A$:
\Verb|einsum('jl,ij->il', D, A)|, again three nested loops. The flops count of
the naive algorithm is 120, while the optimized flops count is reported to be
56, providing a theoretical speed-up of about two. Furthermore, the two-step
contraction allows using the fast function GEMM of a BLAS implementation. More
examples of tensor contraction paths from the FE form evaluation context are
shown in Section~\ref{sec:wfete}.

\section{FE Evaluation and Einsum}
\label{sec:fee}

In this Section we recall some very basic FEM concepts that will be needed in
subsequent text, details can be found e.g. in \cite{Zienkiewicz_Taylor_2000}.
In particular, the concepts are reviewed using vector functions (scalar
functions correspond to a single component) and two example weak forms: the
vector dot product/mass matrix weak form and the convective term weak form from
Navier-Stokes equations. We propose a unified notation that allows
straightforward translation to einsum expressions as well as a trivial
differentiation of those expressions.

\subsection{Notation}
\label{sec:feen}

Let us denote $\ul{u}, \ul{v}$ two vector functions from suitable function
spaces (e.g. $H^{1}(\Omega)$ in the case of elasticity, $H^{\rm div}(\Omega)$
in the case of fluid flow problems). For simplicity, we assume both functions
to be from the same space of functions $V_0(\Omega)$: a space of functions that
are zero on the boundary $\dOmega$. In the FEM and related methods, the
continuous functions are approximated by functions defined using a finite set
of degrees of freedom (DOFs) and a finite element basis. A component $u_i$,
$i = 1, \dots, D$ can be expressed as the linear combination
\begin{equation}
  \label{eq:fe1}
  u_i(\ul{x}) \approx u^h_i(\ul{x}) = \du^{k}_i \phi_k(\ul{x}) \;,
\end{equation}
where $\du^k_i$, $k = 1, \dots, N$ are the discrete DOFs and $\phi_k(\ul{x})$
are the basis functions --- for example piece-wise polynomials with a small
support in the case of FEM. The discrete functions $\ul{u}^h$ $\ul{v}^h$ are
substituted in the weak form of a PDE in question and a discrete problem is
obtained, in terms of weak form integrals. The integrals need to be evaluated
over the solution domain $\Omega$, resp. its approximation $\Omega^h$, that is
covered by non-overlapping cells --- the elements. The total integral is then
the sum of the element contributions.

The actual elements $T_c$, that can have various shapes, sizes and orientations
in space, are often mapped to a reference (unit) element $T$, where the
integration using a numerical quadrature takes place. Let us denote $\ul{x}$
the coordinates in space and $\ul{\xi}$ the coordinates in the reference
element, then the mapping is defined as
$F^c: \ul{\xi} \rightarrow \ul{x};\ \ul{x} \in T_c;\ x_i(\ul{\xi}) = \dx^{k}_i
\phi_k(\ul{\xi})$,
where $\ul{\dx}^{k}$, $k = 1, \dots, N$ are the coordinates of $T_c$ vertices,
$\phi_k(\ul{\xi})$ are the basis functions restricted to the reference
element\footnote{For the sake of notation simplicity, we assume isoparametric
  elements here, i.e. the same basis functions for the variables and the
  reference element mapping. We also use the same basis for each vector
  component.} and the Jacobian matrix of $F^c$ is
$J^c_{ij}(\ul{\xi}) = \{\pdiff{x_i}{\xi_j}\}$. Applying $F^c$ to
$\ul{\xi} \in T_c$, we have also
\begin{equation}
  \label{eq:fe1a}
  u_i(\ul{\xi}) \approx u^h_i(\ul{\xi}) = \du^{k}_i \phi_k(\ul{\xi}) \;.
\end{equation}
This allows evaluating weak form integrals on the reference element.

Thus the basis functions have well defined restrictions $\phi_k(\ul{\xi})$ to
the elements --- the forms can be evaluated independently in the individual
elements, in a (parallel) loop. An example of such a loop is in
Algorithm~\ref{alg:feloopv}. This particular loop evaluates a weak form integral
in each cell using a numerical quadrature for a given DOF vector $\du$. For
linear forms, it is equivalent to the matrix-vector action of the form, i.e.
$\dM \du$. The same loop structure can be seen also in
Algorithm~\ref{alg:feloopm}, which evaluates the weak form derivative w.r.t.
$\du$, i.e. the matrix $\dM$ for linear forms. Both functions return arrays of
values for the given set of cells, the actual assembling to a global
vector/sparse matrix is not shown.

\begin{algorithm}[htp!]
  \caption{Basic finite element loop algorithm for evaluating local element
    contributions to a global residual vector, i.e. the action of the form on a
    DOF vector $\du$. The quadrature weights are incorporated in the element
    reference mapping Jacobian $J(\ul{\xi})$.}
  \label{alg:feloopv}
  \begin{algorithmic}[1]
    \Require $F$ \Comment{A weak form evaluation function.}

    \Require $cells, qps$ \Comment{Lists of FE mesh cells and reference cell
      quadrature points.}

    \Require $v, u$ \Comment{FE variables}

    \Require $d$ \Comment{material parameters, reference
      mappings (Jacobian $J$) and other data}

    \Function {eval\_residual}{$cells$, $qps$, $F$, $v$, $u$, $d$}

    \State $n_c, n_{dr} \gets get\_shape(cells, v, u, d)$ \Comment{get result
      array shape}

    \State $r \gets zeros((n_c, n_{dr}))$ \Comment{allocate result array}

    \For{$i_c$ in $cells$} \Comment{loop over cells}

    \For{$\ul{\xi}$ in $qps$} \Comment{loop over cell quadrature
      points}

    \State $r[i_c, :] \gets r[i_c, :] + d.J(\xi) \cdot F(i_c, \ul{\xi}, v, u, d)$
    \Comment{evaluate and integrate $F$}

    \EndFor
    \EndFor
    \State \textbf{return} $r$
    \EndFunction
  \end{algorithmic}
\end{algorithm}


\begin{algorithm}[htp!]
   \caption{Basic finite element loop algorithm for
     evaluating local element contributions to a global matrix, i.e. the
     derivative of the form w.r.t. a DOF vector $\du$. The quadrature
     weights are incorporated in the element reference mapping Jacobian
     $J(\ul{\xi})$.}
   \label{alg:feloopm}
  \begin{algorithmic}[1]




    \Function {eval\_matrix}{$cells$, $qps$, $F$, $v$, $u$, $d$}

    \State $n_c, n_{dr}, n_{dc} \gets get\_shape(cells, v, u, d)$ \Comment{get result
      array shape}

    \State $M \gets zeros((n_c, n_{dr}, n_{dc}))$ \Comment{allocate result array}

    \For{$i_c$ in $cells$} \Comment{loop over cells}

    \For{$\ul{\xi}$ in $qps$} \Comment{loop over cell quadrature
      points}

    \State $M[i_c, :, :] \gets M[i_c, :, :] + d.J(\xi) \cdot \partial F(i_c,
    \ul{\xi}, v, u, d) / \partial u$
    \Comment{evaluate and integrate $\pdiff{F}{u}$}

    \EndFor
    \EndFor
    \State \textbf{return} $M$
    \EndFunction
  \end{algorithmic}
\end{algorithm}


In the following we will be interested in both modes of evaluation. The global
matrices are required e.g. when using a direct linear system solver, while the
matrix action/residual form can be used in connection with iterative solvers that
do not need the global sparse matrix explicitly, only its application to a
vector.

In compiled languages, such as Fortran or C++, the integrals over a group of
elements can be calculated in a loop. In Python (and other interpreted
languages), the loops are very slow --- that is why NumPy and other packages
allowing fast vectorized calculations were introduced. Considering this, our
aim is to evaluate a FE expression by a single call vectorized over many
elements, unlike the aforementioned listings. Below we expand on the already
introduced notation to allow formulating the evaluation loops using einsum
functions in a straightforward and efficient way. The notation facilitates
using scalar $H^{1}(\Omega)$ basis functions with vector variables, which saves
both time and memory during calculations.

Traditionally the DOFs of all components will be stored in a vector --- a 1D
array: (\ref{eq:fe1a}) in matrix notation, for $D=3$, has the following form:
\begin{equation}
  \label{eq:fe1b}
  \begin{split}
    \ul{u}^h(\ul{\xi}) & =
    \smatrix{ccccccccc}{[}
    \phi_1 & \dots & \phi_N & 0 & \dots & 0 & 0 & \dots & 0 \\
    0 & \dots & 0 & \phi_1 & \dots & \phi_N & 0 & \dots & 0 \\
    0 & \dots & 0 & 0 & \dots & 0 & \phi_1 & \dots & \phi_N
    \ematrix{]}
    \smatrix{c}{[}
    \du_1^1 \\
    \vdots \\
    \du_1^N \\
    \du_2^1 \\
    \vdots \\
    \du_2^N \\
    \du_3^1 \\
    \vdots \\
    \du_3^N \\
    \ematrix{]} \\
    & \equiv
    \smatrix{ccc}{[}
    \bphis & \bzero & \bzero \\
    \bzero & \bphis & \bzero \\
    \bzero & \bzero & \bphis
    \ematrix{]}
    \smatrix{c}{[}
    \bu_1 \\
    \bu_2 \\
    \bu_3
    \ematrix{]}
    \equiv \bphiv(\ul{\xi}) \bu \equiv \sum_{i=1}^{DN}(\bphiv)_i (\bu)_i \;,
  \end{split}
\end{equation}
where $\bphis$ (argument $\ul{\xi}$ omitted) is the vector of scalar basis
functions evaluated in $\ul{\xi}$ and $\bphiv(\ul{\xi})$ is the $D \times DN$
matrix, whose columns can be interpreted as vector basis functions $(\bphiv)_i$
corresponding to items of the DOF vector $\bu$ of size $DN$. Because all
components of $\ul{u}$ in (\ref{eq:fe1b}) are approximated using the same
scalar basis functions, it can be rearranged as follows:
\begin{equation}
  \label{eq:fe1c}
  \ul{u}^h(\ul{\xi}) =
  \smatrix{c}{[}
  \bu_1^T \\
  \bu_2^T \\
  \bu_3^T
  \ematrix{]}
  \bphis^T(\ul{\xi})
  \equiv
  \bU^T \bphis^T(\ul{\xi}) \;,
\end{equation}
where $\bU$ is the $N \times D$ matrix with columns formed by component DOF
vectors $\bu_i$. In our weak form to einsum transpiler in Section \ref{sec:wfet},
we use the compact expression (\ref{eq:fe1c}) as well as a tensor
(three-dimensional array of shape $D \times D \times N$) form of the basis
functions matrix $\bphiv(\ul{\xi})$ of (\ref{eq:fe1b})
\begin{equation}
  \label{eq:fe1d}
  \begin{split}
  \bphiv^{(3)}(\ul{\xi}) &=
  \smatrix{ccc}{[}
  \smatrix{ccc}{[}
  \phi_1 & \dots & \phi_N  \\
  0 & \dots & 0 \\
  0 & \dots & 0
  \ematrix{]}
  &
  \smatrix{ccc}{[}
  0 & \dots & 0 \\
  \phi_1 & \dots & \phi_N  \\
  0 & \dots & 0
  \ematrix{]}
  &
  \smatrix{ccc}{[}
  0 & \dots & 0 \\
  0 & \dots & 0 \\
  \phi_1 & \dots & \phi_N
  \ematrix{]}
  \ematrix{]} \\
  & \equiv \{\delta_{ij} \phi_k(\ul{\xi})\} \;,
  \end{split}
\end{equation}
where $i, j = 1, \dots, D$ and $k = 1, \dots, N$. This allows distinguishing
the indices of $u_i^h$ components from the DOF components $\du_j^k$, by
rewriting (\ref{eq:fe1a}) as
\begin{equation}
  \label{eq:fe1e}
  u^h_i(\ul{\xi}) = \du^{k}_j \delta_{ij} \phi_k(\ul{\xi}) \;.
\end{equation}
The expression (\ref{eq:fe1e}) will be used when differentiating a form w.r.t.
the DOFs and enables using scalar $H^{1}(\Omega)$ basis functions with vector
variables as mentioned earlier in this Section. In the relation (\ref{eq:fe1b})
the result of differentiation w.r.t. the DOFs is obvious: it is the matrix
$\bphiv$. But to avoid forming this rather sparse matrix explicitly,
(\ref{eq:fe1e}) can be used, accompanied with a subsequent reshape operation
$(j, k) \rightarrow J$, $j = 1, \dots, D$, $k = 1, \dots, N$,
$J = 1, \dots, DN$.

\subsection{Example Weak Forms}
\label{sec:ewf}

One of the most common forms is the one corresponding to the vector dot
product/mass matrix, used e.g. for $l_2$ projections, in elastodynamics and
elsewhere. Evaluating this form in an element $T_C$, in the context of vector
variables, means calculating
\begin{equation}
  \label{eq:fe2a}
  m_c(\ul{v}, \ul{u}) = \int_{T_c} v_i(\ul{x}) u_i(\ul{x})
  = \int_{T} v_i(\ul{\xi}) u_i(\ul{\xi})
  |J^c(\ul{\xi})| \;.
\end{equation}
Substituting the FE approximation (\ref{eq:fe1a}), applied to both $v_i$ and
$u_i$, into (\ref{eq:fe2a}) yields
\begin{equation}
  \label{eq:fe3a}
  m_c^h(\ul{v}^h, \ul{u}^h) = \int_{T} \dv^{k}_i \phi_k(\ul{\xi})
  \du^m_i \phi_m(\ul{\xi})
  |J^c(\ul{\xi})| = \dv^{k}_i (\int_{T} \phi_k(\ul{\xi})
  \phi_m(\ul{\xi})
  |J^c(\ul{\xi})|) \du^m_i \;.
\end{equation}
The expression (\ref{eq:fe3a}) is suitable for calculating the dot product of
two known vector functions $\ul{v}^h$, $\ul{u}^h$, i.e. a scalar value, but to
evaluate either a weak form residual or matrix, either the traditional matrix
relation (\ref{eq:fe1b}) can be used
\begin{equation}
  \label{eq:fe3b}
  m_c^h(\ul{v}^h, \ul{u}^h) = \bv^T (\int_{T} \bphiv^T(\ul{\xi}) \bphiv(\ul{\xi})
  |J^c(\ul{\xi})|) \bu = (\bv)_I \bM_{IJ} (\bu)_J \;,
\end{equation}
which requires working with $\bphiv^T(\ul{\xi})$, or the form (\ref{eq:fe1e})
can be employed:
\begin{equation}
  \label{eq:fe3c}
  m_c^h(\ul{v}^h, \ul{u}^h) =  \dv^{k}_a (\int_{T} \delta_{ia} \phi_k(\ul{\xi})
  \delta_{ib} \phi_m(\ul{\xi})
  |J^c(\ul{\xi})|) \du^m_b
  = \dv^{k}_a \dM_{akbm} \du^m_b
 \;.
\end{equation}
followed by a simple reshape operation $ak \rightarrow I$, $bm \rightarrow J$,
which does not need to copy/move data, if the reshaping of $\dM_{akbm}$
corresponds to its memory storage scheme (e.g. row-major contiguous).

The form (\ref{eq:fe3a}) has a very simple structure, so it can be readily seen
that the mass matrix corresponding to vector variables is formed by $D$ blocks
of the scalar mass matrix
($\int_{T} \phi_k(\ul{\xi}) \phi_m(\ul{\xi}) |J^c(\ul{\xi})|$), but the tensor
form (\ref{eq:fe3c}) enables a straightforward (tangent) matrix evaluation of
more complex forms too, for example the convective term below.

The convective term weak form of the Navier-Stokes equations is non-linear, if
the advection velocity coincides with the flow velocity, but its structure is
that of a multi-linear term. In an element $T_C$ it is defined as
\begin{equation}
  \label{eq:fe2}
  c_c(\ul{v}, \ul{u}, \ul{u})
  = \int_{T_c} v_i(\ul{x}) \pdiff{u_i}{x_j}(\ul{x}) u_j(\ul{x})
  = \int_{T} v_i(\ul{\xi}) \pdiff{u_i}{x_j}(\ul{\xi}) u_j(\ul{\xi})
  |J^c(\ul{\xi})| \;,
\end{equation}
where $\pdiff{u_i}{x_j}(\ul{\xi})$ can by expressed in terms of $\ul{\xi}$
using
$\pdiff{u_i}{x_j}(\ul{\xi}) = \pdiff{u_i}{\xi_k}(\ul{\xi})
J^{-1}_{kj}(\ul{\xi})$.
Substituting the tensor FE approximation (\ref{eq:fe1e}) into (\ref{eq:fe2})
yields (the argument $\ul{\xi}$ is omitted for brevity)
\begin{equation}
  \label{eq:fe3}
  c_c^h(\ul{v}^h, \ul{u}^h, \ul{u}^h)
  = \dv^{k}_a \int_{T} \delta_{ia} \phi_k
  \du^l_b \delta_{ib} \pdiff{\phi_l}{x_j}
  \du^m_c \delta_{jc} \phi_m
  |J^c| \;,
\end{equation}
which allows trivial differentiation w.r.t. $\bu$, we just omit $\du^l_b$ and
then $\du^m_c$ from the expression and sum the two terms together:
\begin{equation}
  \label{eq:fe4}
  \begin{split}
    \pdiff{c_c^h}{\bu}
    & \stackrel{\mathrm{reshape}}{=} \dv^{k}_a \int_{T}
    (\delta_{ia} \phi_k
    \delta_{ib} \pdiff{\phi_l}{x_j}
    \du^m_c \delta_{jc} \phi_m
    + \delta_{ia} \phi_k
    \du^l_b \delta_{ib} \pdiff{\phi_l}{x_j}
    \delta_{jc} \phi_m)
    |J^c| \\
    & \quad = \quad \dv^{k}_a \int_{T}
    (\delta_{ia} \phi_k
    \delta_{ib} \pdiff{\phi_l}{x_j}
    \du^m_j \phi_m
    + \delta_{ia} \phi_k
    \du^l_i \pdiff{\phi_l}{x_j}
    \delta_{jc} \phi_m)
    |J^c|
 \;,
  \end{split}
\end{equation}
where in the second identity we replaced back $\du^m_c \delta_{jc}$ by
$\du^m_j$ and $\du^l_b \delta_{ib}$ by $\du^l_i$, i.e. used the compact
expression (\ref{eq:fe1c}).

The weak form to einsum transpiler described in Section~\ref{sec:wfet} allows
automatic transformation of the description corresponding to the left-hand side
of (\ref{eq:fe2}) to an einsum-ready descriptions of the discretized form
(\ref{eq:fe3}) and (\ref{eq:fe4}).

\section{Weak Form to Einsum Transpiler}
\label{sec:wfet}

In this Section we propose a generalized einsum-like notation suitable for
describing weak form integrals and introduce a transpiler that can translate
expressions in this notation to regular einsum expressions and operands.

The generalized einsum-like syntax that we propose is both close to the
mathematical notation and explicit in what tensor axes are contracted, i.e. the
weak form definitions are self-documenting, compare e.g. with the very
versatile \cite{Hale_Brunetti_Bordas_Maurini_2018} Unified Form
Language~\cite{Alnaes_Logg_Olgaard_Rognes_Wells_2014} from the FEniCS project,
where named symbolic operators such as \Verb|inner| or \Verb|grad| are used.

\subsection{Overview}
\label{sec:wfeto}

Let us denote by $p$ a scalar unknown variable and by $q$ the corresponding
test function. Similarly, $\ul{u} \equiv \{u_i\}$, $\ul{v} \equiv \{v_i\}$ term
a vector unknown and test variable pair.

In the proposed implementation, we generalize the einsum notation with symbols
for derivatives and other quantities summarized in Tab.~\ref{tab:wfc}. Using
those symbols it is possible to define all the (multi-)linear terms available in
SfePy in a unified and easy manner. Several example definitions are shown in
Tab.~\ref{tab:ewfd} --- five of those weak form expressions are the subject of the
performance analysis in the next Section.

\begin{table}[ht!]
  \centering
  \begin{tabular}{rp{0.55\linewidth}l}
    \toprule
    symbol & meaning
    & example \\
    \midrule
    \Verb|0| & scalar
    & $p$ \\
    \Verb|i| & $i$-th vector component
    & $u_i$ \\
    \Verb|i.j| & gradient: derivative of $i$-th vector component w.r.t. $j$-th
                 coordinate component
    & $\pdiff{u_i}{x_j}$ \\
    \Verb|i:j| & symmetric gradient
    & $\frac{1}{2} (\pdiff{u_i}{x_j} + \pdiff{u_j}{x_i})$ \\
    \Verb|s(i:j)->I| & vector storage of symmetric second order tensor,
                       $I$ is the vector component
    & Cauchy strain tensor $e_{ij}(\ul{u})$ \\
    \bottomrule
  \end{tabular}
  \caption{The generalized einsum-like notation.}
  \label{tab:wfc}
\end{table}

\begin{table}[ht!]
  \centering
  \renewcommand*{\arraystretch}{1.2}
  \begin{tabular}{rcl}
    \toprule
    description & definition
    & weak form expression \\
    \midrule
    vector dot product & \Verb|('i,i', v, u)|
    & $\int_{T} v_i u_i$ \\
    weighted vector dot product & \Verb|('ij,i,j', M, v, u)|
    & $\int_{T} v_i M_{ij} u_j$ \\
    weak Laplacian & \Verb|('0.i,0.i', v, u)|
    & $\int_{T} \pdiff{v}{x_i} \cdot \pdiff{u}{x_i}$ \\
    Navier-Stokes convection & \Verb|('i,i.j,j', v, u, u)|
    & $\int_{T} v_i \pdiff{u_i}{x_j} u_j$ \\
    Stokes coupling & \Verb|('i.i,0', v, p)|
    & $\int_{T} \pdiff{v_i}{x_i} p$ \\
    divergence operator & \Verb|('i.i', v)|
    & $\int_{T} \pdiff{v_i}{x_i}$ \\
    transposed Stokes coupling & \Verb|('i.i,0', u, q)|
    & $\int_{T} q \pdiff{u_i}{x_i}$ \\
    linear elasticity & \Verb|('IK,s(i:j)->I,s(k:l)->K', D, v, u)|
    & $\int_{T} D_{ijkl}\ e_{ij}(\ul{v}) e_{kl}(\ul{u})$ \\
    Cauchy stress & \Verb|('IK,s(k:l)->K', D, u)|
    & $D_{ijkl}\ e_{kl}(\ul{u})$ \\
    \bottomrule
  \end{tabular}
  \caption{Examples of multi-linear weak form definitions.}
  \label{tab:ewfd}
\end{table}

The generalized einsum-like expressions are transformed to regular einsum
expressions using a simple transpiler with the following features:
\begin{itemize}
\item several einsum evaluation backends, see Section~\ref{sec:wfetb};
\item arbitrary memory layout of the operands;
\item easy automatic differentiation, thanks to (multi-)linearity of the
  considered weak forms;
\item various evaluation modes (see below).
\end{itemize}
The following evaluation modes, a feature available in SfePy, are supported:
\begin{itemize}
\item 'weak' modes for expressions involving a test variable:
  \begin{itemize}
  \item 'residual' mode corresponding to the operator/matrix application to a
    DOF vector without the global sparse matrix assembly;
  \item 'matrix' mode returning the local element matrices --- derivatives of
    the expression w.r.t. a given variable;
  \end{itemize}
\item 'eval' mode returning the integral value in case all variables passed to
  the evaluation function have associated DOFs, i.e. no test variables are in
  the argument list.
\end{itemize}
All weak forms shown in Tab.~\ref{tab:ewfd} support all the above evaluation
modes automatically, with the exception of the Cauchy stress expression: this
form has no test variables and cannot be called in a `weak` mode, only
evaluated.

The transpiler allows performing simple transformations of the expressions and
operands, namely:
\begin{itemize}
\item Memory layout application: the required memory layout of einsum operands
  can be specified using a letter symbol for each operand axis, listed in
  Tab.~\ref{tab:mls}.
\item Expression slicing: to reduce memory usage, an expression can be
  evaluated in a loop one cell at a time.
\item Operand conversion: operands can be retyped to dask arrays with chunks
  defined in terms of cell slices.
\item Transformations of expression indices and special operand storage: see
  the linear elasticity weak form example in Section~\ref{sec:wfete}.
\end{itemize}
The weak form evaluation function implementing the above features is outlined
in Algorithm~\ref{alg:wfef}. In the actual implementation, results of various
algorithm steps (e.g. the expression transpilation) are cached to speed up
subsequent evaluations.

\begin{table}[ht!]
  \centering
  \begin{tabular}{rl}
    \toprule
    letter & axis \\
    \midrule
    c & cells \\
    q & quadrature points \\
    v & variable component \\
    g & gradient component \\
    d & local DOF (basis, node) \\
    0 & all material axes \\
    \bottomrule
  \end{tabular}
  \caption{Memory layout specification letters.}
  \label{tab:mls}
\end{table}

\begin{algorithm}[htp!]
  \caption{Outline of the weak form evaluation function based on the weak form
    to einsum transpiler.}
  \label{alg:wfef}
  \begin{algorithmic}[1]
    \Require $ge$ \Comment{the generalized einsum-like expression of a weak form}

    \Require $data$ \Comment{FE variables, material parameters, reference
      mappings etc.}

    \Require $layout$ \Comment{operands memory layout}

    \Require $backend$ \Comment{einsum expression evaluation backend}

    \Function {evaluate}{ge, data, backend}

    \State $ee \gets transpile(ge)$ \Comment{transpile $ge$ to einsum
      expression (backend independent)}

    \State $eval\_einsum \gets get\_function(backend)$
    \Comment{backend-dependent evaluation function}

    \State $ops \gets get\_operands(data)$ \Comment{operands from FE data}

    \State $shape \gets get\_output\_shape(ee, ops)$

    \State $ee, ops \gets apply\_layout(ee, ops)$ \Comment{apply layout, does
      not change output shape}

    \State $ee, ops \gets transform(ee, ops, backend)$ \Comment{do
      backend-dependent transformations}

    \State $paths \gets get\_paths(ee, ops, backend)$ \Comment{get optimized
      contraction paths}
    \State \textbf{return} $eval\_einsum(shape, ee, ops, paths)$
    \EndFunction
  \end{algorithmic}
\end{algorithm}


\subsection{Backends}
\label{sec:wfetb}

A backend terms an expression evaluation function tailored for the specific
library. The following backends are currently supported, the abbreviations in
parentheses are used in Section~\ref{sec:res}:
\begin{itemize}
\item \Verb|'numpy'| (np): the expression evaluation function uses
  \Verb|numpy.einsum()|.
\item \Verb|'numpy_loop'| (npl): \Verb|numpy.einsum()| is applied to a sliced
  expression sequentially one cell at a time.
\item \Verb|'opt_einsum'| (oe): the expression evaluation function uses
  \Verb|opt_einsum.contract()|, more path optimization options are available than
  with \Verb|`numpy`|;
\item \Verb|'opt_einsum_loop'| (oel): like \Verb|'numpy_loop'| but with
  \Verb|opt_einsum.contract()|.
\item \Verb|'jax'| (jx): the expression evaluation function uses
  \Verb|jax.numpy.einsum()| and is JIT-compiled using \Verb|jax.jit()|.
\item \Verb|'jax_vmap'| (jxv): the expression evaluation function uses
  \Verb|jax.numpy.einsum()|, is JIT-compiled using \Verb|jax.jit()| and
  vectorized using \Verb|jax.vmap()|, which leads to a similar behavior to the
  \Verb|'*_loop'| backends.
\item \Verb|'dask_single'| (das), \Verb|'dask_threads'| (dat): the expression
  evaluation function uses \Verb|dask.array.einsum()|, explicitly without or
  with multiple threads.
\item \Verb|'opt_einsum_dask_single'| (oedas), \Verb|'opt_einsum_dask_threads'|
  (oedat): like \Verb|'opt_einsum'|, but the operands are converted to dask
  arrays with chunks defined in terms of cell slices and dask is used by
  opt\_einsum to do the contraction, either without or with multiple threads.
  The \Verb|'auto'| chunk size is used by default, i.e. suitable sizes are
  determined by dask.
\end{itemize}

\subsection{Expression Examples}
\label{sec:wfete}

Several examples below illustrate the functioning of the transpiler. The
default layout, compatible with SfePy, is used, i.e. the row-major storage of
arrays with the \Verb|cqgvd0| array axes order. The following symbols defined
in Tab.~\ref{tab:suee} will be used in the explanation.
\begin{table}[ht!]
  \centering
  \begin{tabular}{rp{0.82\linewidth}}
    \toprule
    variable.array & meaning \\
    \midrule
    v.det & a reference mapping Jacobian $J$ multiplied by quadrature weights \\
    u.bf & basis functions $\psi_k$ corresponding to the approximation of $u$ \\
    u.bfg & $\pdiff{\psi_k}{\xi_l}$ corresponding to the approximation of $u$ \\
    u.I & the identity matrix \\
    u.dofs & the DOFs of $u$ in the matrix form $\bU$ \\
    u.Psg & a tensor of shape $(D, D, D(D+1)/2)$ expressing the vector storage
            of the symmetric gradient \\
    m1\_m2.arg & an array m2 of a material m1 \\
    \bottomrule
  \end{tabular}
  \caption{Symbols used in the expression examples.}
  \label{tab:suee}
\end{table}

\subsubsection{Weak Laplacian Expression}
\label{sec:wle}
The weak Laplacian integral defined by the expression \Verb|('0.i,0.i', v, u)|
gets translated in the matrix mode to
\Verb|einsum('cq,cqjd,cqje->cde', v.det, v.bfg, u.bfg)|, where \Verb|det| is
the reference element mapping Jacobian and \Verb|bfg| is the matrix of basis
functions gradients, as defined in Tab.~\ref{tab:suee}. For a particular case of
1024 tri-quadratic hexahedral elements, the data sizes are:
\begin{Verbatim}
{'c': 1024, 'q': 27, 'j': 3, 'd': 27, 'e': 27}
cde (1024, 27, 27) =
  v.det     cq      (1024, 27)
  v.bfg     cqjd    (1024, 27, 3, 27)
  u.bfg     cqje    (1024, 27, 3, 27)
path: [(0, 1), (0, 1)]
\end{Verbatim}
That is, the integral is over 1024 cells with 27 quadrature points each,
\Verb|j| denotes the three components of the gradients and there are 27 DOFs
per cell. The last line shows the contraction path resulting from the greedy
optimization algorithm implemented in NumPy's \Verb|einsum()| function: a~pair
$(i, j)$ denotes a contraction of operands $i$, $j$, then the operand $i$ is
removed from the list of operands and the operand $j$ is replaced by the
contraction result. This process is repeated until all the contractions are
evaluated and a single operand, the result, remains.

Using either the \Verb|'numpy_loop'| or the \Verb|'opt_einsum_loop'| backend
causes the expression to be compiled to
\Verb|einsum('q,qjd,qje->de', v.det[c], v.bfg[c], u.bfg[c])| that is evaluated
for each cell $c$. In Section~\ref{sec:res} we demonstrate that this slicing
approach is an efficient method when applied to FE approximation orders greater
or equal to three, regardless the weak form.

\subsubsection{Vector Dot Product Expression}
\label{sec:vdpe}
The dot product of vector variables, discussed in Section~\ref{sec:ewf} is
defined by the expression \Verb|('i,i', v, u)|, which is transpiled to \\
\Verb|einsum('cq,qd,ir,qe,is->crdse', v.det, v.bf, v.I, u.bf, u.I)| in the
matrix mode, with data sizes for 1024 tri-linear hexahedral elements
\begin{Verbatim}
{'c': 1024, 'q': 8, 'd': 8, 'i': 3, 'r': 3, 'e': 8, 's': 3}
crdse (1024, 3, 8, 3, 8) =
  v.det     cq      (1024, 8)
  v.bf      qd      (8, 8)
  v.I       ir      (3, 3)
  u.bf      qe      (8, 8)
  u.I       is      (3, 3)
path: [(2, 4), (0, 1), (0, 2), (0, 1)]
\end{Verbatim}
and to \Verb|einsum('cq,qd,ir,qe,cie->crd', v.det, v.bf, v.I, u.bf, u.dofs)| in
the vector mode, with data sizes
\begin{Verbatim}
{'c': 1024, 'q': 8, 'd': 8, 'i': 3, 'r': 3, 'e': 8}
crd (1024, 3, 8) =
  v.det     cq      (1024, 8)
  v.bf      qd      (8, 8)
  v.I       ir      (3, 3)
  u.bf      qe      (8, 8)
  u.dofs    cie     (1024, 3, 8)
path: [(3, 4), (0, 3), (0, 2), (0, 1)]
\end{Verbatim}
Note that the same expressions work also with scalar variables, but the
transpilation result is different.

\subsubsection{Linear Elasticity Expression}
\label{sec:lee}
The last example covers the linear elasticity weak form and illustrates the
transformations of expression indices and a special operand storage. The form
is defined using the second order Cauchy strain tensor (a.k.a. symmetric
gradient)
\begin{displaymath}
  e_{ij}(\ul{u}) = \frac{1}{2} (\pdiff{u_i}{x_j} + \pdiff{u_j}{x_i})
\end{displaymath}
and the fourth order tensor $D_{ijkl}$ with elastic coefficients. Exploiting
symmetry of those tensors allows a substantial memory saving: in 3D, $e_{ij}$
is commonly stored as a vector of $6$ components instead of a $3 \times 3$
matrix, and $D_{ijkl}$ is stored as a $6 \times 6$ matrix, instead of the full
tensor with 81 items. This storage and shape transformation is supported by our
transpiler via the \Verb|s(i:j)->I| syntax, meaning a symmetric storage subset is
taken from the tensor indices $i$, $j$ and the result is put into a 1D array
with the index $I$. Hence the linear elasticity weak form is defined simply
using the expression \Verb|('IK,s(i:j)->I,s(k:l)->K', D, v, u)|, that is
compiled to \\
{\Verb|einsum('cq,cqje,rjI,cqIK,cqlf,slK->cresf', v.det, v.bfg, v.Psg, m_D.arg, u.bfg, u.Psg)|}
in the matrix mode, with data sizes for 1024 tri-linear hexahedral elements
\begin{Verbatim}
{'c': 1024, 'q': 8, 'j': 3, 'e': 8, 'r': 3, 'I': 6, 'K': 6, 'l': 3, 'f': 8, 's': 3}
cresf (1024, 3, 8, 3, 8) =
  v.det     cq      (1024, 8)
  v.bfg     cqje    (1024, 8, 3, 8)
  v.Psg     rjI     (3, 3, 6)
  m_D.arg   cqIK    (1024, 8, 6, 6)
  u.bfg     cqlf    (1024, 8, 3, 8)
  u.Psg     slK     (3, 3, 6)
path: [(0, 1), (0, 1), (0, 1, 2, 3)]
\end{Verbatim}
\noindent The above contraction path corresponds again to the greedy
optimization algorithm implemented in NumPy's \Verb|einsum()| function. Better
performance can be obtained in this case e.g. by using the dynamic programming
optimization algorithm of opt\_einsum's \Verb|contract()| function, which
returns the following path:
\begin{Verbatim}
path: [(0, 4), (2, 3), (1, 3), (1, 2), (0, 1)]
\end{Verbatim}
Note also the special tensor \Verb|Psg| of shape $(D, D, D(D+1)/2)$ which
extracts the basis function gradient components and stores them in a vector.
This tensor has only $D$ nonzeros (ones), however this sparsity is not exploited.
Consequently, the current performance of the linear elastic form einsum
expression is not optimal. Nevertheless, the ability to express this term using
a single simple expression is very convenient.

\section{Performance Evaluation Results}
\label{sec:res}

Five representative weak forms from Tab.~\ref{tab:ewfd} were chosen for
studying elapsed time and memory consumption of the einsum evaluation backends
from Section \ref{sec:wfetb}, namely the vector dot product, weighted vector
dot product, weak Laplacian, Navier Stokes convection and linear elasticity
terms. For comparison, their counterparts from SfePy were used, all implemented
using a hand-crafted C code, called from Python via SfePy's lightweight Cython
\cite{behnel2010cython} wrapper. The reference implementation evaluates the
forms cell-by-cell, and thus requires a very small memory additional to that
occupied by the operands and the resulting array.

The einsum backend performance measurements were executed on a Linux
workstation with the AMD Ryzen Threadripper 1920X 12-Core Processor, 32~GB RAM
and Python version 3.8.5, NumPy 1.20.1, opt\_einsum 3.3.0, JAX 0.2.9, Dask
2021.02.0 and SfePy 2021.1 (development version) installed.

We were interested mostly in the single-thread performance of the einsum
backends because that was directly influenced by the contraction path
optimality in terms of the number of floating point operations (flops) and some
of the calculations were memory-demanding --- the largest tasks used all the
available RAM. To ensure single-threaded execution of all backends, the
relevant environment variables were set appropriately and the CPU affinity
parameter was set to a single value\footnote{It is a known issue
  (https://github.com/google/jax/issues/1539) that JAX cannot run using a
  single thread, so the affinity workaround was used.}. Vice-versa, the
environment variables and affinity were unset in the potentially multi-threaded
runs in Appendix~\ref{sec:pswfct}.

A 1D bar made of a varying number of 3D hexahedral cells served as the finite
element mesh. Note that the mesh topology is irrelevant for matrix mode linear
weak form evaluations --- it plays some role only for the nonlinear convective
term form and for residual mode evaluations: the mesh cell connectivity is used
to index the DOF vector. On this mesh, the finite element approximations of
orders from 1 to 5 were defined. The used cell and DOF counts for the five
forms are summarized in Tab.~\ref{tab:cdc}: the scalar weak Laplacian form
employs variables with a single component per quadrature point, the other,
vector, forms contain variables with three components per quadrature point,
i.e. the largest vector problem involved 12,582,924 DOFs. For the sake of the
performance measurements, the same numbers of quadrature points as given in the
table were applied to integrating all the five weak forms, although in practice
those numbers would lead to inexact integration e.g. in the convective term
case.
\begin{table}[ht!]
  \centering
  {
    \begin{tabular}{rrrrrrrrr}
\toprule
  \#cells & order & \#QP & \#DOFs/comp. & repeat & $|r_s|$ [MB] & $|M_s|$ [MB] & $|r_v|$ [MB] & $|M_v|$ [MB] \\
\midrule
    1,024 &     1 &    8 &        4,100 &      6 &          0.1 &          0.5 &          0.2 &          4.7 \\
          &     2 &   27 &       18,441 &      6 &          0.2 &          6.0 &          0.7 &         53.7 \\
          &     3 &   64 &       49,168 &      5 &          0.5 &         33.6 &          1.6 &        302.0 \\
          &     4 &  125 &      102,425 &      5 &          1.0 &        128.0 &          3.1 &      1,152.0 \\
          &     5 &  216 &      184,356 &      4 &          1.8 &        382.2 &          5.3 &      3,439.9 \\
    2,048 &     1 &    8 &        8,196 &      6 &          0.1 &          1.0 &          0.4 &          9.4 \\
          &     2 &   27 &       36,873 &      6 &          0.4 &         11.9 &          1.3 &        107.5 \\
          &     3 &   64 &       98,320 &      5 &          1.0 &         67.1 &          3.1 &        604.0 \\
          &     4 &  125 &      204,825 &      4 &          2.0 &        256.0 &          6.1 &      2,304.0 \\
    4,096 &     1 &    8 &       16,388 &      6 &          0.3 &          2.1 &          0.8 &         18.9 \\
          &     2 &   27 &       73,737 &      6 &          0.9 &         23.9 &          2.7 &        215.0 \\
          &     3 &   64 &      196,624 &      5 &          2.1 &        134.2 &          6.3 &      1,208.0 \\
    8,192 &     1 &    8 &       32,772 &      6 &          0.5 &          4.2 &          1.6 &         37.7 \\
          &     2 &   27 &      147,465 &      5 &          1.8 &         47.8 &          5.3 &        430.0 \\
          &     3 &   64 &      393,232 &      4 &          4.2 &        268.4 &         12.6 &      2,415.9 \\
   16,384 &     1 &    8 &       65,540 &      6 &          1.0 &          8.4 &          3.1 &         75.5 \\
          &     2 &   27 &      294,921 &      5 &          3.5 &         95.6 &         10.6 &        860.0 \\
   32,768 &     1 &    8 &      131,076 &      6 &          2.1 &         16.8 &          6.3 &        151.0 \\
          &     2 &   27 &      589,833 &      5 &          7.1 &        191.1 &         21.2 &      1,719.9 \\
   65,536 &     1 &    8 &      262,148 &      6 &          4.2 &         33.6 &         12.6 &        302.0 \\
          &     2 &   27 &    1,179,657 &      5 &         14.2 &        382.2 &         42.5 &      3,439.9 \\
  131,072 &     1 &    8 &      524,292 &      6 &          8.4 &         67.1 &         25.2 &        604.0 \\
  262,144 &     1 &    8 &    1,048,580 &      5 &         16.8 &        134.2 &         50.3 &      1,208.0 \\
  524,288 &     1 &    8 &    2,097,156 &      5 &         33.6 &        268.4 &        100.7 &      2,415.9 \\
1,048,576 &     1 &    8 &    4,194,308 &      5 &         67.1 &        536.9 &        201.3 &      4,831.8 \\
\bottomrule
\end{tabular}

  }
  \caption{The numbers of cells $n_c$, FE approximation orders, numbers of
    quadrature points, DOF counts per component, number of evaluation
    repetitions and the array sizes of the scalar and vector weak form
    residuals $|r_s|$, $|r_v|$ and matrices $|M_s|$, $|M_v|$.}
  \label{tab:cdc}
\end{table}

Note that only the form evaluations (einsum function calls), leading to arrays
of dense vectors/matrices, were measured, not the subsequent FE assembling into
the global residual vector or sparse matrix forming the linear system, except
the comparison results in Section~\ref{sec:rip}. The residuals $r_s$, $r_v$ in
Tab.~\ref{tab:cdc} are arrays of shapes $(n_c, n_{dr})$, where $n_{dr}$ is the
number of DOFs in one cell (the number of shape functions), the matrices $M_s$,
$M_v$ have shapes $(n_c, n_{dr}, n_{dc})$, i.e. the local element matrices are
$n_{dr} \times n_{dc}$ and $n_{dr}$, $n_{dc}$ depend on the approximation order
and the type of the weak form, see also Algorithms~\ref{alg:feloopv},
\ref{alg:feloopm} in Section~\ref{sec:feen}.

In relation to the reference implementation performance in
Section~\ref{sec:rip}, we first explore the influence of the operand memory
layouts on the performance of the weak Laplacian and Navier-Stokes convection
forms in Section~\ref{sec:itml} and then report both relative and absolute
results for all the five forms in Section~\ref{sec:pswfc}, considering only
the default, SfePy-compatible, layout. All data presented below are available
in the form of Pandas DataFrames \cite{McKinney_2010} online
\cite{zenodo-time-tensors-data}.

\subsection{Measuring Memory Consumption}
\label{sec:mmc}

A time-based memory usage including child processes was tracked using
memory\_profiler, version 0.58.0~\cite{memory_profiler2021github}. The
individual evaluation functions were tagged in the data, and the memory
consumption of a function was defined as the maximum minus minimum in the
tagged interval of that function. The sampling periods from $0.001$ seconds to
$0.1$ seconds were used, depending on the problem size. Due to the time-based
sampling nature of the memory tracking, values for very small problems may be
skewed. In such cases the memory requirements were negligible with respect to
the available memory. On the other hand, memory consumption values for large
problems, which are of most interest, were measured accurately.

\subsection{Choice of Performance Indicators}
\label{sec:cpi}

The first evaluation of a weak form using the Algorithm~\ref{alg:wfef} is
slower then subsequent evaluations because of the necessary first-time steps.
Each weak form was evaluated using a single backend/optimization combination
several times according to the repeat parameter for various mesh sizes and
function approximation orders, see Tab.~\ref{tab:cdc}. To omit the additional
setup time of the first evaluation, which is amortized e.g. in practical
time-dependent problems, the mean elapsed times reported below are the means
without the worst case, denoted $\twwmean$, the worst case being the largest
elapsed time among each set of repeated evaluations. This is important
especially for the JAX-based backends because of the JIT-compilation during the
first call. On the other hand, the memory consumption is given as the maximum
over the repeated evaluations, denoted $\mmax$, because it determines whether a
problem can be calculated within available RAM or not. Although not included in
the paper, the complete data \cite{zenodo-time-tensors-data} contain all the
other statistics (means, minima, maxima, ranks etc.)

\subsection{Reference Implementation Performance}
\label{sec:rip}

\begin{figure}[ht!]
  \centering
  \includegraphics[width=0.48\linewidth]{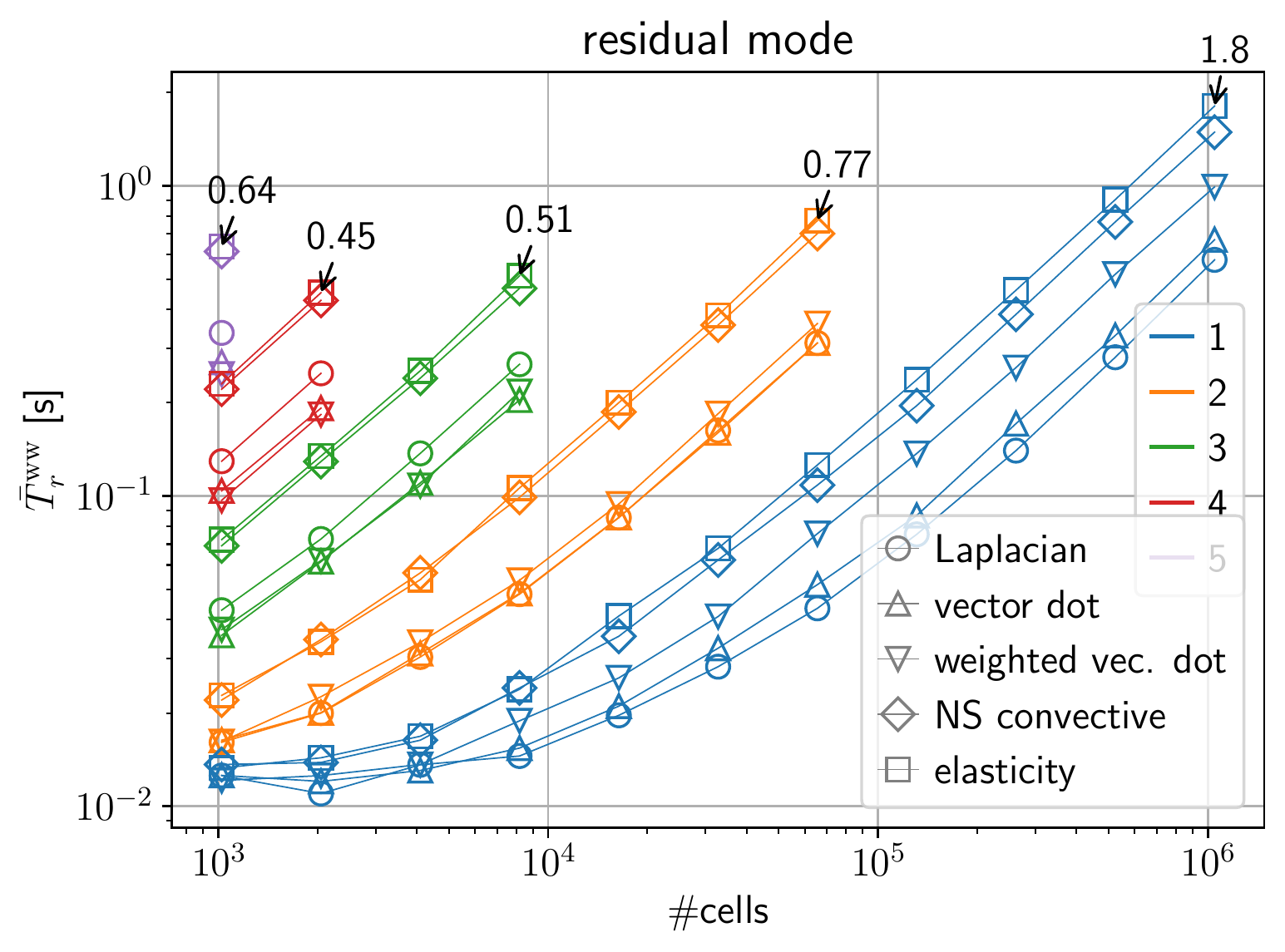}
  \includegraphics[width=0.48\linewidth]{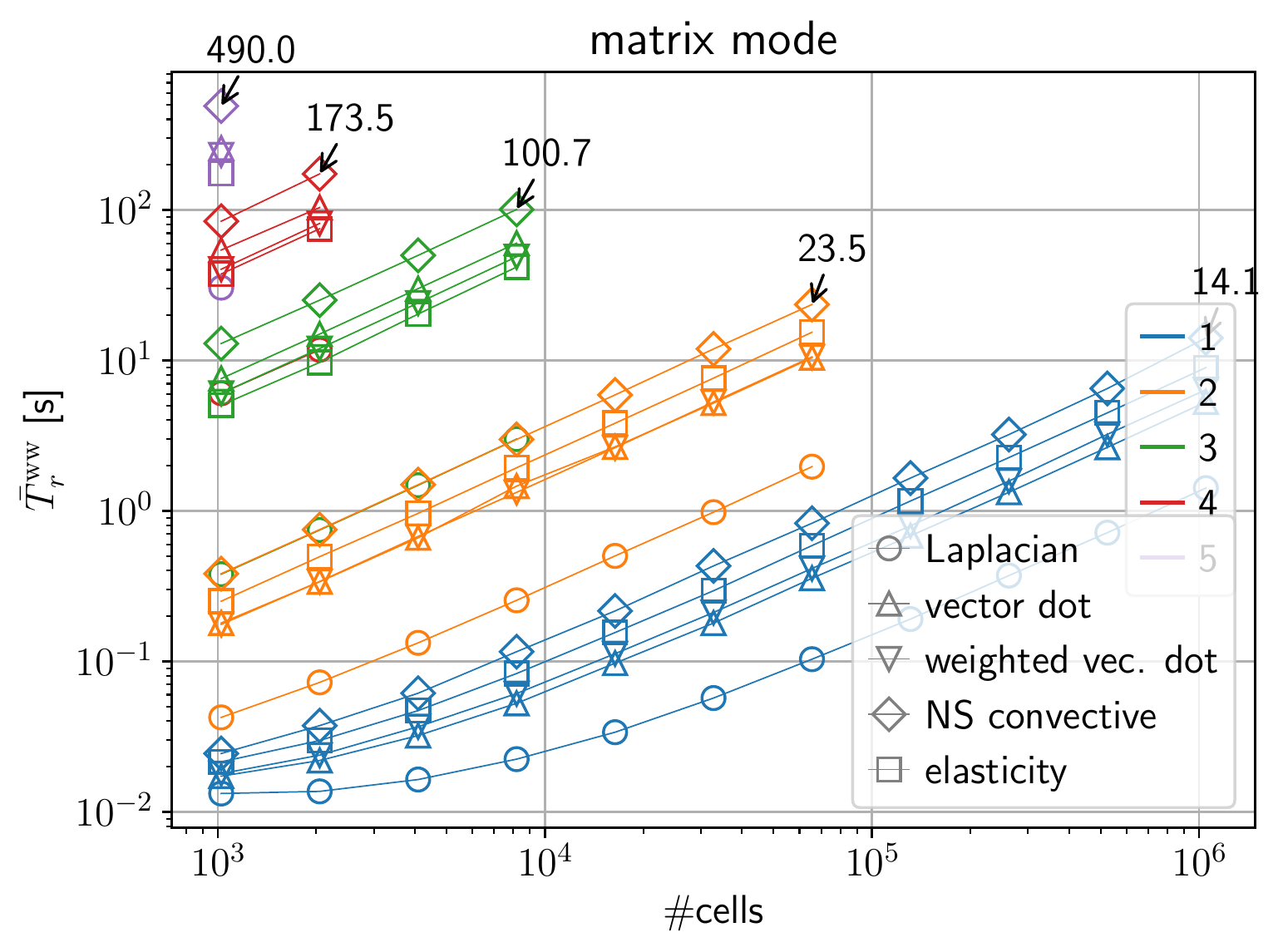}
  \includegraphics[width=0.48\linewidth]{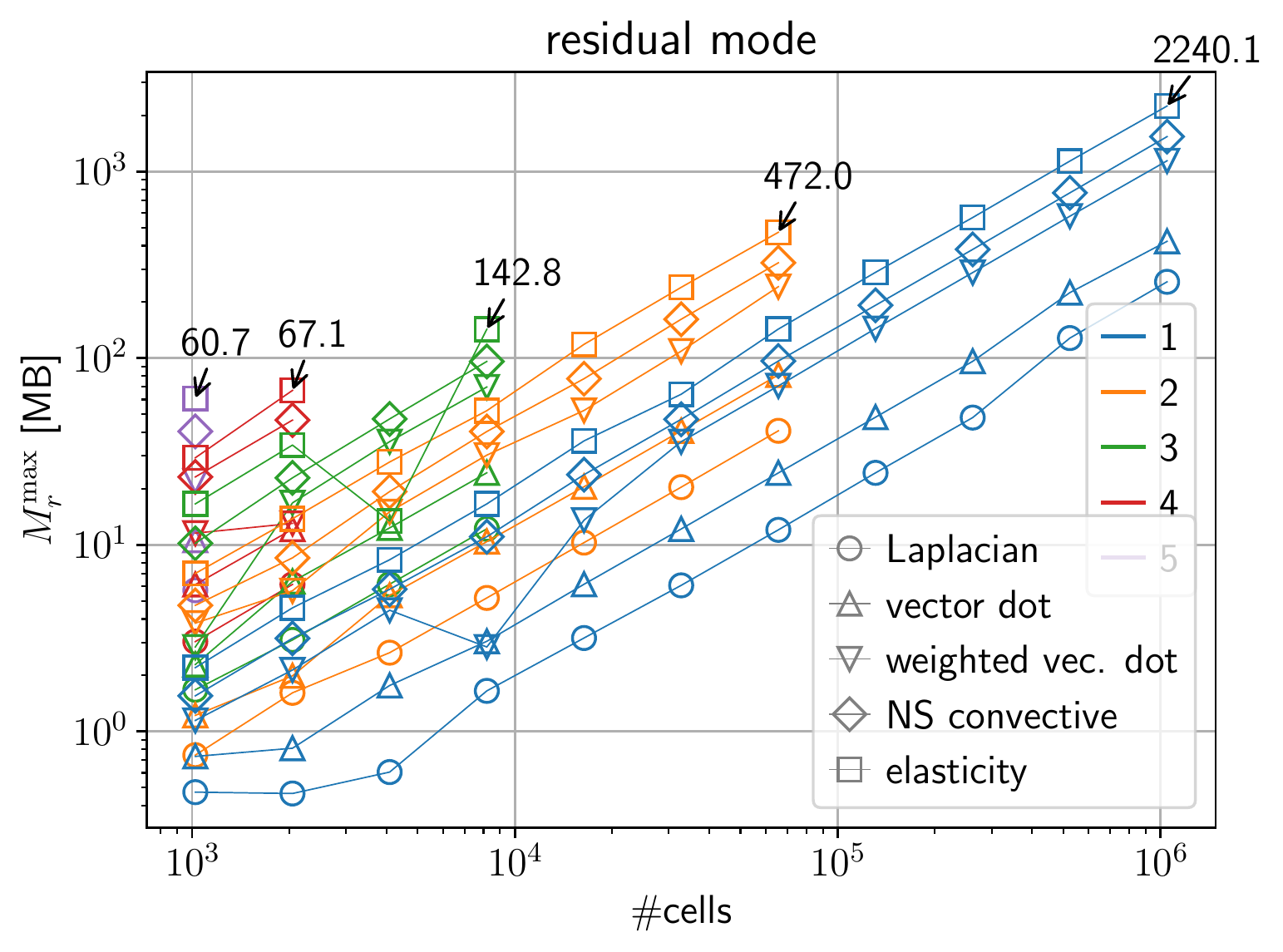}
  \includegraphics[width=0.48\linewidth]{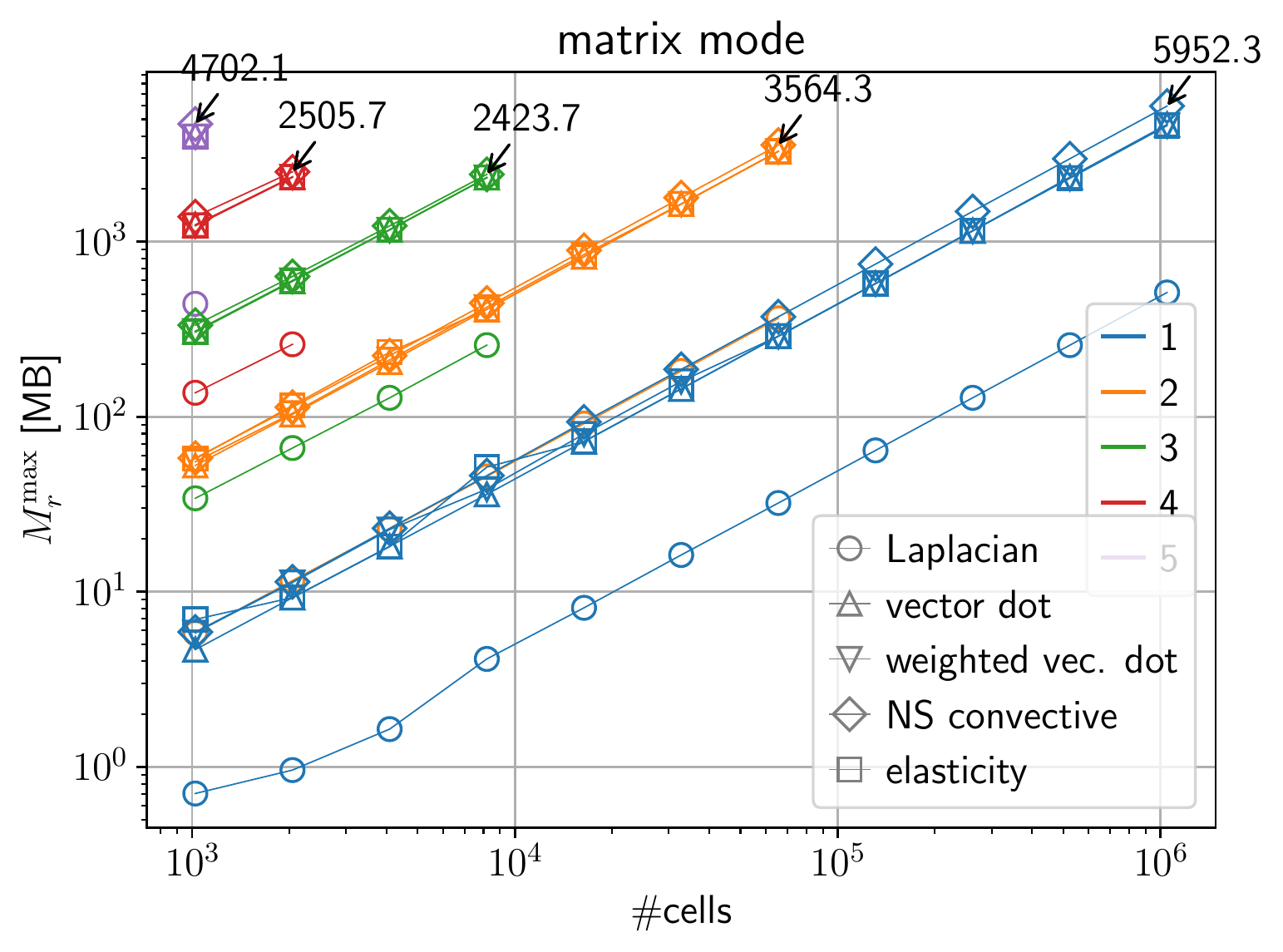}
  \caption{Reference implementation performance in evaluating the five selected
    weak forms denoted by markers. Left: residual mode, right: matrix mode,
    top: the elapsed time mean without the worst case $\twwmeanr$, bottom: the
    maximum memory consumption with respect to the repeat parameter $\mmaxr$.
    The highest values for each color-coded approximation order are annotated
    by arrows.}
  \label{fig:rip}
\end{figure}

The reference implementation performance in evaluating the five selected weak
forms is illustrated in Fig.~\ref{fig:rip}. The reference results shown, i.e.
the elapsed time mean without the worst case $\twwmeanr$~[s] and the maximum
memory consumption $\mmaxr$~[MB], are with respect to the repeat parameter in
Tab.~\ref{tab:cdc}.

The results demonstrate a linear dependence of the elapsed times as well as the
memory consumption on the number of mesh cells. For small meshes and the
approximation order 1 it is apparent --- especially in the residual mode, where
the resulting arrays are much smaller than in the matrix mode --- that there is
a small fixed time overhead and that this overhead is amortized for more than
$10^4$ cells. Similarly, the memory sampling measurements are not very accurate
for very fast evaluations/small data even with the sampling period of 0.001
seconds, however, for larger data (e.g. the matrix mode) the measurements are
reliable enough for comparison purposes. The relative data in subsequent
sections are w.r.t. the results shown in Fig.~\ref{fig:rip}.

\paragraph{Comparison with FEniCS}

\begin{figure}[ht!]
  \centering
  Laplacian \\
  \includegraphics[width=0.48\linewidth]{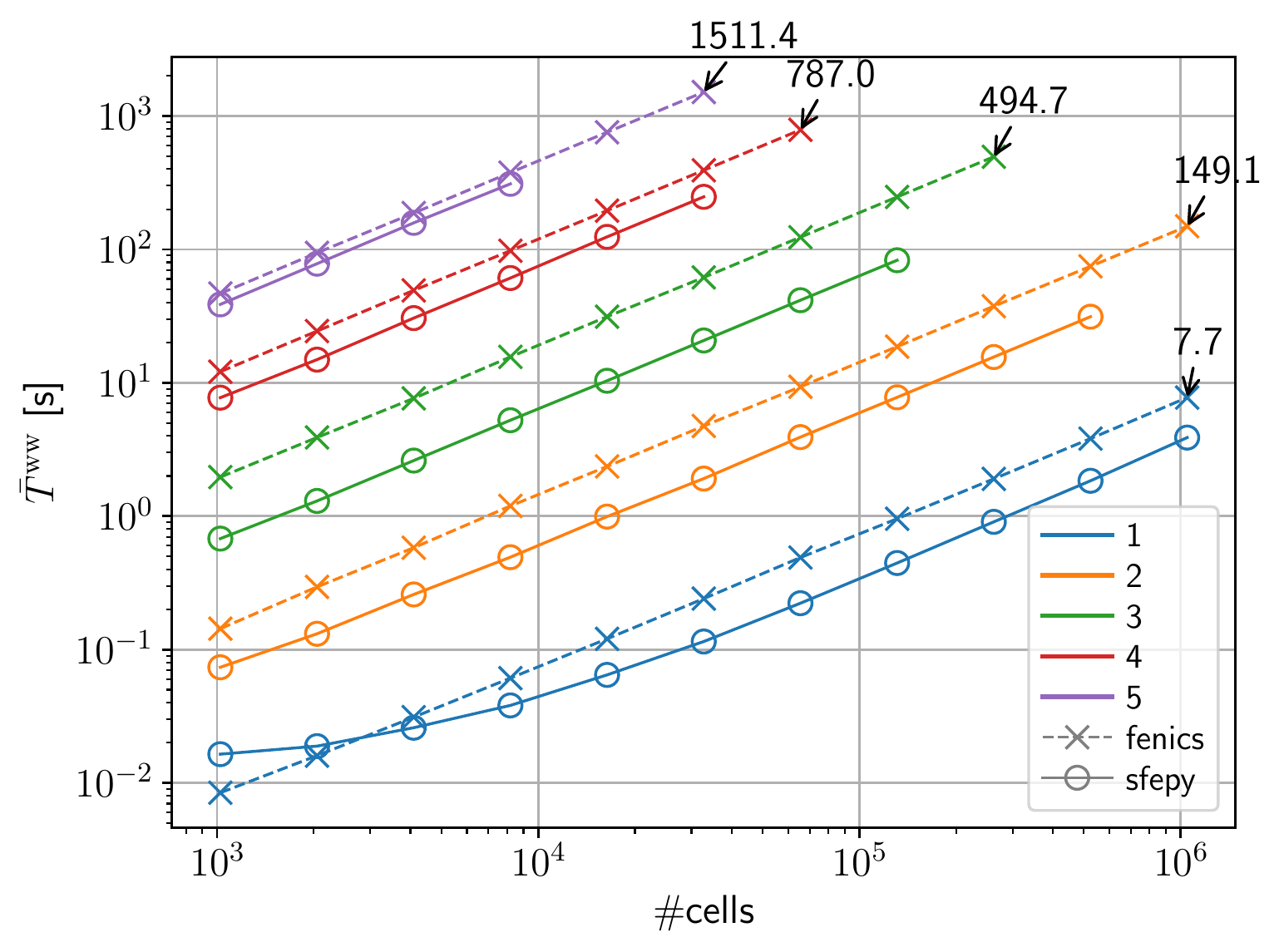}
  \includegraphics[width=0.48\linewidth]{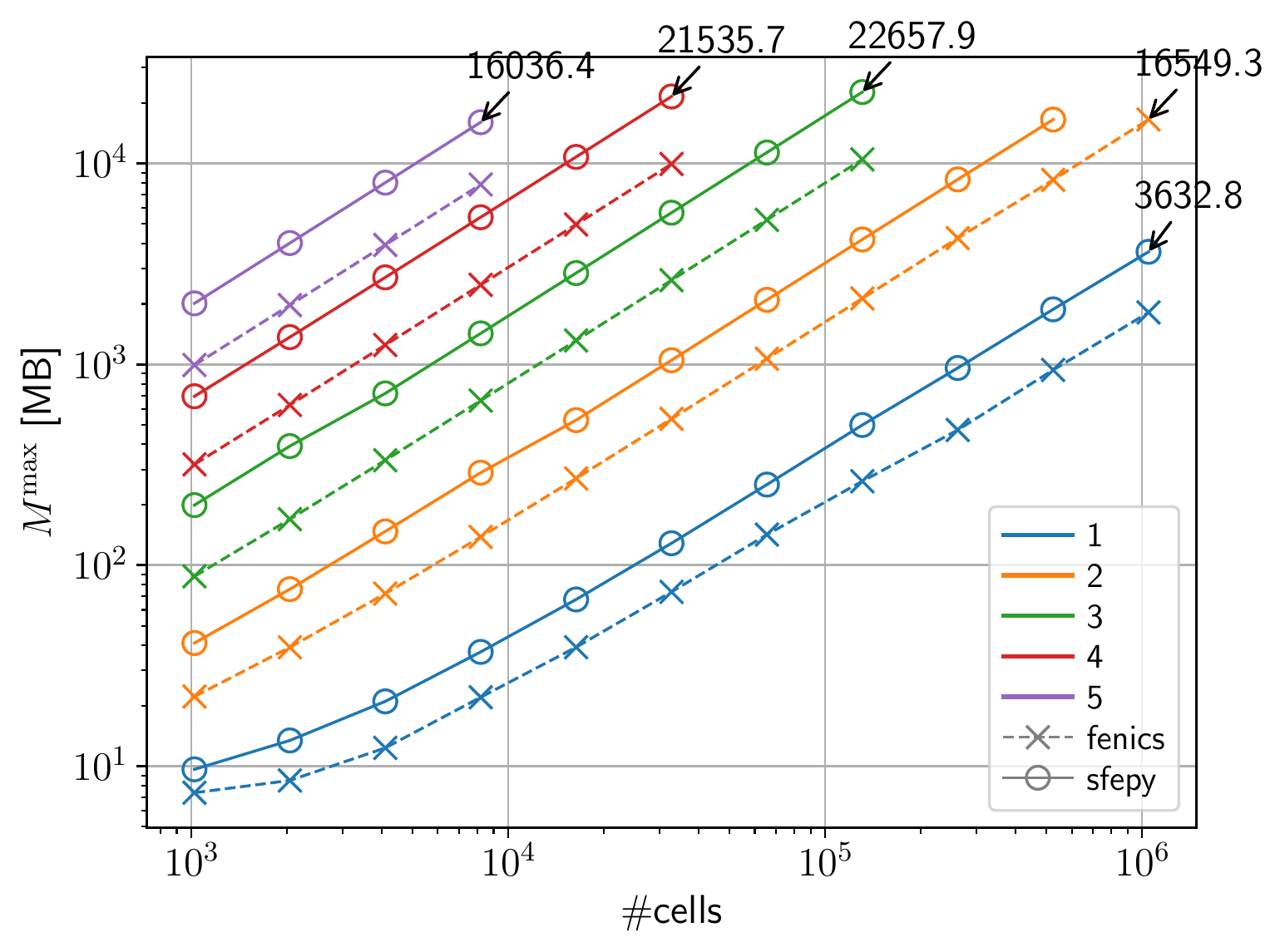}
  NS convective \\
  \includegraphics[width=0.48\linewidth]{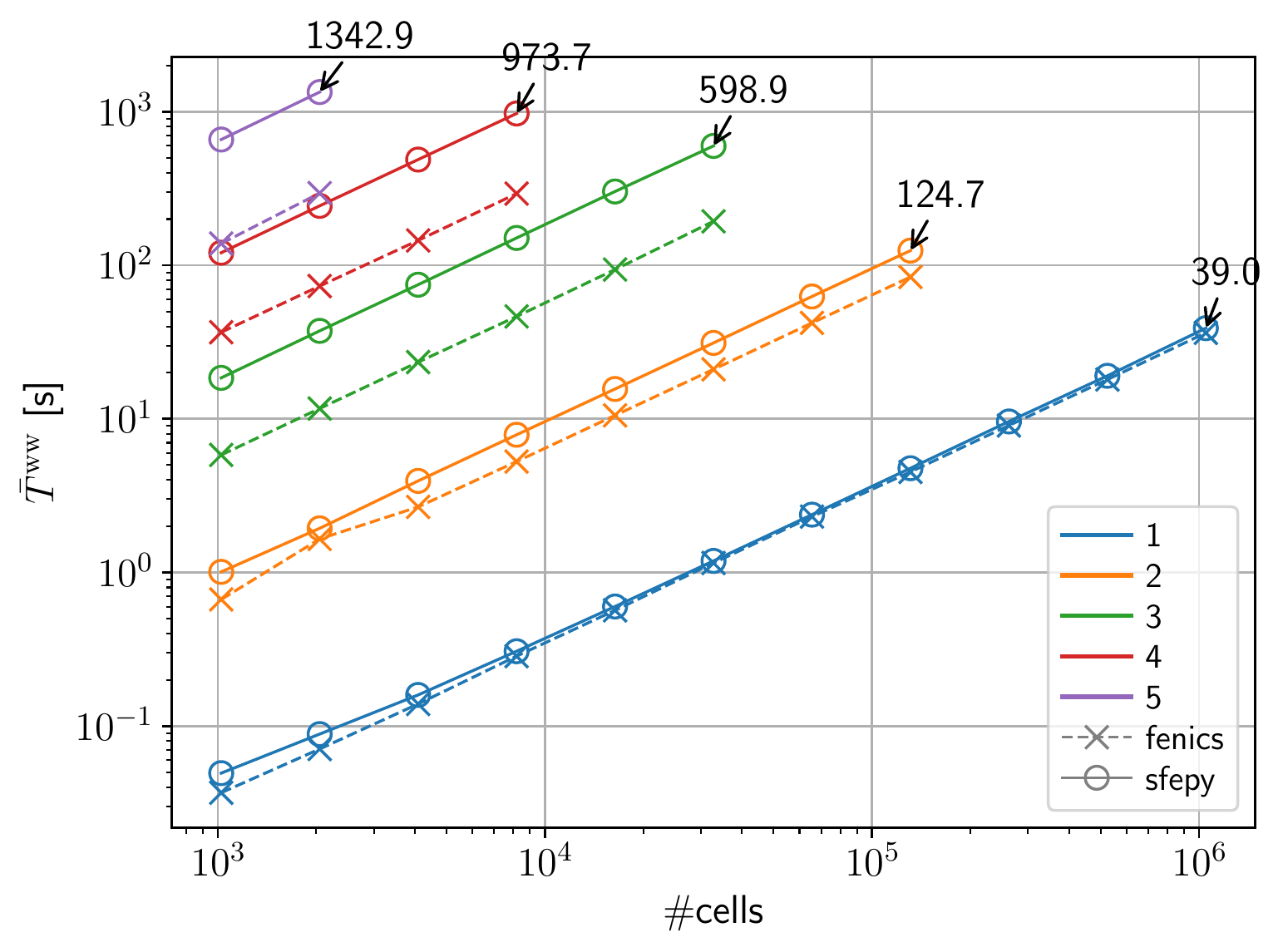}
  \includegraphics[width=0.48\linewidth]{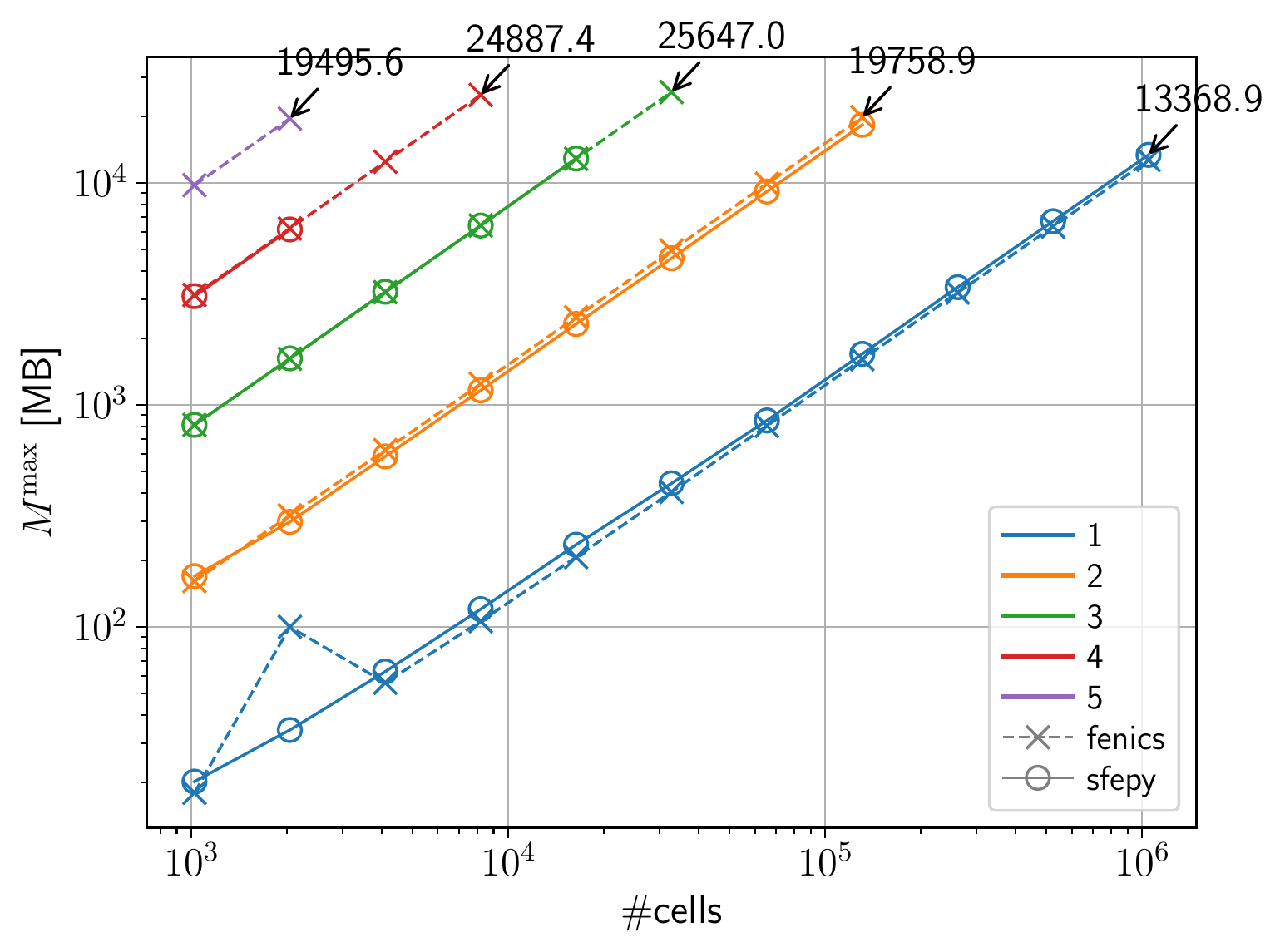}
  \caption{Performance of matrix mode evaluations of the SfePy reference
    implementation (circles) in comparison with FEniCS (crosses) for various
    approximation orders. Left: the elapsed time mean without the worst case
    $\twwmean$, right: the memory consumption $\mmax$. Top: the weak Laplacian,
    bottom: the convective term weak form. The highest values for each
    color-coded approximation order are annotated by arrows.}
  \label{fig:ripf}
\end{figure}

The aim of this auxiliary study is not to provide a comprehensive comparison,
indeed the results discussed below were obtained using a single
software/hardware configuration, but to provide a broader context of the our
main results in the following sections. We compare the serial performance of
the reference implementation in SfePy with a widely acknowledged and used
package FEniCS~\cite{Logg_Mardal_Wells_2012}, version 2019.1.0. The weak
Laplacian and convective term weak forms were used for the comparison. The
results presented in Fig.~\ref{fig:ripf} show the elapsed time mean without the
worst case $\twwmean$~[s] with respect to five repetitions and the maximum
memory consumption $\mmax$~[MB] (max. - min. consumption over all the
repetitions together). Unlike results in Fig.~\ref{fig:rip}, the measurements
correspond to combined weak form data allocation, evaluation and assembling to
a sparse matrix, i.e. the whole procedure of FEM application to a weak form.
The memory sampling measurements were done with the sampling period of 0.001
seconds. The results are summarized in Tab.~\ref{tab:ripf}, where the ratios
SfePy/FEniCS are given for all approximation orders, using medians over the
various numbers of mesh cells. Missing data are due to the time-out of 2000
seconds and/or insufficient memory.

\begin{table}[ht!]
  \centering
  {\small
    \begin{tabular}{lrrrrrrrrrr}
\toprule
{} & \multicolumn{5}{l}{Laplacian} & \multicolumn{5}{l}{NS convective} \\
{} &               1 &               2 &               3 &               4 &               5 &               1 &               2 &               3 &               4 &               5 \\
\midrule
med($\bar T^{\rm ww}_{\rm sfepy} / \bar T^{\rm ww}_{\rm fenics}$) & $\enspace 0.50$ & $\enspace 0.42$ & $\enspace 0.34$ & $\enspace 0.63$ & $\enspace 0.83$ & $\enspace 1.06$ & $\enspace 1.49$ & $\enspace 3.20$ & $\enspace 3.33$ & $\enspace 4.65$ \\
med($M^{\rm max}_{\rm sfepy} / M^{\rm max}_{\rm fenics}$)         & $\enspace 1.75$ & $\enspace 1.96$ & $\enspace 2.16$ & $\enspace 2.17$ & $\enspace 2.04$ & $\enspace 1.06$ & $\enspace 0.93$ & $\enspace 1.00$ & $\enspace 0.99$ &               - \\
\bottomrule
\end{tabular}

  }
  \caption{The SfePy/FEniCS ratios of the elapsed time mean without the worst
    case $\twwmean$ and the memory consumption $\mmax$ for the weak Laplacian
    and convective term weak forms for various approximation orders. The values
    correspond to medians over various numbers of mesh cells, see
    Fig.~\ref{fig:ripf}.}
  \label{tab:ripf}
\end{table}

The evaluation and assembling of the matrix corresponding to the weak Laplacian
is faster in SfePy, almost three times for the order 3: the $\twwmean$ ratios
in Tab.~\ref{tab:ripf} are decreasing up to the order 3 and then again
increasing. However, SfePy needed about two times more memory than FEniCS,
which caused calculation failures for several higher-order/large number of
cells cases (missing circles in Fig.~\ref{fig:ripf} top left). On the other
hand, in the case of the convective term the memory consumption of both
packages was comparable. The calculation speed was also essentially the same
for the order 1, but SfePy was getting progressively worse with the increasing
order: up to four and half times for the order 5, where SfePy timed-out and so
the corresponding memory consumption data points are missing. Analogous
behavior was observed also for the vector dot form (not shown here for
brevity). To conclude, SfePy assembling performance seemed on par with FEniCS
for low order approximations, and worse for higher order approximations with
about the same memory requirements. Faster SfePy performance when assembling
the weak Laplacian was connected with higher memory consumption. It is worth
noting that the weak form evaluation algorithm based on the developments of
Section~\ref{sec:wfet} offers a significant performance boost w.r.t. the SfePy
reference implementation especially for higher order approximations, as will be
demonstrated in the following text.

\subsection{Influence of Tensor Memory Layouts}
\label{sec:itml}

This section explores the influence of the various possible memory layouts
on the weak form evaluation performance, to establish the position of the
default, SfePy-compatible, layout within other possible layouts. The default
layout elapsed time mean without the worst case is denoted by $\twwmeand$.

The default layout specification is constructed as follows. The scalar basis
function gradient tensor has 4 dimensions (cells, quadrature points, gradient
components, DOFs) and the layout \Verb|cqgd| ($L_1$). The DOFs tensor has 3
dimensions (cells, variable components, cell DOFs) with the layout \Verb|cvd|
($L_2$). Material parameter tensors default layout is \Verb|cq0| ($L_3$), where
\Verb|0| is a placeholder for any number of axes. Compressing the three
specifications into a single string yields the default specification
\Verb|cqgvd0|. Other possible memory layouts of the weak form einsum expression
operands were generated by permuting the layout specification letters of $L_1$,
$L_2$, $L_3$ and compressing them together, leading to $216$ different
layouts\footnote{This is less than $4! 3! 3!$ because the operand layouts $L_i$
  share letters.}, that were applied to the operands of the matrix mode einsum
expression of the weak Laplacian and convective terms.

\subsubsection{Weak Laplacian Layouts}
\label{sec:wll}

The results below were obtained by a parametric study involving 1461
calculations run using soops \cite{soops2021github}, all with the mesh of 8192
cells. The FE approximation order 3 runs had a time-out set to 50 seconds for 4
repetitions, the order 1 and 2 runs used 6 and 5 repetitions, respectively, and
no time-out. The number of runs that timed-out is excluded from the
calculations count.

The dependence of the weak Laplacian einsum expression evaluation performance
on the operand memory layouts is first summarized in Tab.~\ref{tab:flts-l-m},
where the fastest layouts for each FE approximation order and einsum backend
are listed, together with $\rtwwmean$ and $\dtwwmean$. The layouts are
specified using permutations of the reduced default layout \Verb|cqgd|, because
the FE space is scalar (no \Verb|v|) and the weak form has no material
parameters (no \Verb|0|).

\begin{figure}[htp!]
  \centering
  \includegraphics[width=0.48\linewidth]{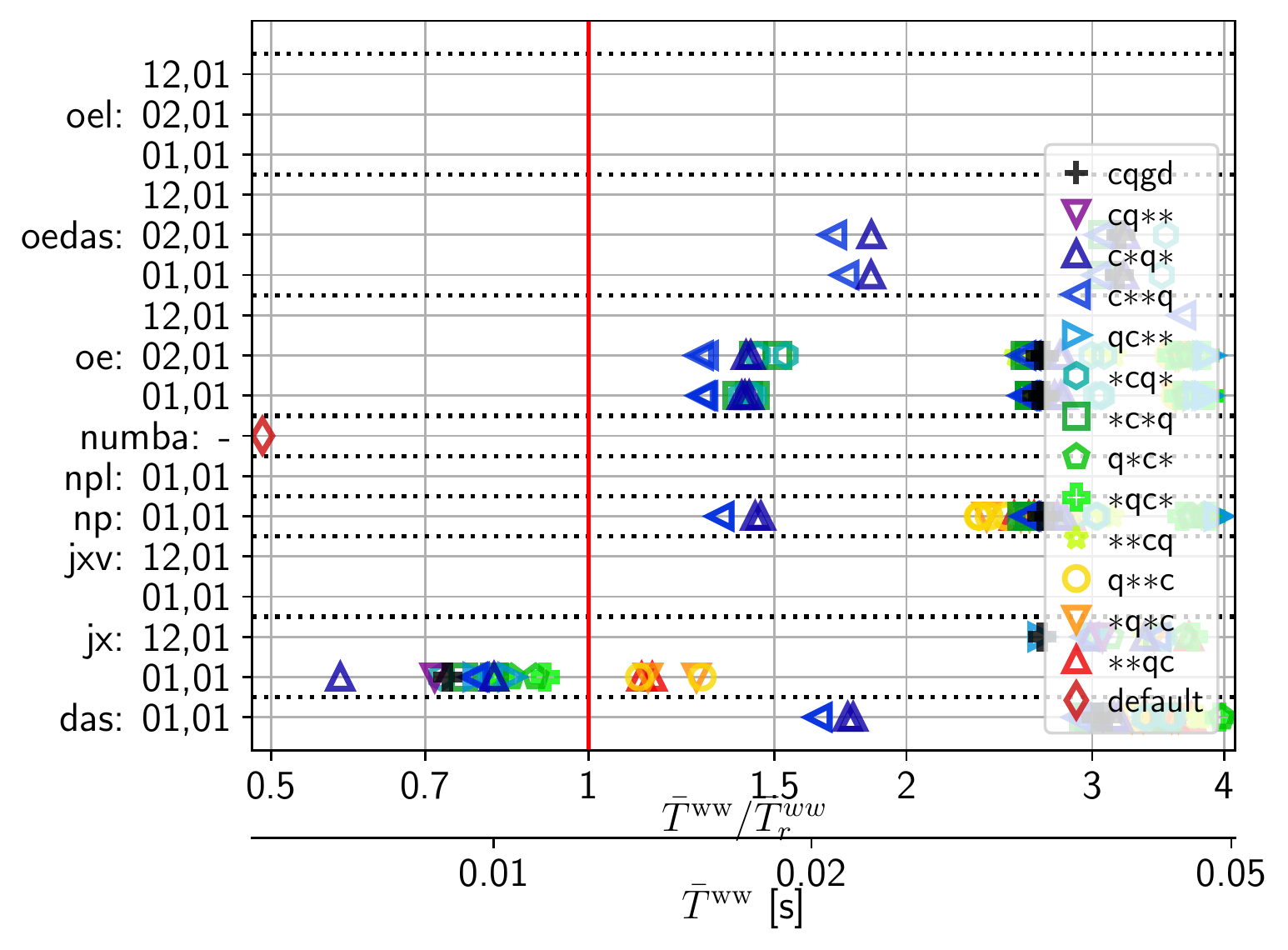}
  \includegraphics[width=0.48\linewidth]{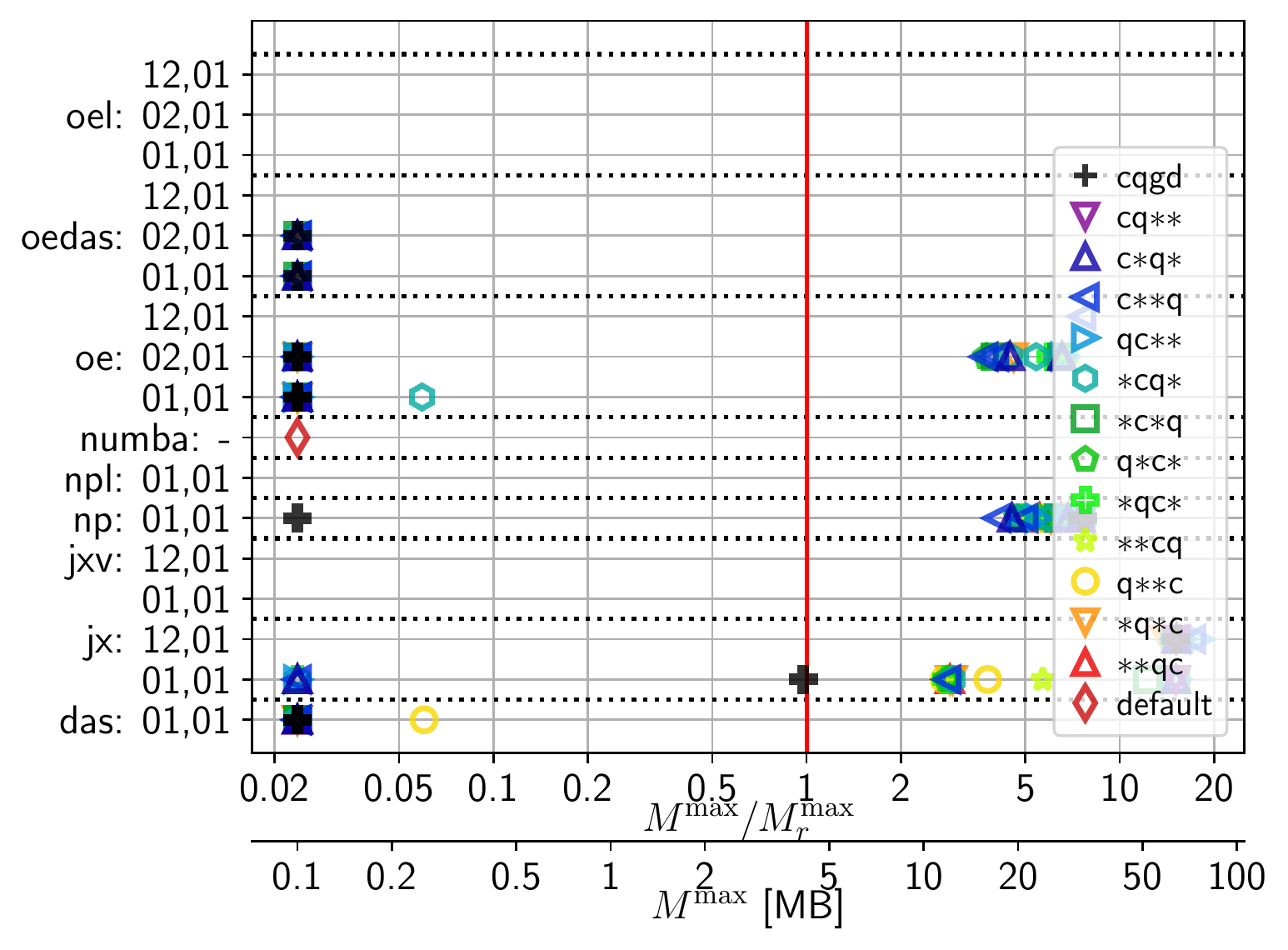}

  \includegraphics[width=0.48\linewidth]{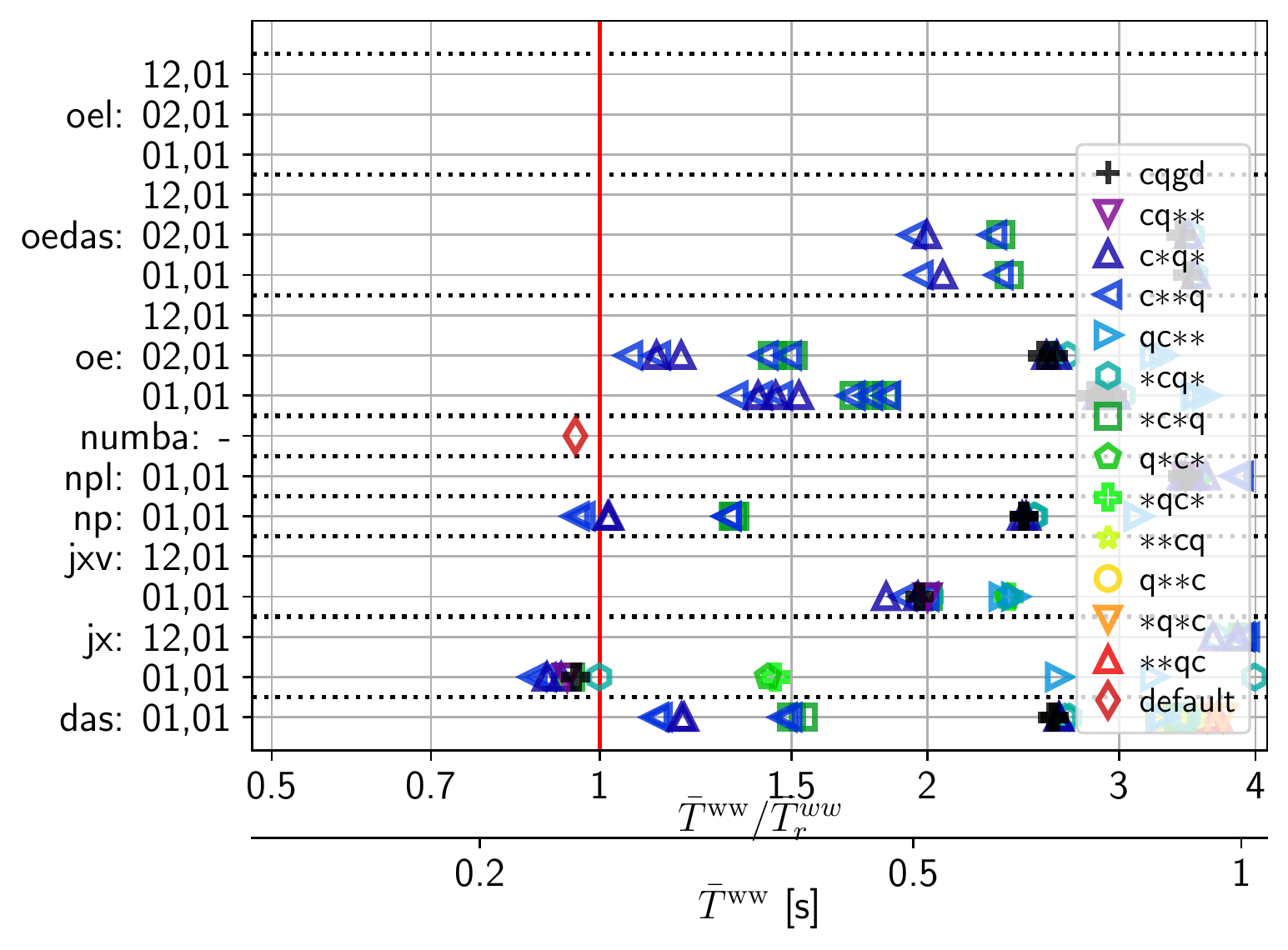}
  \includegraphics[width=0.48\linewidth]{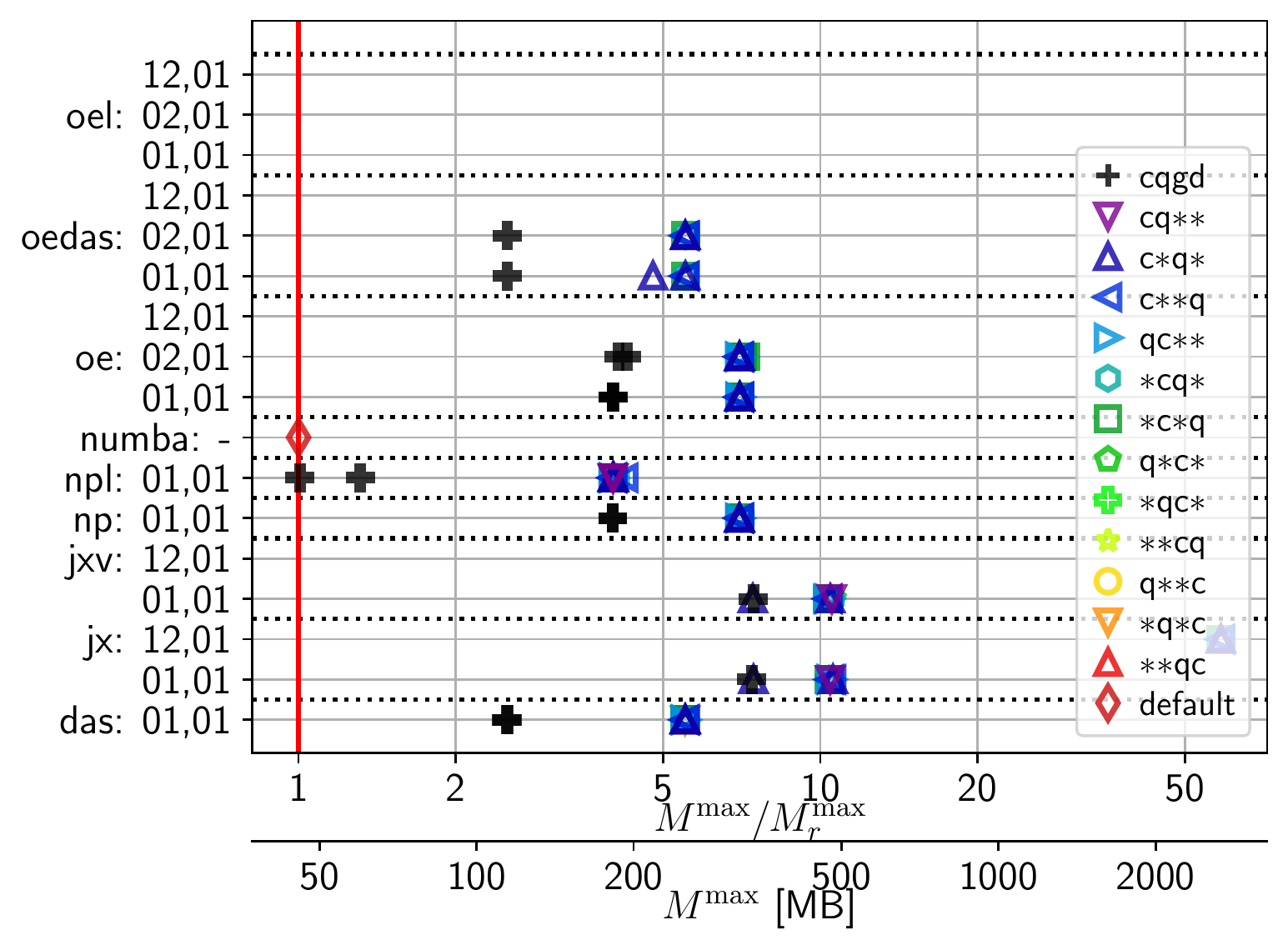}

  \includegraphics[width=0.48\linewidth]{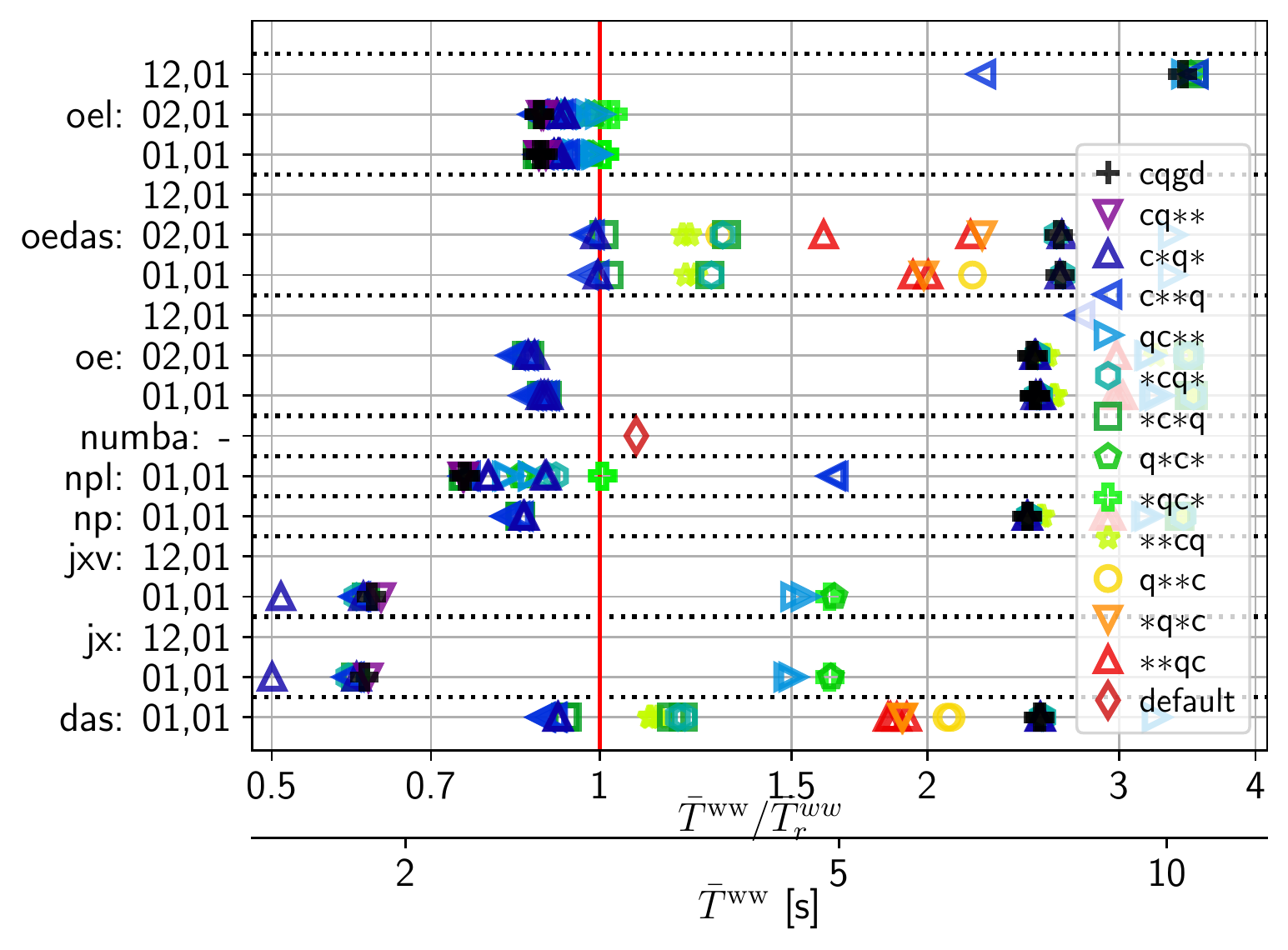}
  \includegraphics[width=0.48\linewidth]{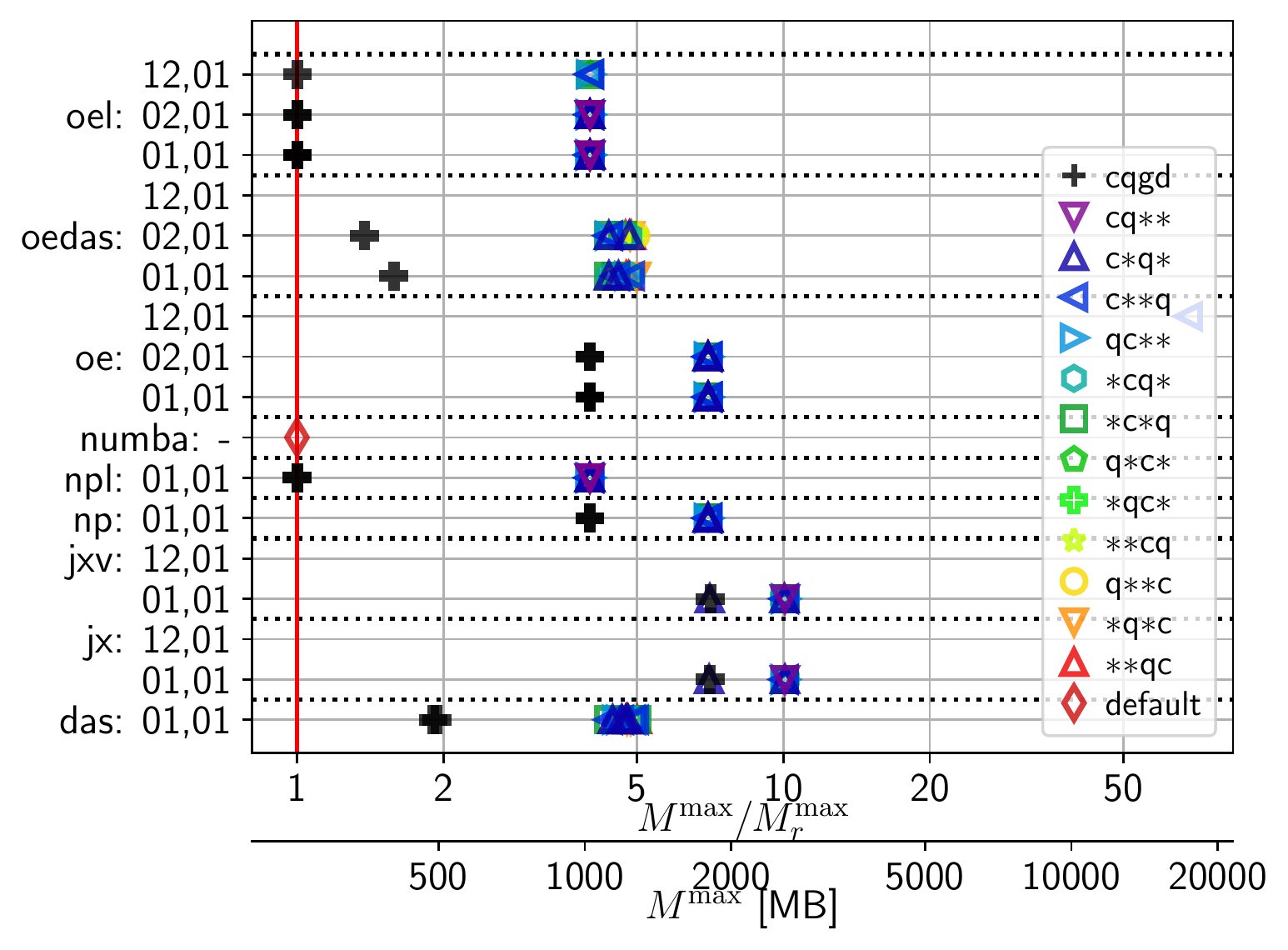}
  \caption{Elapsed time and memory requirements of the weak Laplacian einsum
    expression depending on the operand memory layouts for 8192 cells, FE
    approximation orders 1 (top), 2 (middle), 3 (bottom). Left: (relative)
    elapsed time mean without the worst case ($\rtwwmean$, $\twwmean$). Right:
    (relative) maximum memory consumption ($\rmmax$, $\mmax$). The reference
    values obtained by SfePy are marked by vertical red lines. The libraries
    and contraction paths are indicated in $y$ axis labels. The colors/markers
    indicate positions of cell and quadrature axes in the layout string, the
    SfePy layout is marked by black crosses among einsum expression backends.}
  \label{fig:layouts-laplace-1}
\end{figure}

Elapsed time and memory requirements of the individual backends are shown in
Fig.~\ref{fig:layouts-laplace-1} for the FE approximation orders 1 (top), 2
(middle) and 3 (bottom). Both $\rtwwmean$ with $\twwmean$ and $\rmmax$ with
$\mmax$ are provided. The libraries and contraction paths are indicated in $y$
axis labels. The contraction paths are used instead of the optimization
algorithms because the paths actually significantly influence the performance,
and furthermore the optimization time was negligible. The paths are written in
a condensed way, where e.g. $12.01$ corresponds to the $[(1, 2), (0, 1)]$, see
Section~\ref{sec:wfete}. Note that several optimization algorithms can
lead to a single contraction path, so each (library, path) pair corresponds to
the number of calculations equal to 216 layouts times the number of
optimizations with the given path. However, only the values at most $4\times$
slower than the reference SfePy implementation are shown in
Fig.~\ref{fig:layouts-laplace-1}, with the corresponding memory consumption.
To reduce the color/marker legend size, only positions of \Verb|cq| axes are
indicated in the legend, \Verb|*| denotes any letter from \Verb|gd|.
For comparison, a hand-written weak Laplacian element matrix evaluation loop
implementation, JIT-compiled using Numba (version 0.52.0)
\cite{Lam_Pitrou_Seibert_2015}, was included in this study.

\begin{table}[ht!]
  \centering
  {
    \begin{tabular}{rrlll}
\toprule
order &   lib &     $\bar T^{ww} / \bar T^{ww}_r$ & layout & $\bar T^{ww} / \bar T^{ww}_d$ \\
\midrule
    1 &   das & $\enspace \enspace \enspace 1.65$ &   cdgq &                          0.54 \\
      &    jx & $\enspace \enspace \enspace 0.58$ &   cgqd &                          0.79 \\
      &   jxv &         $\enspace \enspace 12.91$ &   cdgq &                          0.54 \\
      &    np & $\enspace \enspace \enspace 1.33$ &   cdgq &                          0.50 \\
      &   npl &         $\enspace \enspace 57.76$ &   cgqd &                          0.95 \\
      &    oe & $\enspace \enspace \enspace 1.27$ &   cdgq &                          0.48 \\
      & oedas & $\enspace \enspace \enspace 1.70$ &   cdgq &                          0.54 \\
      &   oel &         $\enspace \enspace 60.52$ &   gcdq &                          0.79 \\
    2 &   das & $\enspace \enspace \enspace 1.12$ &   cdgq &                          0.43 \\
      &    jx & $\enspace \enspace \enspace 0.87$ &   cdgq &                          0.87 \\
      &   jxv & $\enspace \enspace \enspace 1.83$ &   cgqd &                          0.91 \\
      &    np & $\enspace \enspace \enspace 0.95$ &   cdgq &                          0.39 \\
      &   npl & $\enspace \enspace \enspace 3.42$ &   cgqd &                          0.98 \\
      &    oe & $\enspace \enspace \enspace 1.06$ &   cdgq &                          0.41 \\
      & oedas & $\enspace \enspace \enspace 1.94$ &   cdgq &                          0.56 \\
      &   oel & $\enspace \enspace \enspace 4.87$ &   cqgd &                          0.99 \\
    3 &   das & $\enspace \enspace \enspace 0.88$ &   cdgq &                          0.35 \\
      &    jx & $\enspace \enspace \enspace 0.50$ &   cgqd &                          0.78 \\
      &   jxv & $\enspace \enspace \enspace 0.51$ &   cgqd &                          0.81 \\
      &    np & $\enspace \enspace \enspace 0.82$ &   cdgq &                          0.33 \\
      &   npl & $\enspace \enspace \enspace 0.75$ &   gcdq &                          0.99 \\
      &    oe & $\enspace \enspace \enspace 0.83$ &   cdgq &                          0.33 \\
      & oedas & $\enspace \enspace \enspace 0.97$ &   cdgq &                          0.36 \\
      &   oel & $\enspace \enspace \enspace 0.87$ &   cgdq &                          0.98 \\
\bottomrule
\end{tabular}

  }
  \caption{The fastest layouts per FE approximation order and backend in the
    case of the weak Laplacian einsum expression, $\dtwwmean$ terms the
    relative speed-up of the fastest layout w.r.t. the default layout.}
  \label{tab:flts-l-m}
\end{table}

\subsubsection{Weak Convective Term Layouts}
\label{sec:wctl}

The results below were obtained by a parametric study involving 3602
calculations run using soops \cite{soops2021github}, all with the mesh of 8192
cells. The FE approximation order 1 and 2 runs used 6 and 5 repetitions,
respectively, and no time-out.

The dependence of the weak convective term expression evaluation performance on
the operand memory layouts is first summarized in Tab.~\ref{tab:flts-c-m},
where the fastest layouts for each FE approximation order and einsum backend
are listed, together with $\rtwwmean$ and $\dtwwmean$. The layouts are
specified using permutations of the reduced default layout \Verb|cqgvd|,
because the weak form has no material parameters (no \Verb|0|).

\begin{figure}[htp!]
  \centering
  \includegraphics[width=0.48\linewidth]{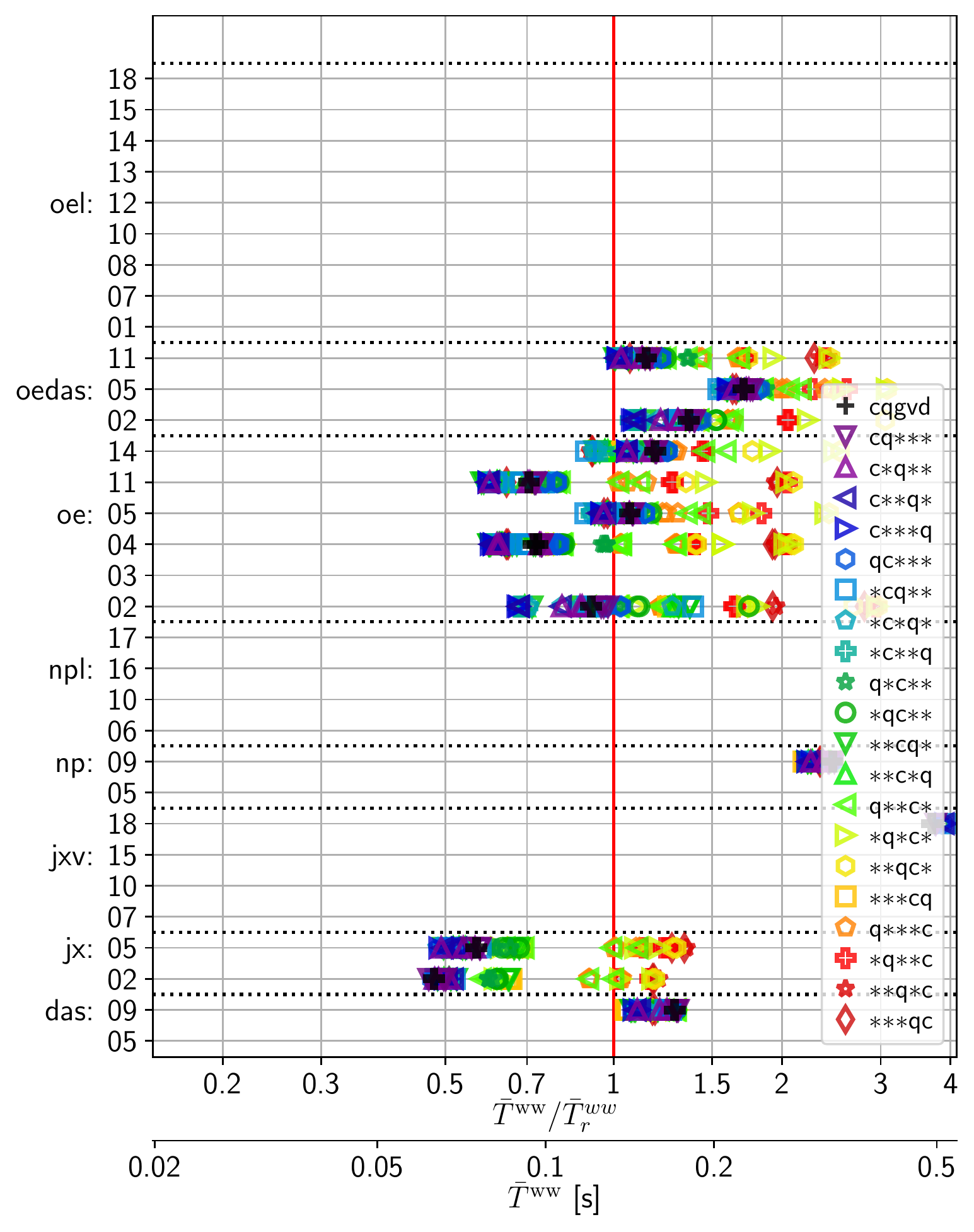}
  \includegraphics[width=0.48\linewidth]{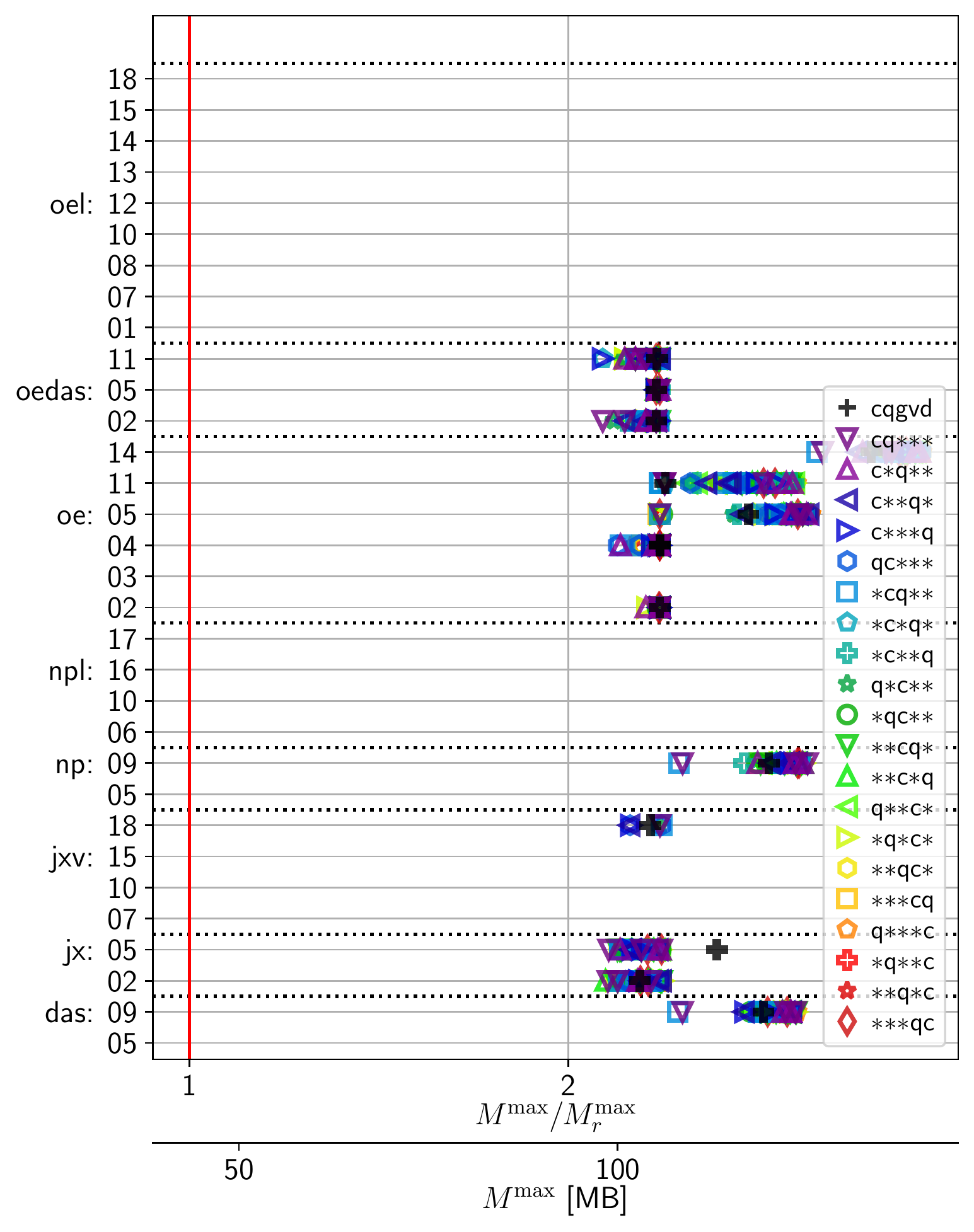}

  \includegraphics[width=0.48\linewidth]{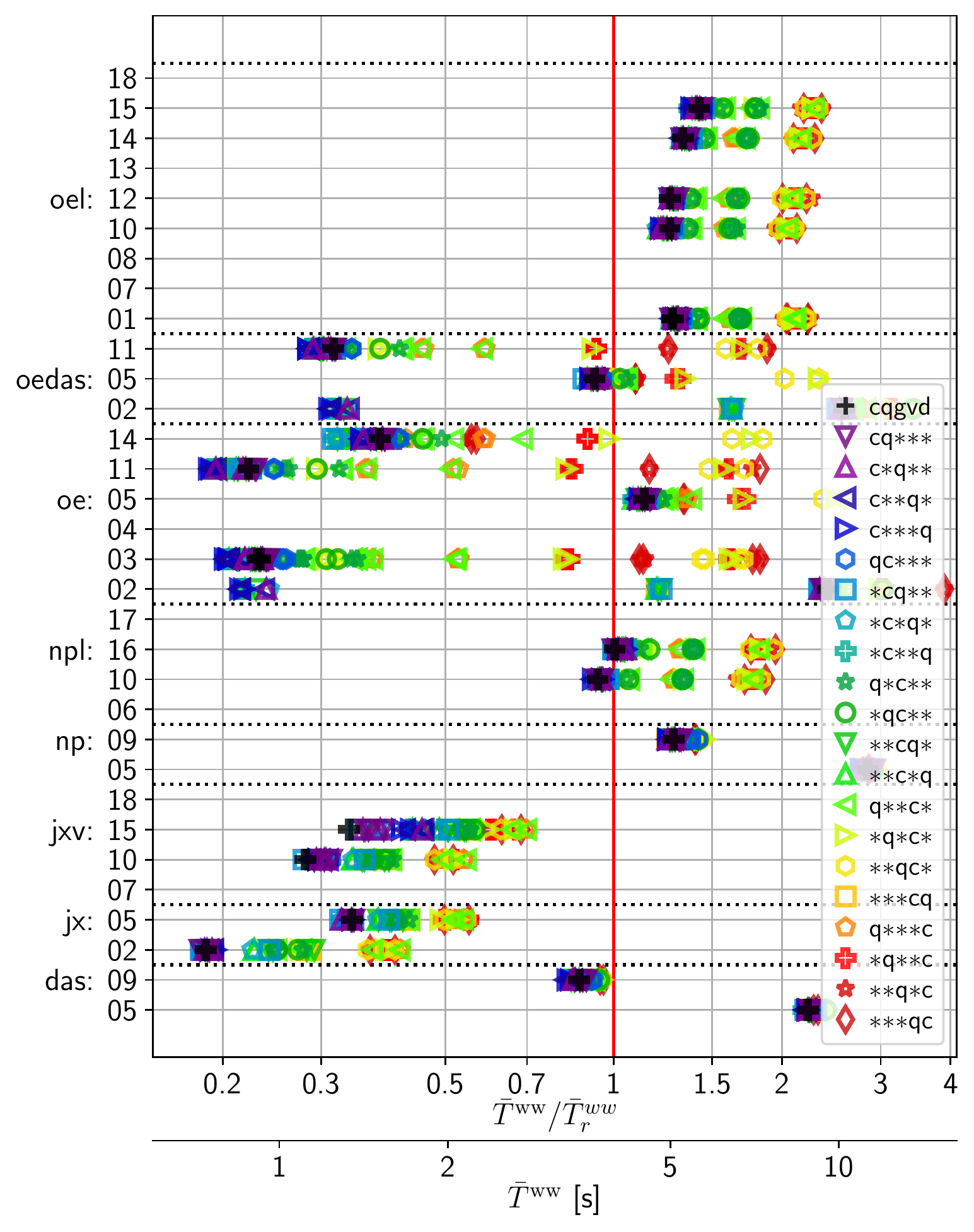}
  \includegraphics[width=0.48\linewidth]{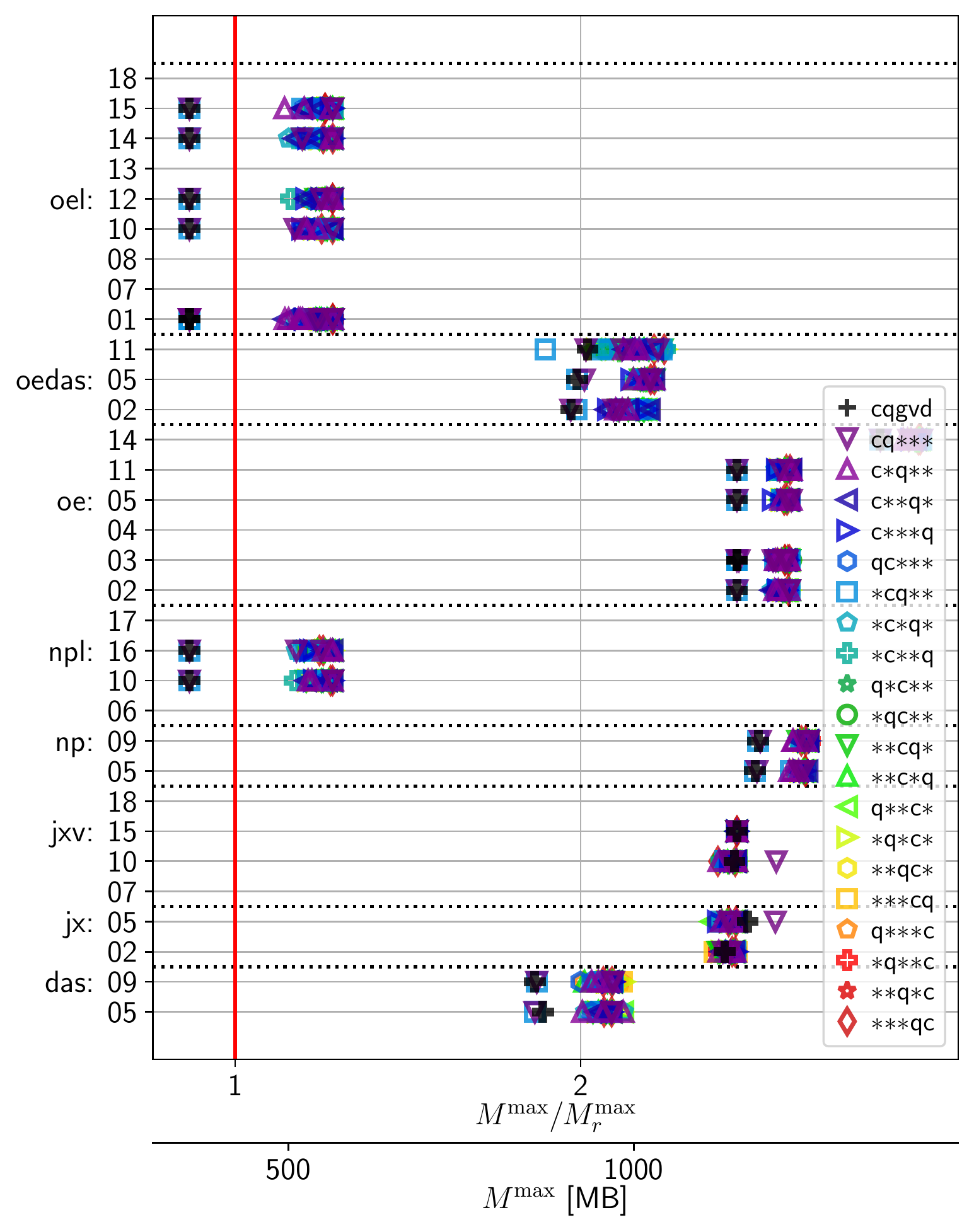}

  \caption{Elapsed time and memory requirements of the weak convective term
    einsum expression depending on the operand memory layouts for 8192 cells,
    FE approximation orders 1 (top), 2 (bottom). Left: (relative) elapsed time
    mean without the worst case ($\rtwwmean$, $\twwmean$). Right: (relative)
    maximum memory consumption ($\rmmax$, $\mmax$). The reference values
    obtained by SfePy are marked by vertical red lines. The libraries and
    contraction paths are indicated in $y$ axis labels, see
    Tab.~\ref{tab:short-paths}. The colors/markers indicate positions of cell
    and quadrature axes in the layout string, the SfePy layout is marked by
    black crosses among einsum expression backends.}
  \label{fig:layouts-convect-1}
\end{figure}

Elapsed time and memory requirements of the individual backends are shown in
Fig.~\ref{fig:layouts-convect-1} for the FE approximation orders 1 (top), 2
(bottom). Both $\rtwwmean$ with $\twwmean$ and $\rmmax$ with $\mmax$ are shown.
The libraries and contraction paths are indicated in $y$ axis labels. Again the
values at most $4\times$ slower than the reference SfePy implementation are
shown, with the corresponding memory consumption. To reduce the color/marker
legend size, only positions of \Verb|cq| axes are indicated in the legend,
\Verb|*| denotes any letter from \Verb|gvd|. Abbreviations given in
Tab.~\ref{tab:short-paths} are used to save space, the plus sign denotes
addition of two einsum expressions due to the convective term nonlinearity.

\begin{table}[ht!]
  \centering
  {
    \begin{tabular}{rrlll}
\toprule
order &   lib &     $\bar T^{ww} / \bar T^{ww}_r$ & layout & $\bar T^{ww} / \bar T^{ww}_d$ \\
\midrule
    1 &   das & $\enspace \enspace \enspace 1.05$ &  vdgcq &                          0.82 \\
      &    jx & $\enspace \enspace \enspace 0.47$ &  cqgdv &                          0.98 \\
      &   jxv & $\enspace \enspace \enspace 3.71$ &  cqgvd &                          1.00 \\
      &    np & $\enspace \enspace \enspace 2.18$ &  gvdcq &                          0.89 \\
      &   npl &         $\enspace \enspace 25.73$ &  cdgqv &                          1.00 \\
      &    oe & $\enspace \enspace \enspace 0.58$ &  vdcqg &                          0.82 \\
      & oedas & $\enspace \enspace \enspace 1.00$ &  dvcqg &                          0.88 \\
      &   oel &         $\enspace \enspace 32.74$ &  cqgdv &                          0.99 \\
    2 &   das & $\enspace \enspace \enspace 0.82$ &  cgdqv &                          0.95 \\
      &    jx & $\enspace \enspace \enspace 0.18$ &  cqgdv &                          0.96 \\
      &   jxv & $\enspace \enspace \enspace 0.28$ &  vcqgd &                          0.99 \\
      &    np & $\enspace \enspace \enspace 1.25$ &  gdvcq &                          0.98 \\
      &   npl & $\enspace \enspace \enspace 0.92$ &  cvdgq &                          0.98 \\
      &    oe & $\enspace \enspace \enspace 0.19$ &  cgvdq &                          0.85 \\
      & oedas & $\enspace \enspace \enspace 0.28$ &  cgvdq &                          0.90 \\
      &   oel & $\enspace \enspace \enspace 1.20$ &  gcdqv &                          0.95 \\
\bottomrule
\end{tabular}

  }
  \caption{The fastest layouts per FE approximation order and backend in the
    case of the weak convective term einsum expression, $\dtwwmean$ terms the
    relative speed-up of the fastest layout w.r.t. the default layout.}
  \label{tab:flts-c-m}
\end{table}

\begin{table}[htp!]
  \centering
  {
  \begin{tabular}{ll}
\toprule
                  contraction paths & abbreviations \\
\midrule
                                  - &             - \\
01,02,12,03,02,01+01,02,04,13,02,01 &            01 \\
01,02,12,03,02,01+06,04,02,03,02,01 &            02 \\
03,34,34,03,01,01+03,12,01,12,02,01 &            03 \\
03,34,34,03,01,01+15,13,14,03,02,01 &            04 \\
03,34,34,03,01,01+34,05,14,23,01,01 &            05 \\
24,34,04,13,02,01+01,02,04,02,01,01 &            06 \\
24,34,04,13,02,01+01,02,04,13,01,01 &            07 \\
24,34,04,13,02,01+05,05,02,03,02,01 &            08 \\
24,34,04,13,02,01+06,04,02,03,02,01 &            09 \\
24,34,04,13,02,01+06,13,14,23,02,01 &            10 \\
24,34,04,13,02,01+15,13,14,13,02,01 &            11 \\
24,34,04,13,02,01+24,25,34,03,12,01 &            12 \\
24,34,24,01,12,01+24,01,03,13,01,01 &            13 \\
24,34,24,03,02,01+34,45,34,23,02,01 &            14 \\
56,35,01,23,01,01+34,01,34,13,02,01 &            15 \\
56,35,01,23,01,01+34,05,14,23,02,01 &            16 \\
56,35,23,01,02,01+34,25,34,01,02,01 &            17 \\
56,35,23,01,02,01+34,25,34,01,20,01 &            18 \\
\bottomrule
\end{tabular}

  }
  \caption{The mapping of convective term einsum expression contraction paths
    to short labels in Fig.~\ref{fig:layouts-convect-1}.}
  \label{tab:short-paths}
\end{table}

\subsubsection{Layouts Discussion}
\label{sec:ld}

Based on the above results, the following observations were made:
\begin{itemize}
\item For each approximation order, there was a single contraction path per
  library/optimization pair, independent of the layout.
\item The layouts were both slower and faster than the default \Verb|cqgvd0|
  layout. Although there was not a single best layout, some patterns can be
  observed in Figs.~\ref{fig:layouts-laplace-1},~\ref{fig:layouts-convect-1}.
  According to Tab.~\ref{tab:flts-l-m}, \Verb|cdgq| was $14\times$, and
  \Verb|cgqd| $6\times$ the fastest layout for the weak Laplacian, while,
  according to Tab.~\ref{tab:flts-c-m}, \Verb|cqgdv| was $3\times$, and
  \Verb|cgvdq| $2\times$ the fastest layout for the weak convective term.
\item In the weak Laplacian case, 37.3\%, of layouts were no more than
  $4\times$ slower and 3.2\% were faster than the reference implementation for
  the approximation order 1 while for the approximation orders 2 it was 27.0\%
  and 1.9\%, respectively. For the approximation order 3, due to the 50 s
  time-out, all layouts times were within the $4\times$ limit and 49.1\% were
  faster than the reference implementation. In the case of the weak convective
  term, 51.2\% resp. 99.1\% of layouts were no more than $4\times$ slower and
  20.1\% resp. 44.8\% were faster than the reference implementation for the
  approximation orders 1 resp. 2.
\item In the weak Laplacian case, the jx backend performed the best in terms of
  the elapsed time, but required much more memory. For the approximation order
  1, it was the only backend faster than the reference implementation, together
  with the hand-written Numba code, due to the low numerical intensity of the
  calculations. From the order 2, np and oe backends started to be competitive
  and for the order 3 also the loop-based npl and oel backends with their low
  memory footprint, as well as oedas.
\item In the weak convective term case, being more computationally intensive,
  many oe backend layouts were faster than the reference implementation already
  for the approximation order 1, together with the again fastest jx layouts.
  For the order 2, also jxv and das, oedas (lower memory usage than oe) layouts
  out-performed the reference implementation.
\end{itemize}

We can conclude that alternative layouts can be used as another simple way of
speeding the FE weak form evaluation calculations using einsum expressions,
additional to optimized contraction paths.

\subsection{Performance of Selected Weak Forms Contractions}
\label{sec:pswfc}

A parametric study run using soops \cite{soops2021github} lead to 6478 finished
simulations\footnote{This includes the reference implementation simulations
  shown in Fig.~\ref{fig:rip}. 26 backend/optimization pairs, 25 mesh
  size/order combinations, 5 terms and 2 evaluation modes, 6500 in total, 22
  failed due to insufficient memory.}, each evaluating a single weak form using
a single backend/optimization combination several times according to the repeat
parameter for various mesh sizes and function approximation orders, see
Tab.~\ref{tab:cdc}. The default \Verb|cqgvd0| layout of the weak form einsum
expression operands and result arrays was used.

First the overall results, i.e. the fastest backend for each weak form, number
of cells and FE approximation order, are presented in Section~\ref{sec:or}.
Because many backends/contraction paths performed similarly, we provide also a
detailed comparison in Sections~\ref{sec:wl}--~\ref{sec:le}, where both the
elapsed time differences among the backends/contraction paths are shown,
alongside the memory consumption. The discussions of evaluation performance are
included directly in the respective sections.

\subsubsection{Overall Results}
\label{sec:or}

\begin{table}[ht!]
  \centering
  {
    \begin{tabular}{rrlllll}
\toprule
  \#cells & order &                Laplacian &                   v. dot &          weighted v. dot &            NS convective &                elasticity \\
\midrule
    1,024 &     1 & $\enspace 0.01$ (oe 0.9) & $\enspace 0.01$ (oe 0.9) & $\enspace 0.01$ (jx 1.0) & $\enspace 0.01$ (jx 1.0) &  $\enspace 0.01$ (jx 1.1) \\
          &     2 & $\enspace 0.01$ (oe 0.9) & $\enspace 0.01$ (oe 0.8) & $\enspace 0.01$ (jx 0.9) & $\enspace 0.02$ (jx 0.8) &  $\enspace 0.03$ (jx 1.5) \\
          &     3 & $\enspace 0.02$ (oe 0.6) & $\enspace 0.02$ (jx 0.5) & $\enspace 0.02$ (jx 0.5) & $\enspace 0.04$ (jx 0.6) &  $\enspace 0.12$ (jx 1.6) \\
          &     4 & $\enspace 0.06$ (oe 0.5) & $\enspace 0.03$ (jx 0.3) & $\enspace 0.03$ (jx 0.4) & $\enspace 0.11$ (jx 0.5) &  $\enspace 0.41$ (jx 1.8) \\
          &     5 & $\enspace 0.15$ (oe 0.4) & $\enspace 0.06$ (jx 0.2) & $\enspace 0.07$ (jx 0.3) & $\enspace 0.30$ (jx 0.5) & $\enspace 0.74$ (npl 1.2) \\
    2,048 &     1 & $\enspace 0.01$ (oe 1.1) & $\enspace 0.01$ (jx 1.0) & $\enspace 0.01$ (jx 1.0) & $\enspace 0.01$ (jx 1.0) &  $\enspace 0.02$ (jx 1.1) \\
          &     2 & $\enspace 0.02$ (np 0.9) & $\enspace 0.02$ (jx 0.8) & $\enspace 0.02$ (jx 0.8) & $\enspace 0.02$ (jx 0.6) &  $\enspace 0.06$ (jx 1.8) \\
          &     3 & $\enspace 0.04$ (np 0.5) & $\enspace 0.02$ (jx 0.4) & $\enspace 0.03$ (jx 0.5) & $\enspace 0.07$ (jx 0.5) &  $\enspace 0.22$ (jx 1.7) \\
          &     4 & $\enspace 0.11$ (oe 0.4) & $\enspace 0.05$ (jx 0.3) & $\enspace 0.06$ (jx 0.3) & $\enspace 0.20$ (jx 0.5) &  $\enspace 0.81$ (jx 1.8) \\
    4,096 &     1 & $\enspace 0.01$ (oe 0.9) & $\enspace 0.01$ (np 1.0) & $\enspace 0.02$ (jx 1.1) & $\enspace 0.01$ (jx 0.9) &  $\enspace 0.02$ (jx 1.1) \\
          &     2 & $\enspace 0.02$ (oe 0.8) & $\enspace 0.02$ (jx 0.6) & $\enspace 0.02$ (jx 0.7) & $\enspace 0.03$ (jx 0.5) &  $\enspace 0.11$ (jx 2.1) \\
          &     3 & $\enspace 0.06$ (oe 0.5) & $\enspace 0.04$ (jx 0.4) & $\enspace 0.05$ (jx 0.4) & $\enspace 0.12$ (jx 0.5) &  $\enspace 0.45$ (jx 1.8) \\
    8,192 &     1 & $\enspace 0.01$ (oe 1.0) & $\enspace 0.01$ (jx 1.0) & $\enspace 0.02$ (jx 0.9) & $\enspace 0.02$ (jx 0.7) &  $\enspace 0.03$ (jx 1.2) \\
          &     2 & $\enspace 0.03$ (oe 0.7) & $\enspace 0.03$ (jx 0.6) & $\enspace 0.04$ (jx 0.7) & $\enspace 0.06$ (jx 0.6) &  $\enspace 0.21$ (jx 2.0) \\
          &     3 & $\enspace 0.12$ (oe 0.4) & $\enspace 0.07$ (jx 0.3) & $\enspace 0.10$ (jx 0.5) & $\enspace 0.23$ (jx 0.5) &  $\enspace 0.82$ (jx 1.6) \\
   16,384 &     1 & $\enspace 0.02$ (np 1.0) & $\enspace 0.02$ (jx 0.9) & $\enspace 0.02$ (jx 0.9) & $\enspace 0.02$ (jx 0.7) &  $\enspace 0.05$ (jx 1.3) \\
          &     2 & $\enspace 0.06$ (np 0.7) & $\enspace 0.05$ (jx 0.5) & $\enspace 0.07$ (jx 0.7) & $\enspace 0.10$ (jx 0.5) &  $\enspace 0.41$ (jx 2.0) \\
   32,768 &     1 & $\enspace 0.03$ (oe 0.9) & $\enspace 0.03$ (jx 0.9) & $\enspace 0.04$ (jx 0.9) & $\enspace 0.04$ (jx 0.6) &  $\enspace 0.09$ (jx 1.4) \\
          &     2 & $\enspace 0.10$ (oe 0.6) & $\enspace 0.08$ (jx 0.5) & $\enspace 0.13$ (jx 0.7) & $\enspace 0.19$ (jx 0.5) &  $\enspace 0.81$ (jx 2.1) \\
   65,536 &     1 & $\enspace 0.04$ (oe 0.9) & $\enspace 0.05$ (jx 0.9) & $\enspace 0.08$ (jx 1.1) & $\enspace 0.06$ (jx 0.6) &  $\enspace 0.17$ (jx 1.3) \\
          &     2 & $\enspace 0.20$ (oe 0.6) & $\enspace 0.17$ (jx 0.5) & $\enspace 0.29$ (jx 0.8) & $\enspace 0.39$ (jx 0.6) &  $\enspace 1.61$ (jx 2.1) \\
  131,072 &     1 & $\enspace 0.07$ (oe 0.9) & $\enspace 0.08$ (oe 0.9) & $\enspace 0.13$ (jx 1.0) & $\enspace 0.12$ (jx 0.6) &  $\enspace 0.33$ (jx 1.4) \\
  262,144 &     1 & $\enspace 0.13$ (oe 1.0) & $\enspace 0.15$ (oe 0.9) & $\enspace 0.30$ (jx 1.1) & $\enspace 0.24$ (jx 0.6) &  $\enspace 0.66$ (jx 1.4) \\
  524,288 &     1 & $\enspace 0.26$ (oe 0.9) & $\enspace 0.30$ (oe 0.9) & $\enspace 0.58$ (jx 1.1) & $\enspace 0.47$ (jx 0.6) &  $\enspace 1.33$ (jx 1.5) \\
1,048,576 &     1 & $\enspace 0.50$ (oe 0.9) & $\enspace 0.58$ (oe 0.9) & $\enspace 1.13$ (jx 1.1) & $\enspace 0.92$ (jx 0.5) &  $\enspace 2.63$ (jx 1.5) \\
\bottomrule
\end{tabular}

  }
  \caption{Summary results of residual evaluation mode. For each weak form, mesh
    size and FE approximation order, the shortest elapsed time mean
    without the worst case $\twwmean$~[s] across all backends is given, as well
    as the corresponding backend and $\rtwwmean$ (in parentheses).}
  \label{tab:fts-r}
\end{table}

\begin{table}[ht!]
  \centering
  {
    \begin{tabular}{rrlllll}
\toprule
  \#cells & order &                 Laplacian &                    v. dot &          weighted v. dot &             NS convective &               elasticity \\
\midrule
    1,024 &     1 &  $\enspace 0.01$ (jx 1.0) &  $\enspace 0.02$ (jx 1.0) & $\enspace 0.02$ (jx 1.0) &  $\enspace 0.02$ (jx 0.8) & $\enspace 0.03$ (jx 1.4) \\
          &     2 &  $\enspace 0.04$ (jx 1.0) &  $\enspace 0.10$ (np 0.5) & $\enspace 0.10$ (np 0.5) &  $\enspace 0.14$ (jx 0.4) & $\enspace 0.37$ (jx 1.5) \\
          &     3 &  $\enspace 0.24$ (jx 0.6) & $\enspace 0.33$ (npl 0.0) & $\enspace 0.73$ (np 0.1) &  $\enspace 0.83$ (jx 0.1) & $\enspace 1.99$ (jx 0.4) \\
          &     4 &  $\enspace 1.72$ (jx 0.3) & $\enspace 1.29$ (npl 0.0) & $\enspace 3.57$ (jx 0.1) & $\enspace 4.38$ (jxv 0.1) &         $13.89$ (jx 0.4) \\
          &     5 &  $\enspace 5.07$ (jx 0.2) & $\enspace 4.78$ (npl 0.0) &         $15.53$ (jx 0.1) &          $17.71$ (jx 0.0) &         $46.30$ (jx 0.3) \\
    2,048 &     1 &  $\enspace 0.01$ (jx 1.0) &  $\enspace 0.02$ (jx 1.0) & $\enspace 0.02$ (np 0.9) &  $\enspace 0.02$ (jx 0.7) & $\enspace 0.05$ (jx 1.7) \\
          &     2 &  $\enspace 0.07$ (jx 0.9) &  $\enspace 0.18$ (jx 0.5) & $\enspace 0.18$ (np 0.5) &  $\enspace 0.27$ (jx 0.4) & $\enspace 0.70$ (jx 1.4) \\
          &     3 &  $\enspace 0.49$ (jx 0.7) & $\enspace 0.65$ (npl 0.0) & $\enspace 1.45$ (jx 0.1) &  $\enspace 1.62$ (jx 0.1) & $\enspace 4.01$ (jx 0.4) \\
          &     4 & $\enspace 3.48$ (npl 0.3) & $\enspace 2.54$ (npl 0.0) & $\enspace 7.18$ (jx 0.1) & $\enspace 8.60$ (jxv 0.0) &         $27.70$ (jx 0.4) \\
    4,096 &     1 &  $\enspace 0.02$ (jx 1.0) &  $\enspace 0.05$ (jx 1.4) & $\enspace 0.03$ (oe 0.9) &  $\enspace 0.04$ (jx 0.6) & $\enspace 0.09$ (jx 1.9) \\
          &     2 &  $\enspace 0.12$ (jx 0.9) & $\enspace 0.39$ (jxv 0.6) & $\enspace 0.34$ (np 0.5) &  $\enspace 0.53$ (jx 0.4) & $\enspace 1.39$ (jx 1.4) \\
          &     3 & $\enspace 1.01$ (jxv 0.7) & $\enspace 1.27$ (npl 0.0) & $\enspace 2.92$ (np 0.1) &  $\enspace 3.31$ (jx 0.1) & $\enspace 8.25$ (jx 0.4) \\
    8,192 &     1 &  $\enspace 0.02$ (jx 1.0) &  $\enspace 0.06$ (jx 1.2) & $\enspace 0.05$ (oe 0.9) &  $\enspace 0.08$ (jx 0.7) & $\enspace 0.19$ (jx 2.2) \\
          &     2 &  $\enspace 0.25$ (jx 1.0) & $\enspace 0.69$ (jxv 0.5) & $\enspace 0.66$ (np 0.5) &  $\enspace 1.05$ (jx 0.4) & $\enspace 2.79$ (jx 1.4) \\
          &     3 &  $\enspace 1.85$ (jx 0.6) & $\enspace 2.55$ (npl 0.0) & $\enspace 5.70$ (jx 0.1) &  $\enspace 6.50$ (jx 0.1) &         $15.97$ (jx 0.4) \\
   16,384 &     1 &  $\enspace 0.04$ (jx 1.1) & $\enspace 0.19$ (jxv 1.9) & $\enspace 0.10$ (oe 0.8) &  $\enspace 0.14$ (jx 0.7) & $\enspace 0.34$ (jx 2.2) \\
          &     2 &  $\enspace 0.47$ (jx 0.9) & $\enspace 1.21$ (jxv 0.5) & $\enspace 1.33$ (np 0.5) & $\enspace 2.04$ (jxv 0.3) & $\enspace 5.44$ (jx 1.4) \\
   32,768 &     1 &  $\enspace 0.06$ (jx 1.1) & $\enspace 0.28$ (jxv 1.5) & $\enspace 0.18$ (oe 0.8) &  $\enspace 0.28$ (jx 0.6) & $\enspace 0.67$ (jx 2.3) \\
          &     2 &  $\enspace 0.95$ (jx 1.0) & $\enspace 2.26$ (jxv 0.4) & $\enspace 2.66$ (np 0.5) & $\enspace 3.94$ (jxv 0.3) &         $11.12$ (jx 1.5) \\
   65,536 &     1 &  $\enspace 0.13$ (jx 1.2) & $\enspace 0.44$ (jxv 1.2) & $\enspace 0.35$ (oe 0.9) &  $\enspace 0.54$ (jx 0.7) & $\enspace 1.32$ (jx 2.2) \\
          &     2 &  $\enspace 1.84$ (jx 0.9) & $\enspace 4.74$ (jxv 0.5) & $\enspace 5.29$ (np 0.5) & $\enspace 7.70$ (jxv 0.3) &         $22.29$ (jx 1.4) \\
  131,072 &     1 &  $\enspace 0.24$ (jx 1.2) & $\enspace 0.79$ (jxv 1.2) & $\enspace 0.68$ (oe 0.9) &  $\enspace 1.07$ (jx 0.6) & $\enspace 2.62$ (jx 2.3) \\
  262,144 &     1 &  $\enspace 0.46$ (jx 1.2) & $\enspace 1.49$ (jxv 1.1) & $\enspace 1.39$ (oe 0.9) &  $\enspace 2.11$ (jx 0.7) & $\enspace 5.23$ (jx 2.3) \\
  524,288 &     1 &  $\enspace 0.95$ (jx 1.3) & $\enspace 2.87$ (jxv 1.1) & $\enspace 2.79$ (oe 0.9) &  $\enspace 4.30$ (jx 0.7) &         $10.55$ (jx 2.4) \\
1,048,576 &     1 &  $\enspace 1.86$ (jx 1.3) & $\enspace 5.64$ (jxv 1.1) & $\enspace 5.56$ (oe 0.9) &       $13.75$ (oedas 1.0) &         $57.33$ (oe 6.4) \\
\bottomrule
\end{tabular}

  }
  \caption{Summary results of matrix evaluation mode. For each weak form, mesh
    size and FE approximation order, the shortest elapsed time mean
    without the worst case $\twwmean$~[s] across all backends is given, as well
    as the corresponding backend and $\rtwwmean$ (in parentheses).}
  \label{tab:fts-m}
\end{table}

The overall results are summarized in Tab.~\ref{tab:fts-r} for the residual
mode and in Tab.~\ref{tab:fts-m} for the matrix mode evaluations. For each weak
form, mesh size and FE approximation order, the shortest elapsed time mean
without the worst case $\twwmean$ across all backends is given, as well as the
corresponding backend and $\rtwwmean$.

In the residual mode, the oe and np backends were the fastest for the weak
Laplacian, oe, followed by np and jx for vector dot product weak form and
mostly jx, with several occurrences of oe and np for the other three weak forms
with the exception of the order 4 and 5 of the linear elastic form, where the
loop-based backends npl and oel performed the best. According to the relative
elapsed times means, the evaluations were mostly faster or comparable with the
reference implementation, except the case of the linear elastic weak form.

In the case of the much more memory demanding matrix mode, the jx backend was
mostly the fastest when evaluating the weak Laplacian, the convective and the
linear elasticity weak forms. Notable exceptions occurred for the largest mesh
size, where the memory-demanding jx was replaced by oedas (convective term) and
oe (elasticity), see also the detailed analysis in the following sections. Note
also that jx was much slower for small problems in the $\tmean$ metric,
because of the JIT initialization in the first evaluation, omitted in
$\twwmean$. The jxv backend, followed by npl, were mostly the fastest in the
case of the vector dot product form, while for the weighted vector dot product,
the oe (most orders 1) and np (most orders 2) backends were the fastest.

Analogical tables summarizing the results of the simulations without the
single-thread limitation are presented in Appendix~\ref{sec:pswfct}. A detailed
comparison of the backends and tensors contraction paths follows in the next
sections.

\subsubsection{Detailed Weak Form Analyses --- Common Settings}
\label{dwfacs}

Two kinds of figures are presented below for the five selected weak forms:
\begin{itemize}
\item To allow comparison among the five weak forms, the highest throughput of
  each backend is presented in Figs.~\ref{fig:atr-1}--\ref{fig:atr-5}. The
  throughput is defined as the the size (in MB) of the resulting array of
  vectors (residual mode) or matrices (matrix mode) divided by the shortest
  elapsed time means without the worst case $\twwmean$ for each backend, i.e.
  $|r| / \twwmean$, resp. $|M| / \twwmean$, corresponding to the highest number
  of megabytes each backend evaluates per second. The backends are indicated by
  markers, while the fastest einsum contraction paths by color. The cell counts
  ($x$ axis) are grouped by the approximation orders.
\item Relative elapsed time means without the worst case $\rtwwmean$ and
  relative maximum memory consumption $\rmmax$ are reported in
  Figs.~\ref{fig:at-1}--\ref{fig:at-5} using the $x$ axis logarithmic scale.
  The backends are indicated by markers, while the einsum contraction paths by
  color, the same as in the throughput figures. The cell counts ($y$ axis) are
  grouped by the approximation orders. Only cases no more than $5\times$ slower
  than the reference implementation are shown.
\end{itemize}

\subsubsection{Weak Laplacian}
\label{sec:wl}

\begin{figure}[p!]
  \centering
  \includegraphics[width=\myrelwidth]{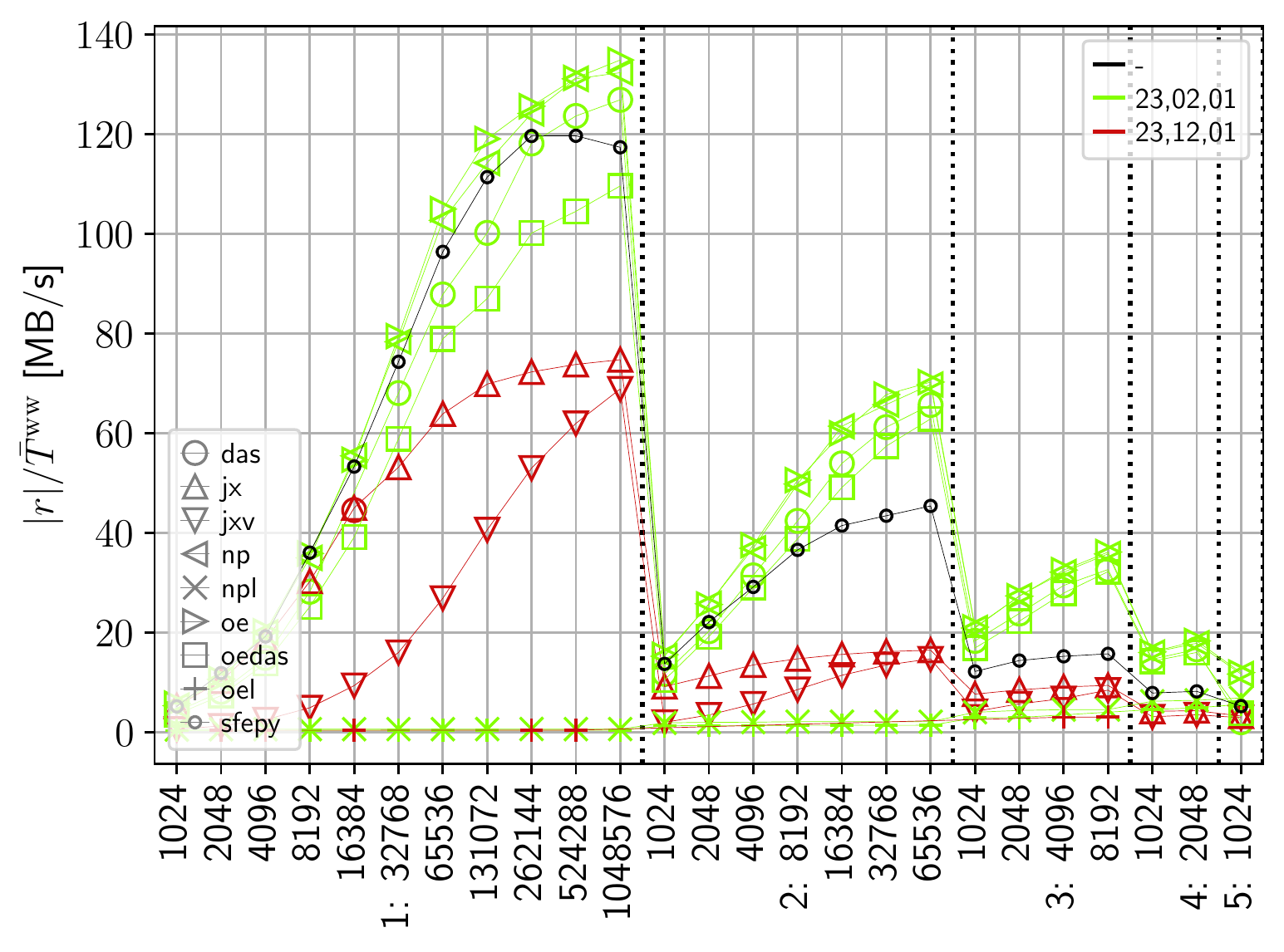}
  \includegraphics[width=\myrelwidth]{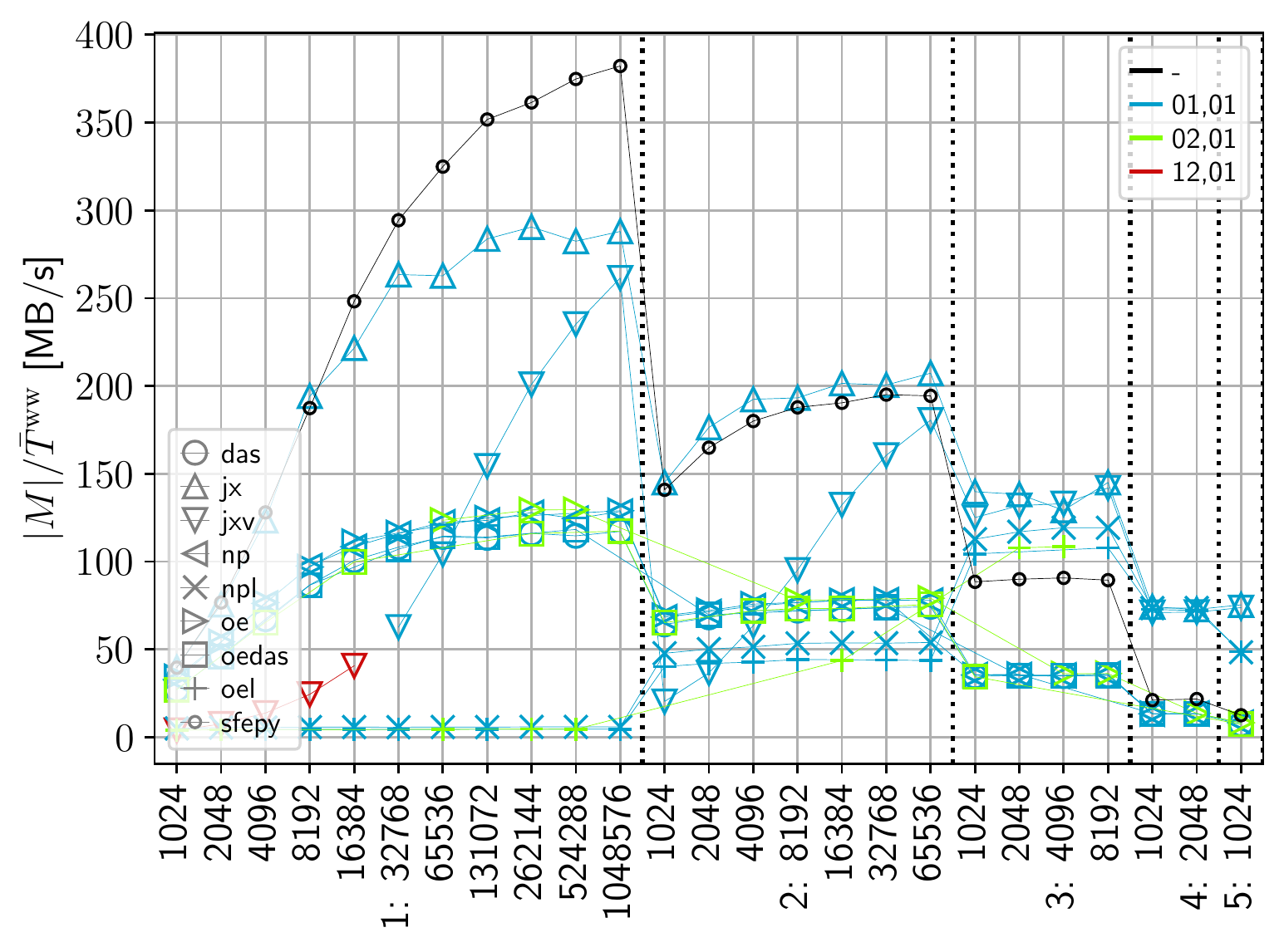}
  \caption{The highest throughput of each backend (number of megabytes
    evaluated per second) for the weak Laplacian einsum expression. Left:
    residual mode evaluations $|r| / \twwmean$, right: matrix mode evaluations
    $|M| / \twwmean$. The backends are indicated by markers, the einsum
    contraction paths by color. The cell counts ($x$ axis) are grouped
    by the approximation orders.}
  \label{fig:atr-1}
\end{figure}

\begin{figure}[p!]
  \centering
  \includegraphics[width=\myrelwidth]{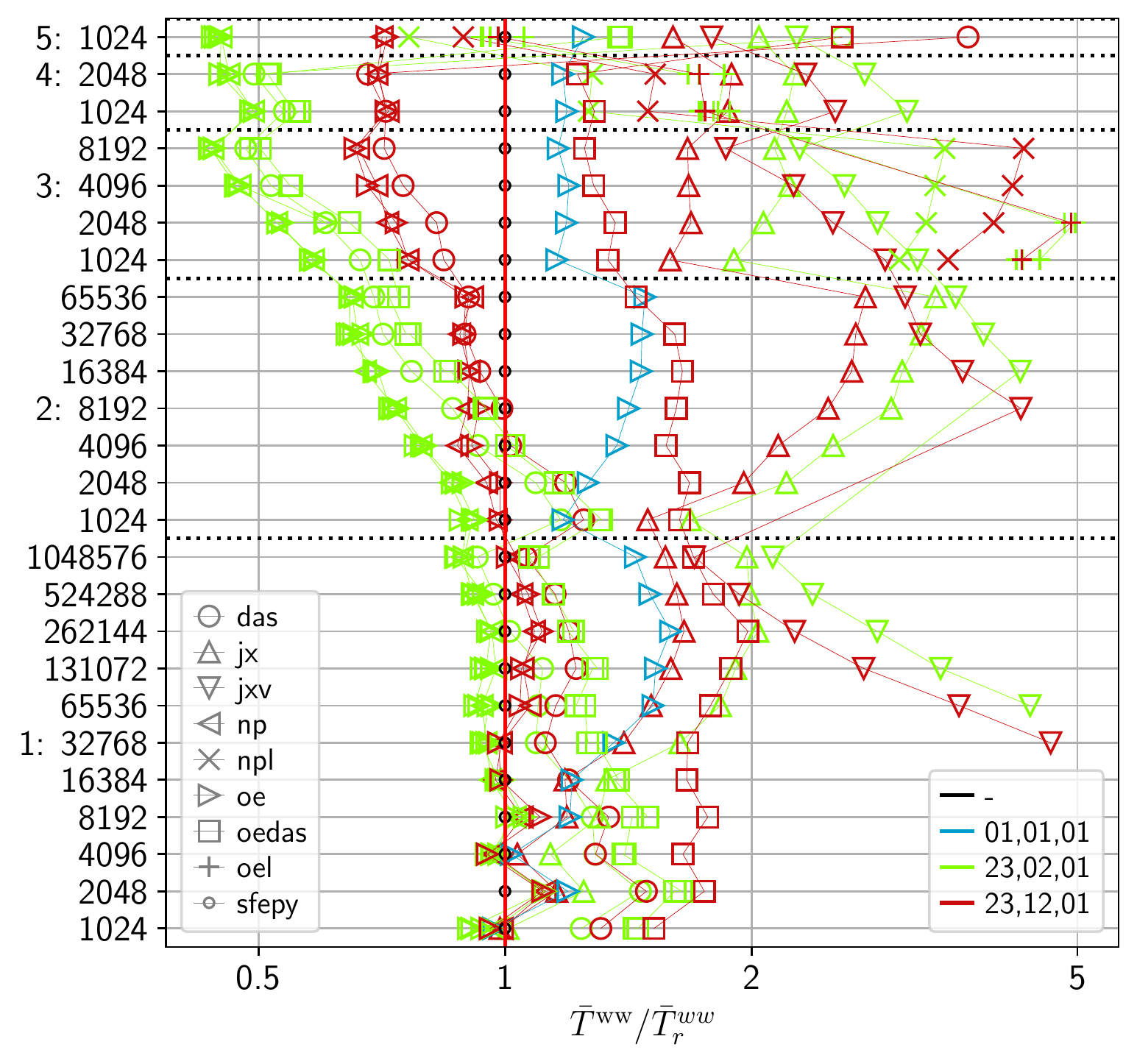}
  \includegraphics[width=\myrelwidth]{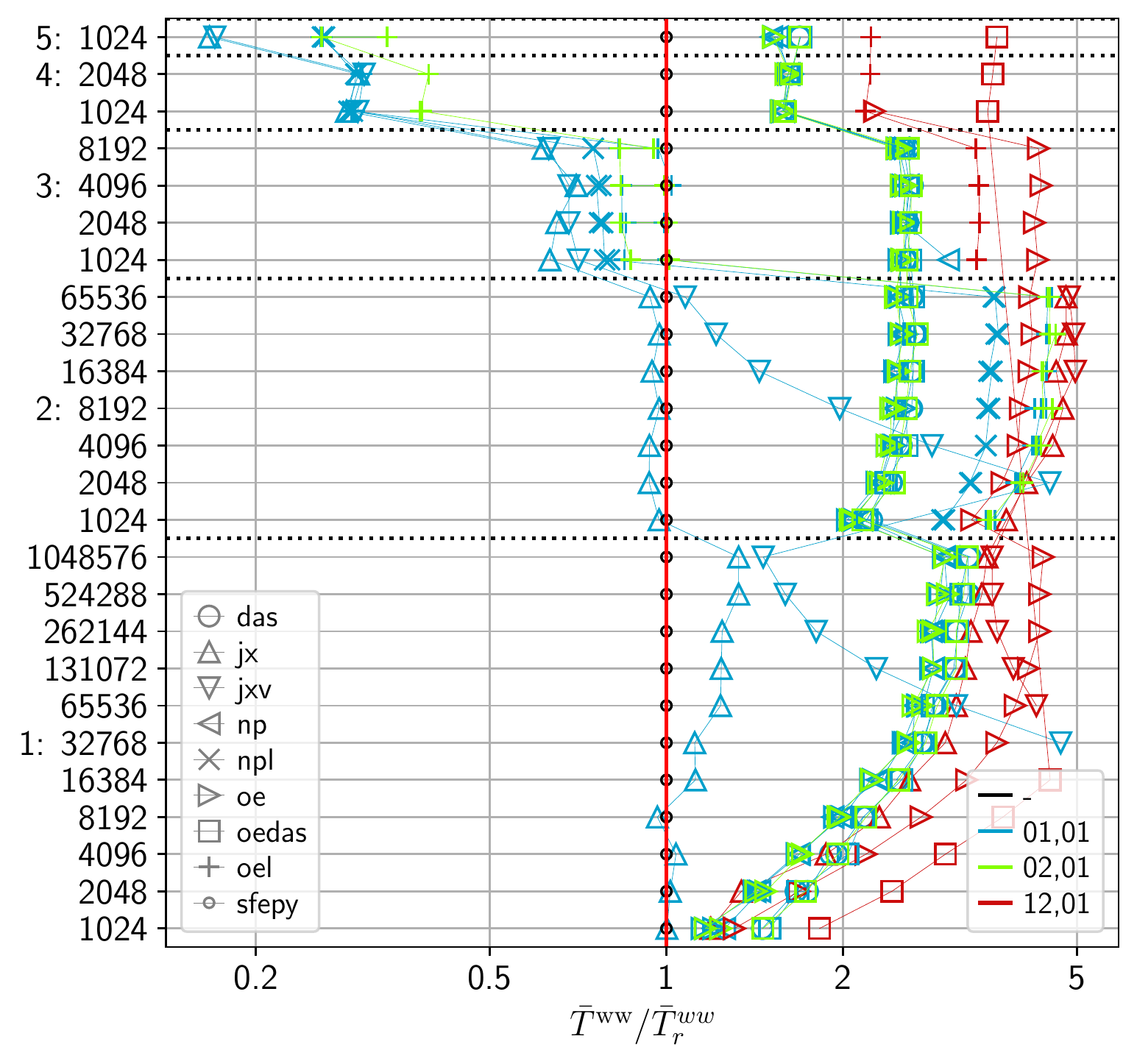}
  \includegraphics[width=\myrelwidth]{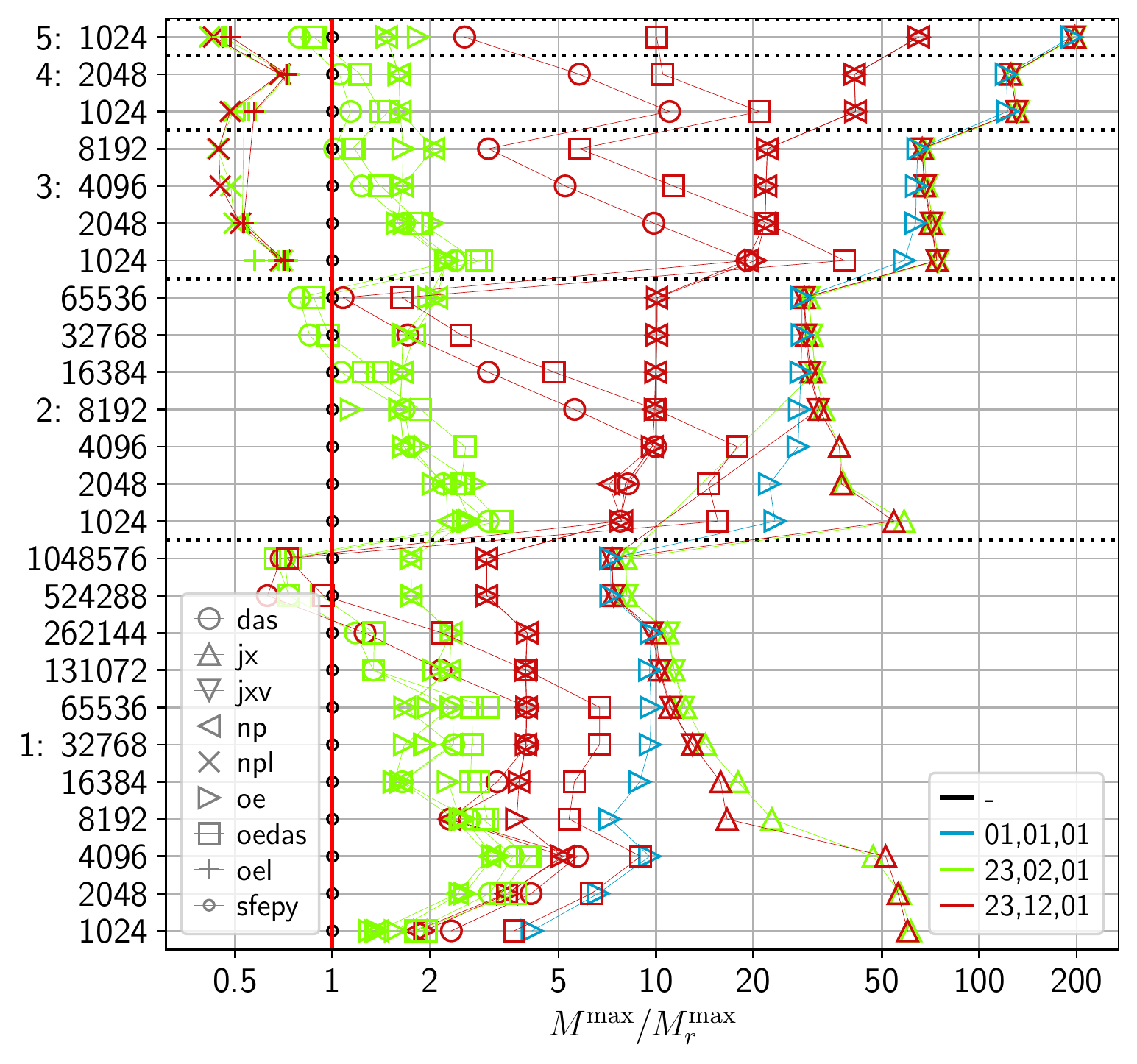}
  \includegraphics[width=\myrelwidth]{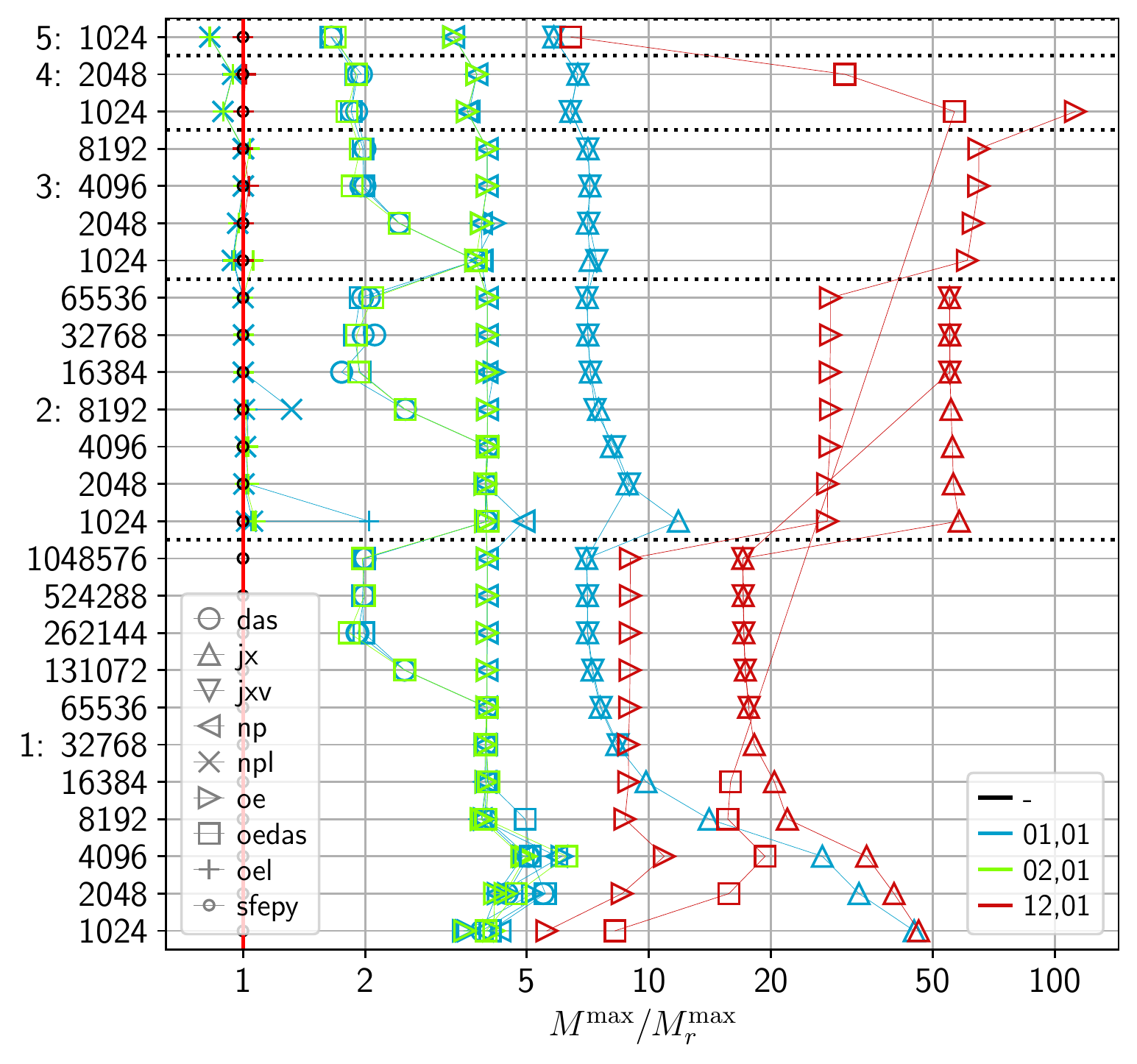}
  \caption{Relative elapsed time and memory requirements of the weak Laplacian
    einsum expression w.r.t. the reference implementation. Left: residual mode,
    right: matrix mode, top: $\rtwwmean$, bottom: $\rmmax$. The backends are
    indicated by markers, the einsum contraction paths by color. The cell
    counts ($y$ axis) are grouped by the approximation orders.}
  \label{fig:at-1}
\end{figure}

The performance of the einsum backends applied to the evaluation of the weak
Laplacian einsum expression is shown in Figs.~\ref{fig:atr-1} (evaluation
throughput) and \ref{fig:at-1} (relative performance).

In the residual mode, the oe backend was the fastest in most cases, followed by
np and das with oedas. It was faster than the reference implementation even for
the order 1. From the order 2, also das and oedas, and for the order 5 also
npl, oel backends out-performed the reference implementation. The memory
consumption was in general higher than the reference one for smaller data,
while for large problems several backends had a lower memory footprint: the
das, oedas backends due to array chunking, and npl, oel thanks to looping over
the mesh cells.

In the matrix mode, the jx backend was the fastest and out-performed the
reference implementation from the order 2. From the order 3, also npl, oel and
jxv backends were faster than the reference implementation. The cell loop
backends (npl, oel) performance seemed to be less dependent on the particular
contraction paths. We attribute this to the fact that the restriction of the
operands to a cell is small and so also the temporary allocations are small ---
compare with the huge memory consumption differences among the contraction
paths of other backends. The npl backend was faster than oel, probably due to a
simpler code path, important again for the small cell-restricted data.


\subsubsection{Weak Convective Term}
\label{sec:wct}

\begin{figure}[p!]
  \centering
  \includegraphics[width=\myrelwidth]{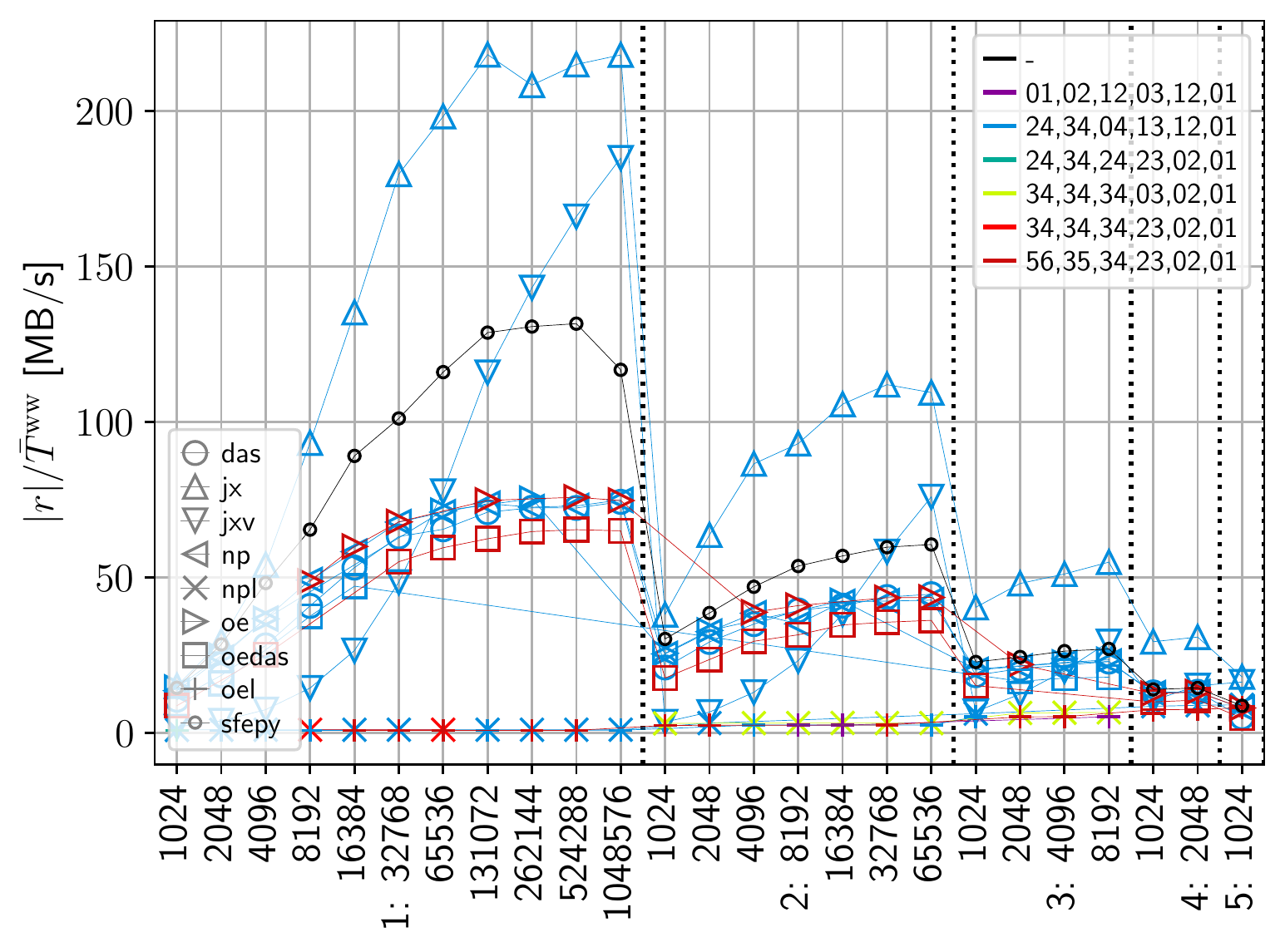}
  \includegraphics[width=\myrelwidth]{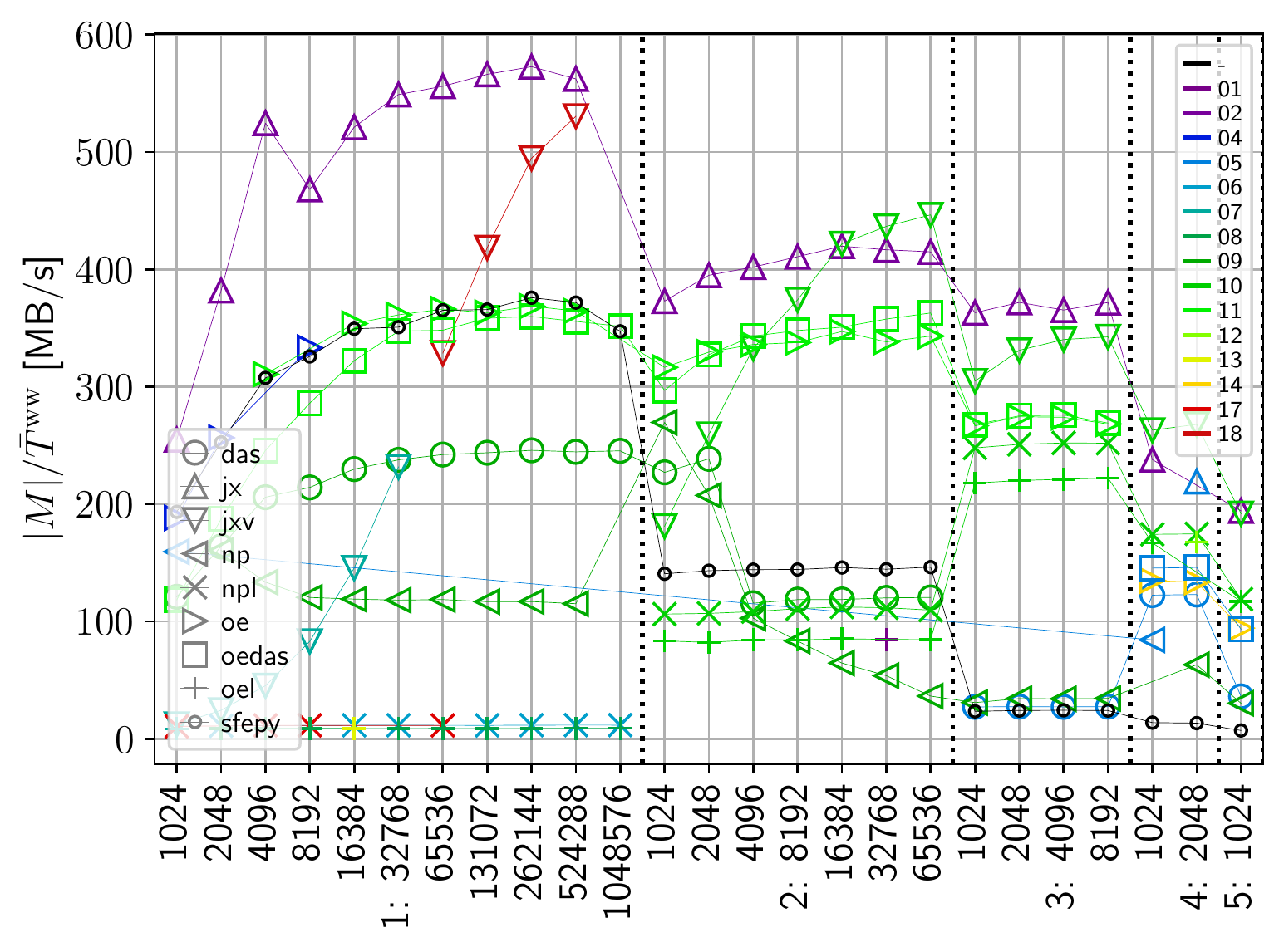}
  \caption{The highest throughput of each backend (number of megabytes
    evaluated per second) for the weak convective term einsum expression. Left:
    residual mode evaluations $|r| / \twwmean$, right: matrix mode evaluations
    $|M| / \twwmean$. The backends are indicated by markers, the einsum
    contraction paths by color. The cell counts ($x$ axis) are grouped
    by the approximation orders.}
  \label{fig:atr-2}
\end{figure}

\begin{figure}[p!]
  \centering
  \includegraphics[width=\myrelwidth]{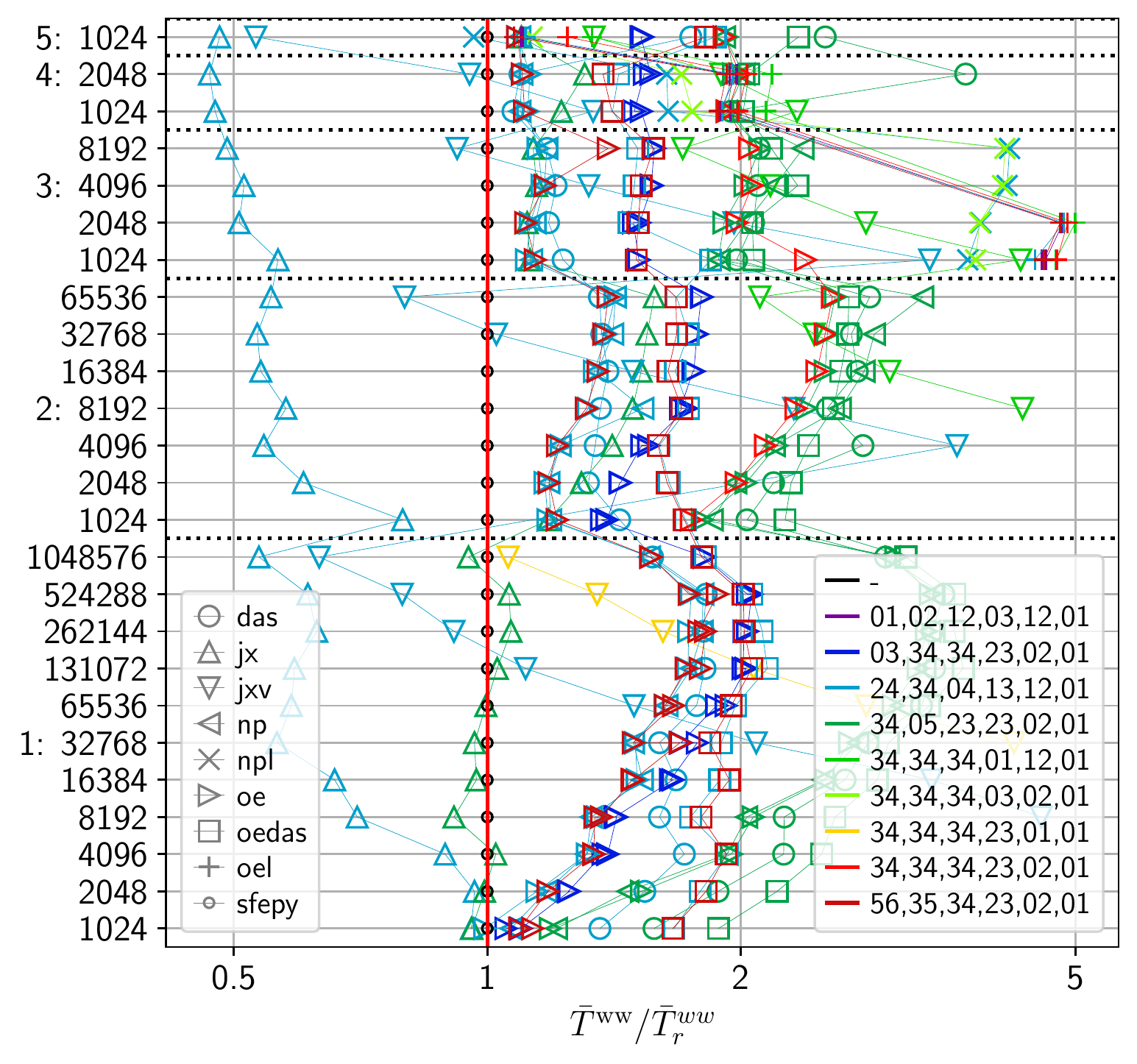}
  \includegraphics[width=\myrelwidth]{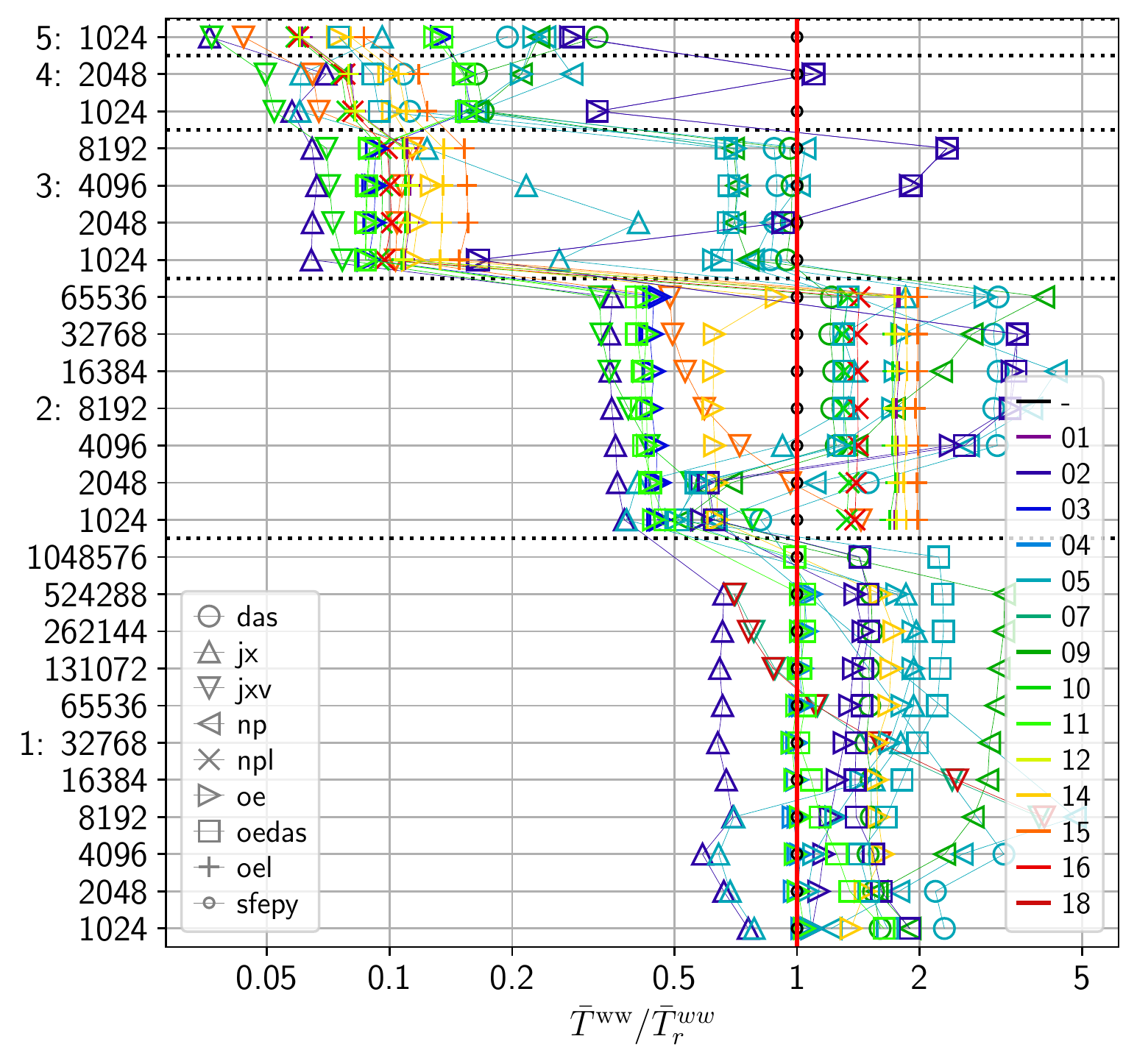}
  \includegraphics[width=\myrelwidth]{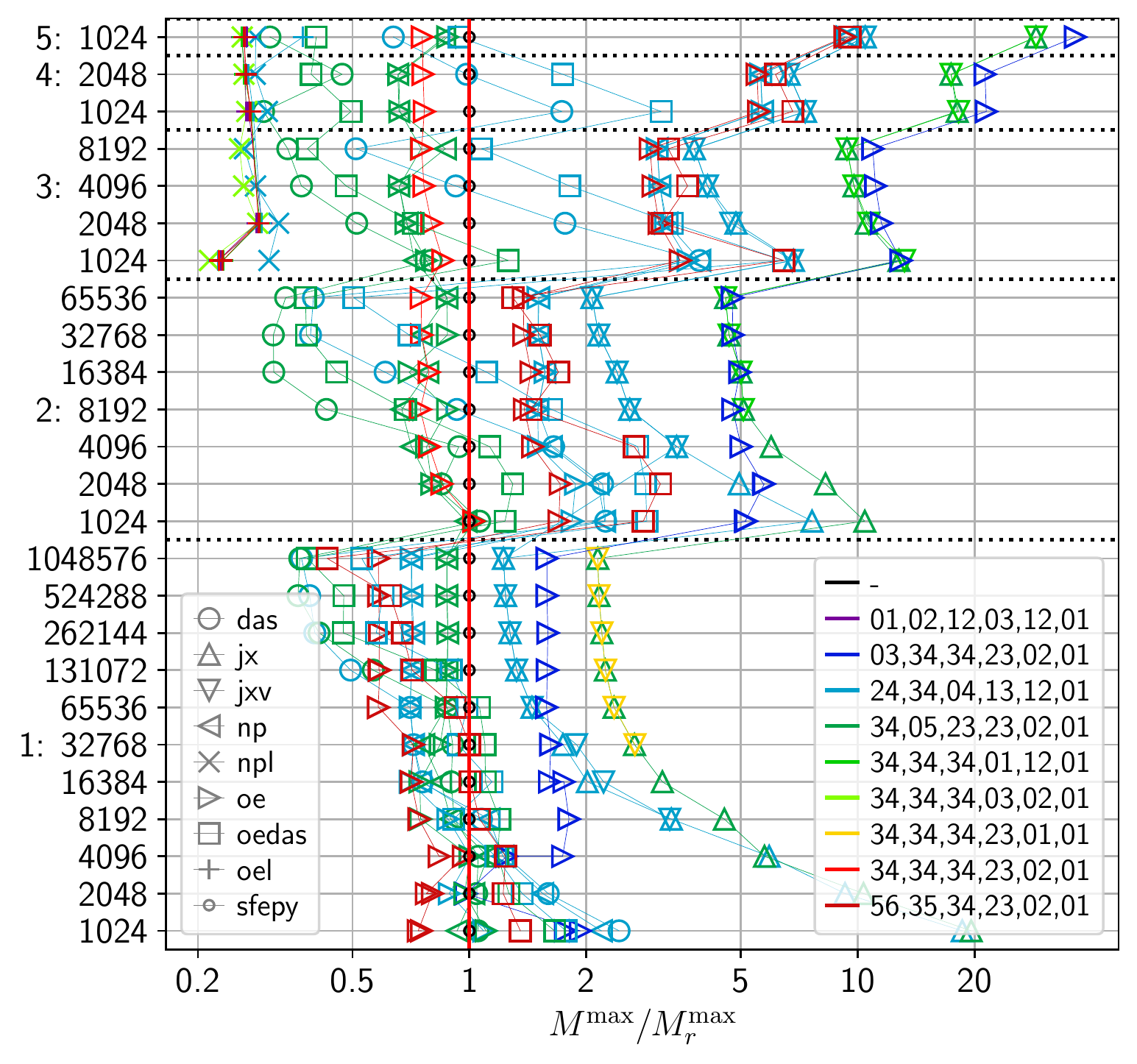}
  \includegraphics[width=\myrelwidth]{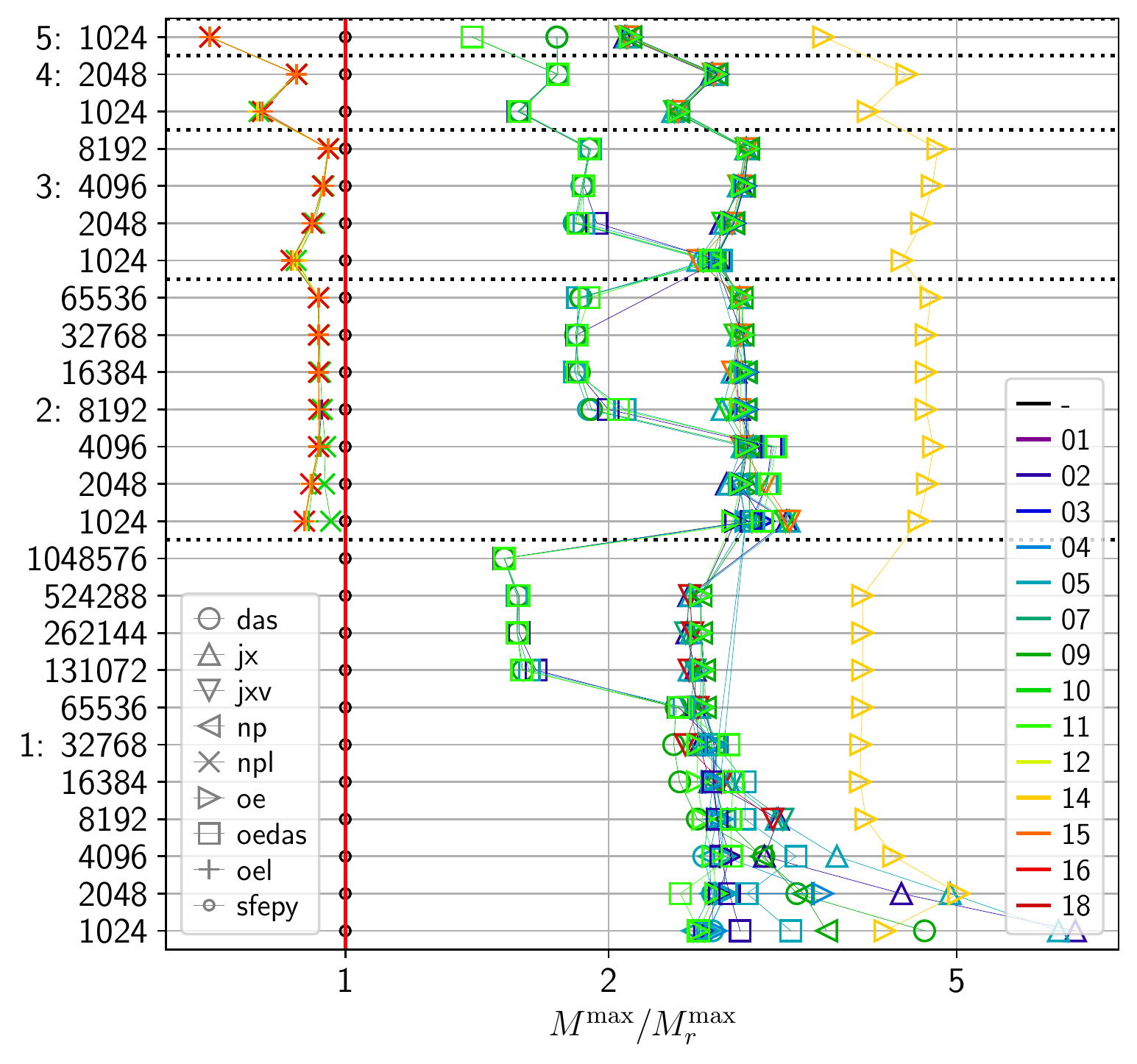}
  \caption{Relative elapsed time and memory requirements of the weak convective
    term einsum expression w.r.t. the reference implementation. Left: residual
    mode, right: matrix mode, top: $\rtwwmean$, bottom: $\rmmax$. The backends
    are indicated by markers, the einsum contraction paths by color. The cell
    counts ($y$ axis) are grouped by the approximation orders.}
  \label{fig:at-2}
\end{figure}

The performance of the einsum backends applied to the evaluation of the weak
convective term einsum expression is shown in Figs.~\ref{fig:atr-2} (evaluation
throughput) and \ref{fig:at-2} (relative performance).

In the residual mode, the jx backend was the fastest by a large margin,
followed by jxv. The jx backend out-performed the reference implementation,
while jxv only in some cases and all the other backends were slower (mostly no
more than $2\times$). However, the fastest contraction paths memory consumption
of jx, jxv was up to $10\times$ more than reference consumption. For higher
order approximations, npl, oel, das, oedas but also eo and np approached the
reference times and had lower memory consumption.

In the matrix mode, the jx backend was again the fastest, but its lead over
jxv, oedas or oe was small from the order 2. The jx backend was faster than the
reference implementation from the order 1, jxv, oedas and oe from the order 2 and
essentially all backends from the order 3. The npl, oel backends performed very
well from the order 3, and unlike other backends (especially jx, jxv) had a
lower memory consumption than the reference implementation. Overall the memory
consumption stayed less than $5\times$ reference, in contrast to the weak
Laplacian in the previous section.


\subsubsection{Vector Dot Product}
\label{sec:vdp}

\begin{figure}[p!]
  \centering
  \includegraphics[width=\myrelwidth]{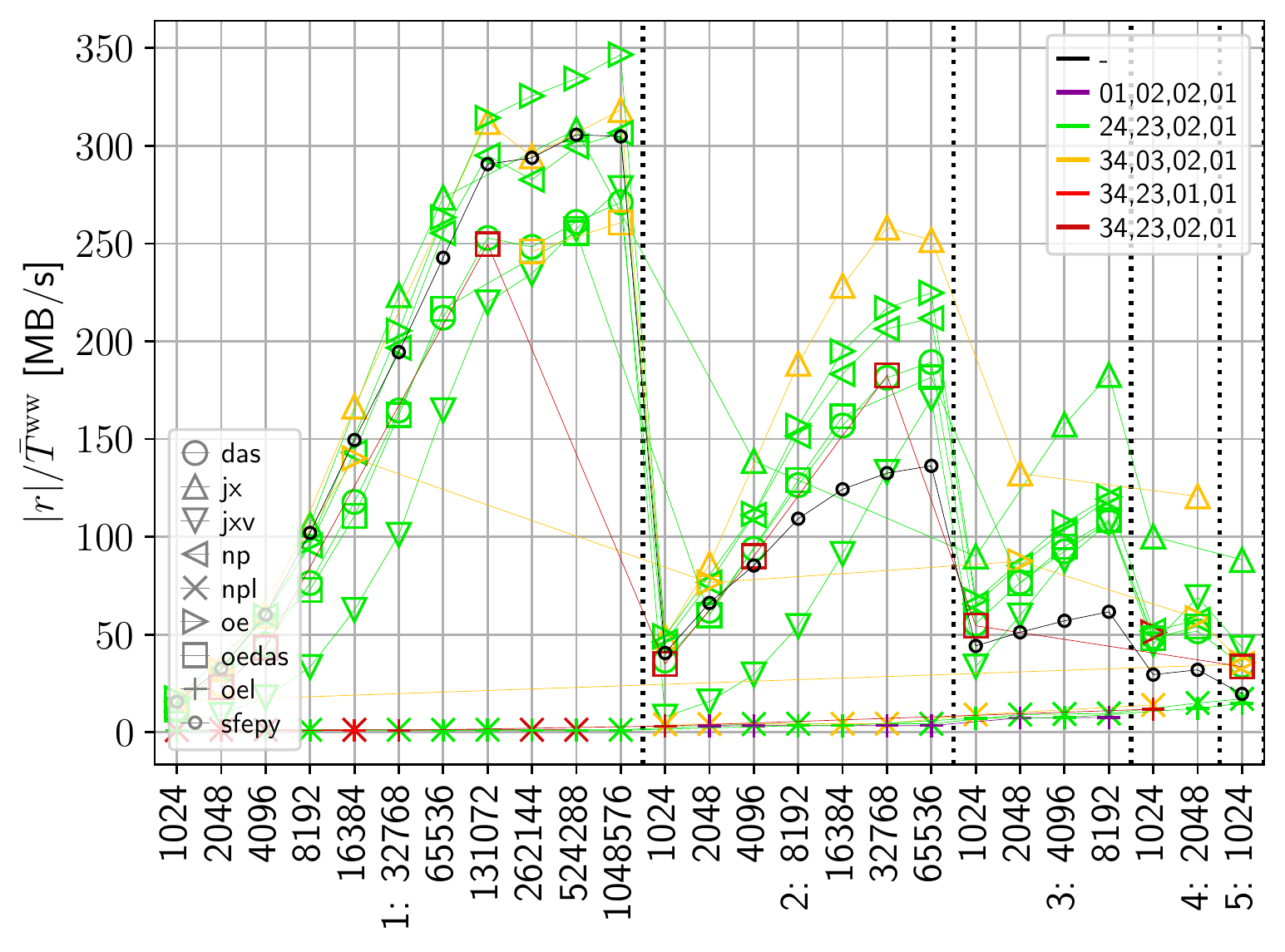}
  \includegraphics[width=\myrelwidth]{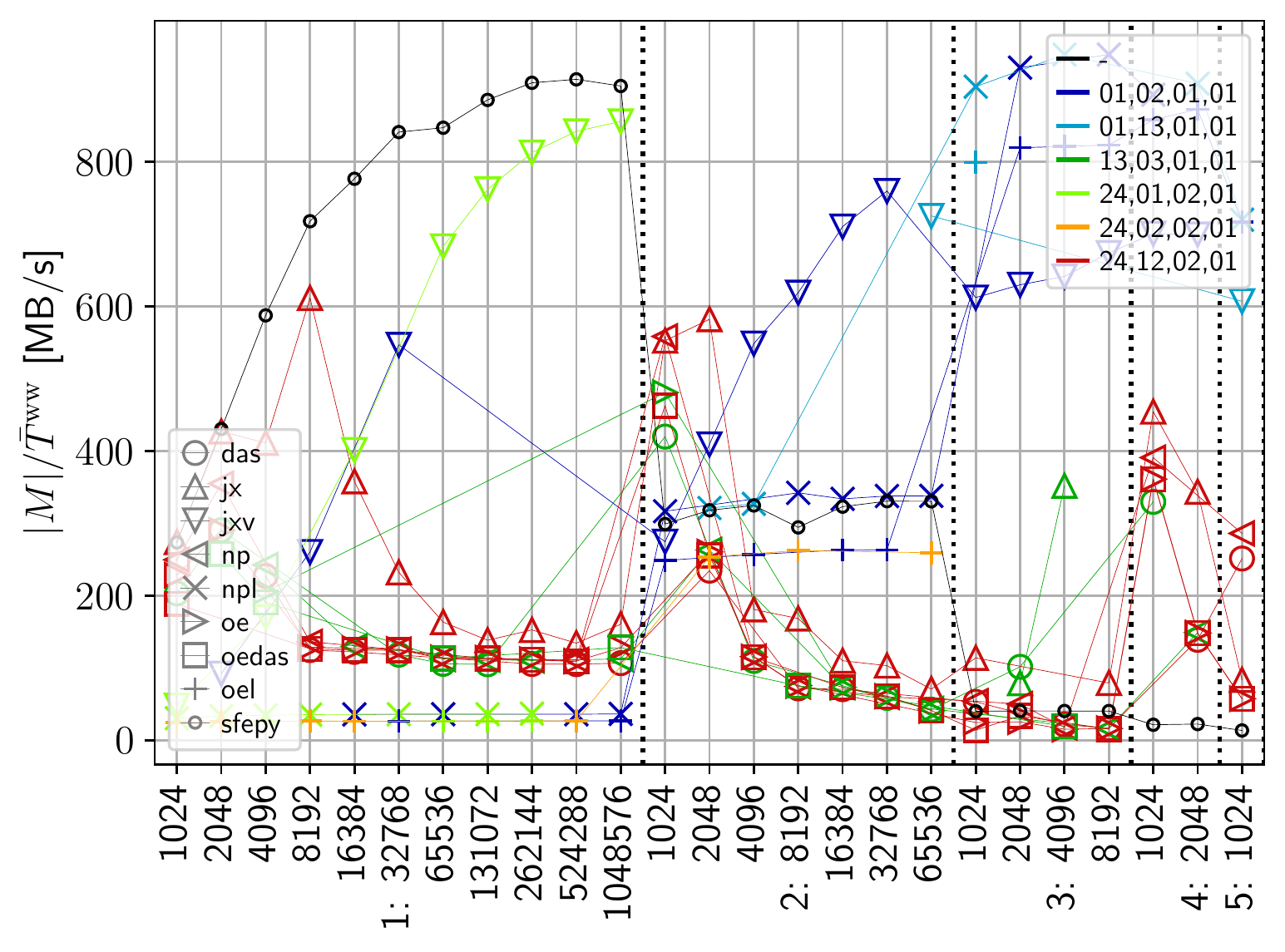}
  \caption{The highest throughput of each backend (number of megabytes
    evaluated per second) for the vector dot product einsum expression. Left:
    residual mode evaluations $|r| / \twwmean$, right: matrix mode evaluations
    $|M| / \twwmean$. The backends are indicated by markers, the einsum
    contraction paths by color. The cell counts ($x$ axis) are grouped
    by the approximation orders.}
  \label{fig:atr-3}
\end{figure}

\begin{figure}[p!]
  \centering
  \includegraphics[width=\myrelwidth]{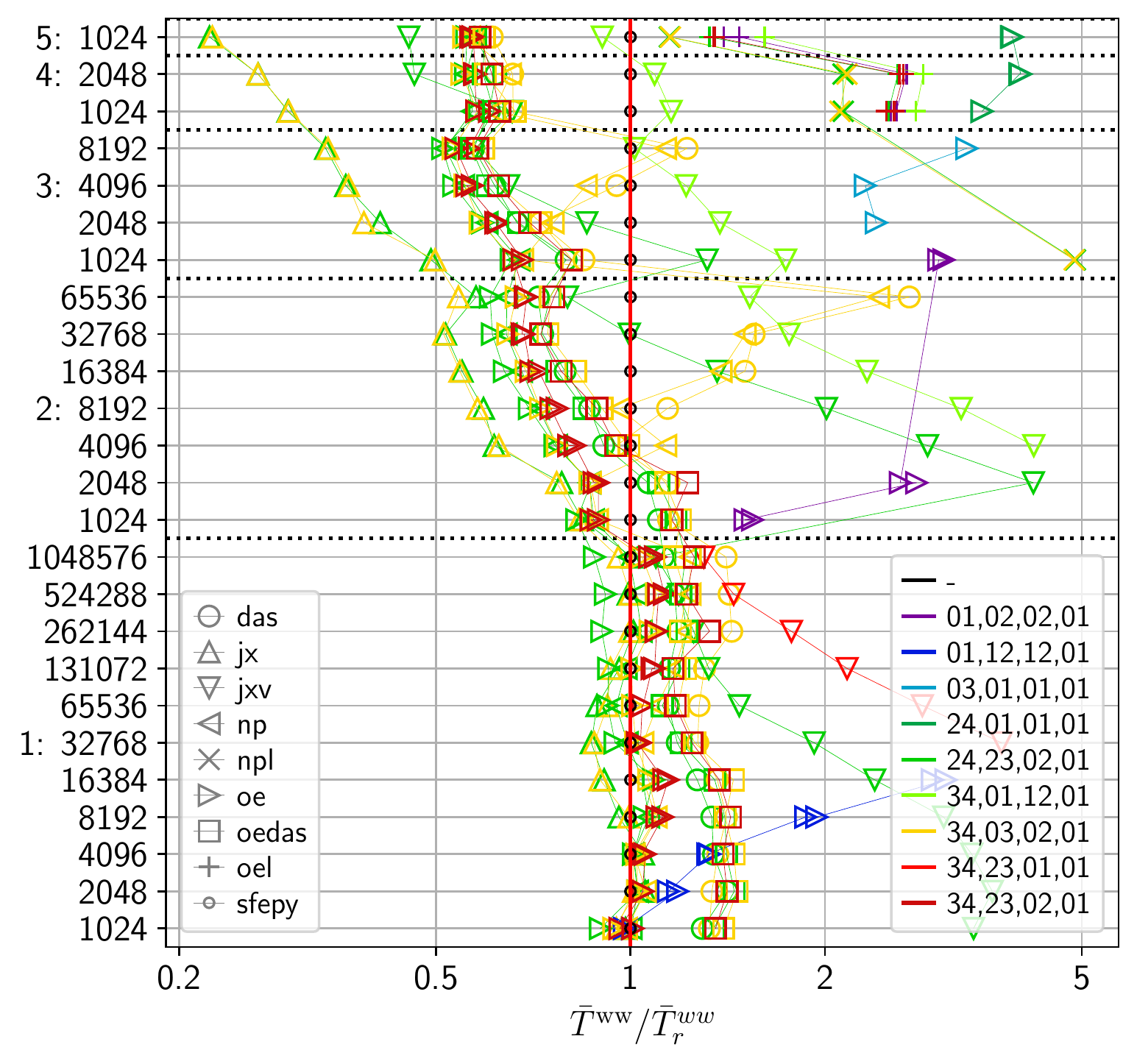}
  \includegraphics[width=\myrelwidth]{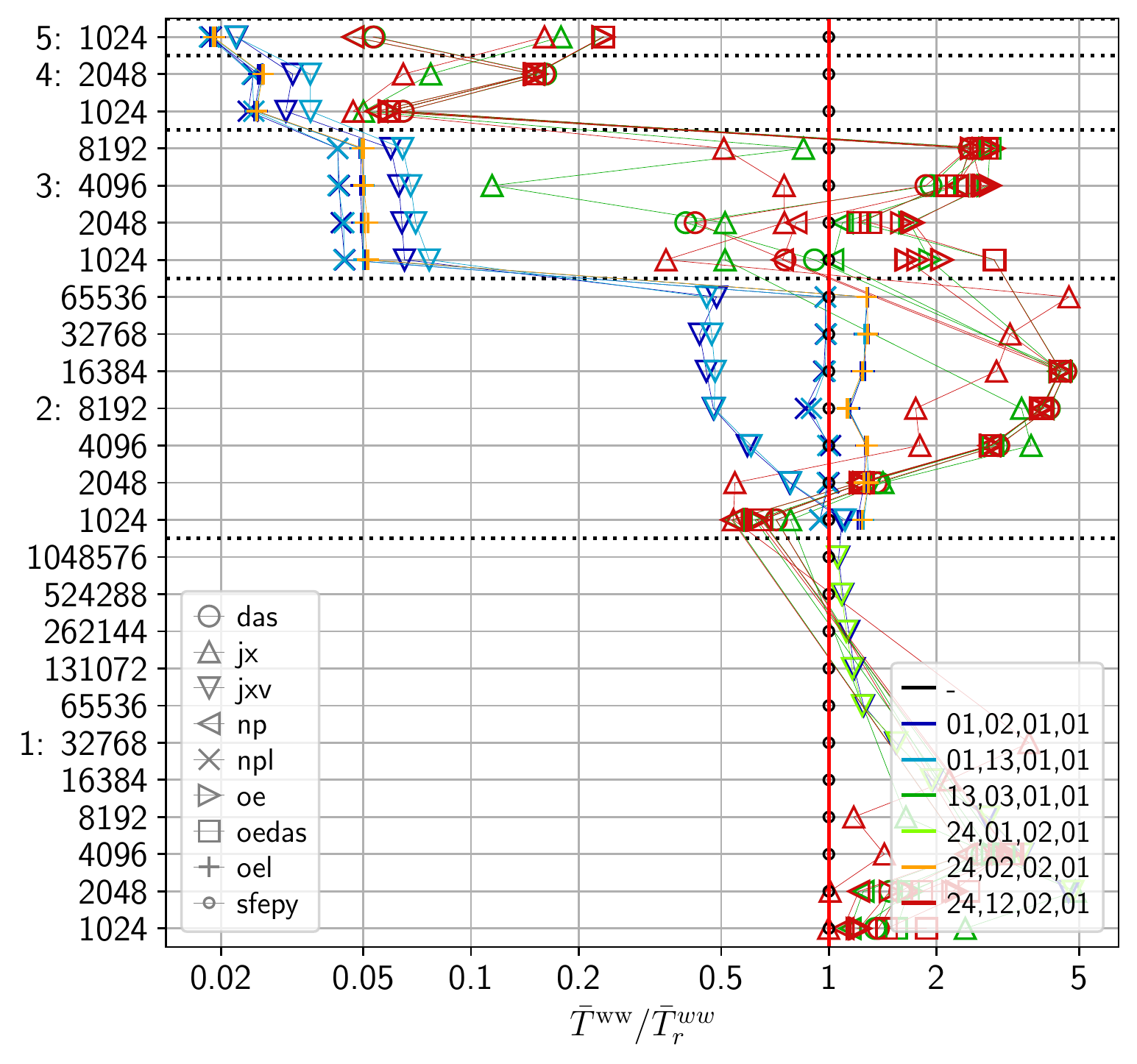}
  \includegraphics[width=\myrelwidth]{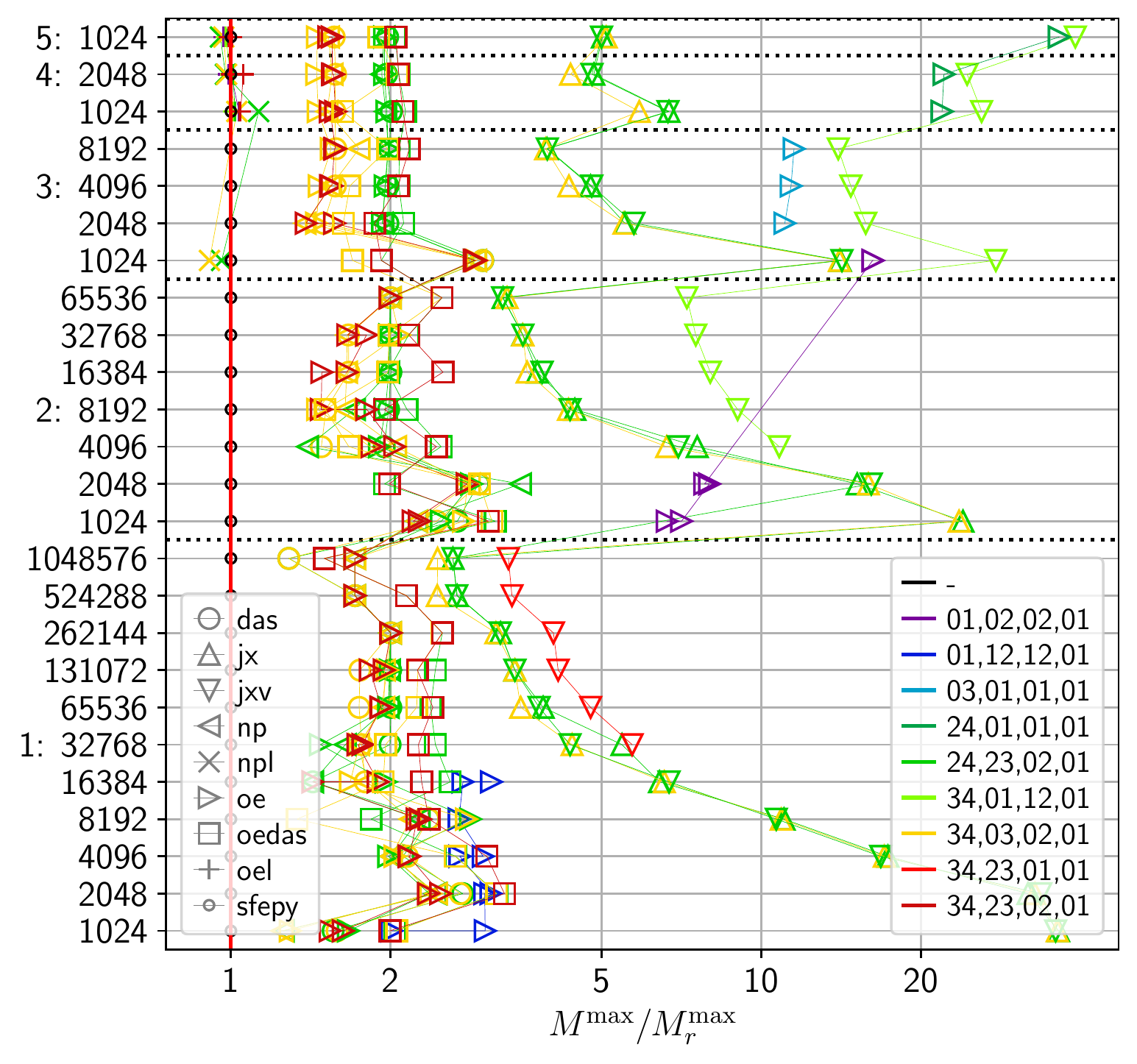}
  \includegraphics[width=\myrelwidth]{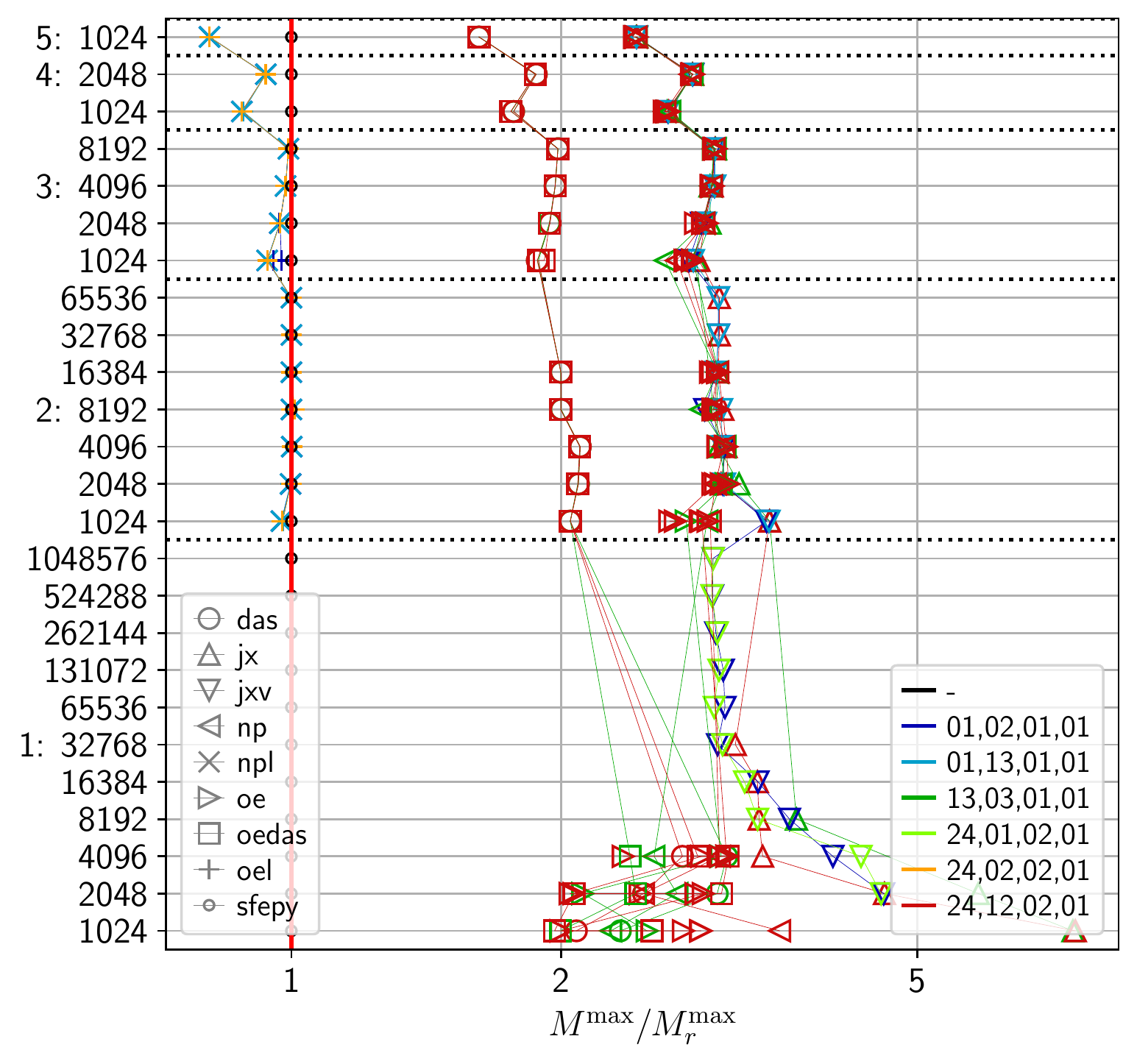}
  \caption{Relative elapsed time and memory requirements of the vector dot product
    einsum expression w.r.t. the reference implementation. Left: residual mode,
    right: matrix mode, top: $\rtwwmean$, bottom: $\rmmax$. The backends are
    indicated by markers, the einsum contraction paths by color. The cell
    counts ($y$ axis) are grouped by the approximation orders.}
  \label{fig:at-3}
\end{figure}

The performance of the einsum backends applied to the evaluation of the vector
dot product einsum expression is shown in Figs.~\ref{fig:atr-3} (evaluation
throughput) and \ref{fig:at-3} (relative performance).

In the residual mode, the jx backend was the fastest, together with oe for the
order 1, where it out-performed the reference implementation. From the order 2
also np, oedas, das and from the order 3 jxv backends were faster than the
reference code. The memory consumption mostly stayed less than $5\times$
reference, except jx, jxv and oe for some contraction paths.

In the matrix mode, jxv replaced jx as the overall fastest backend for orders
1, 2, but from the order 3 it was replaced by npl and oel. The jxv, npl
backends were faster than the reference implementation from the order 2, jx
from the order 3 and das, oedas from the order 4. The memory requirements
stayed within $3\times$ reference, npl and oel backends had memory consumption
lower than reference from the order 3.


\subsubsection{Weighted Vector Dot Product}
\label{sec:wvdp}

\begin{figure}[p!]
  \centering
  \includegraphics[width=\myrelwidth]{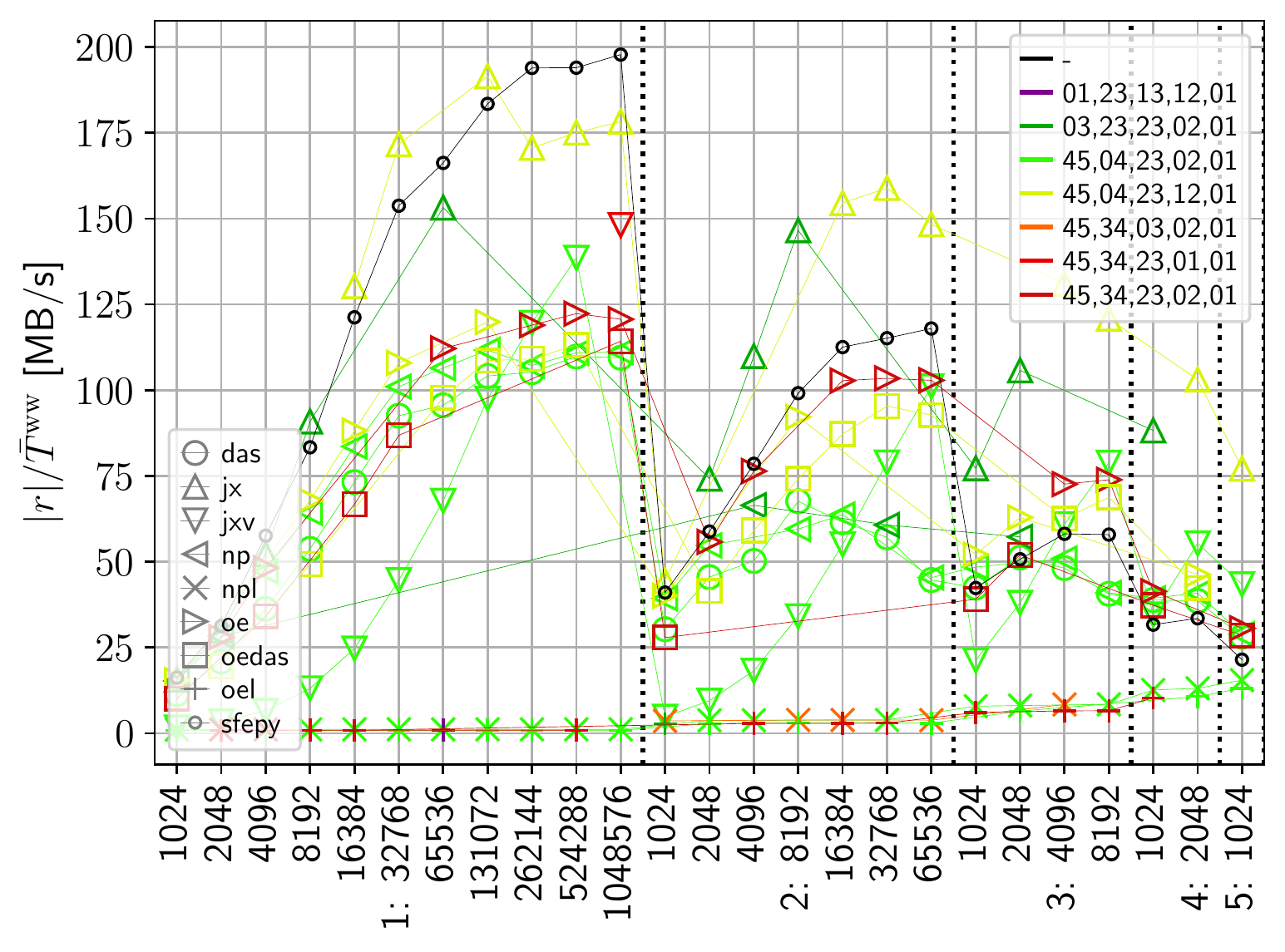}
  \includegraphics[width=\myrelwidth]{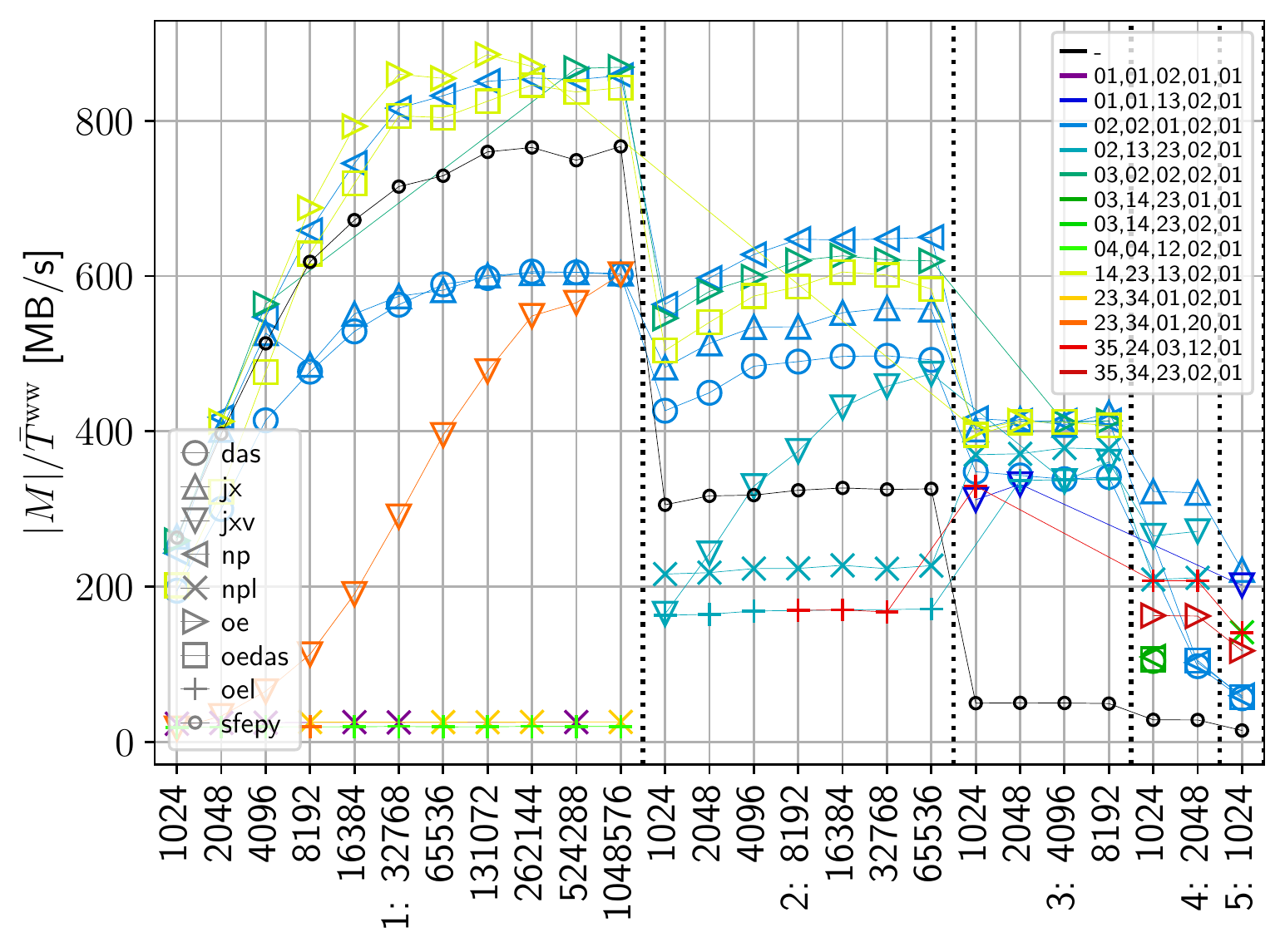}
  \caption{The highest throughput of each backend (number of megabytes
    evaluated per second) for the weighted vector dot product einsum
    expression. Left: residual mode evaluations $|r| / \twwmean$, right: matrix
    mode evaluations $|M| / \twwmean$. The backends are indicated by markers,
    the einsum contraction paths by color. The cell counts ($x$ axis) are
    grouped by the approximation orders.}
  \label{fig:atr-4}
\end{figure}

\begin{figure}[p!]
  \centering
  \includegraphics[width=\myrelwidth]{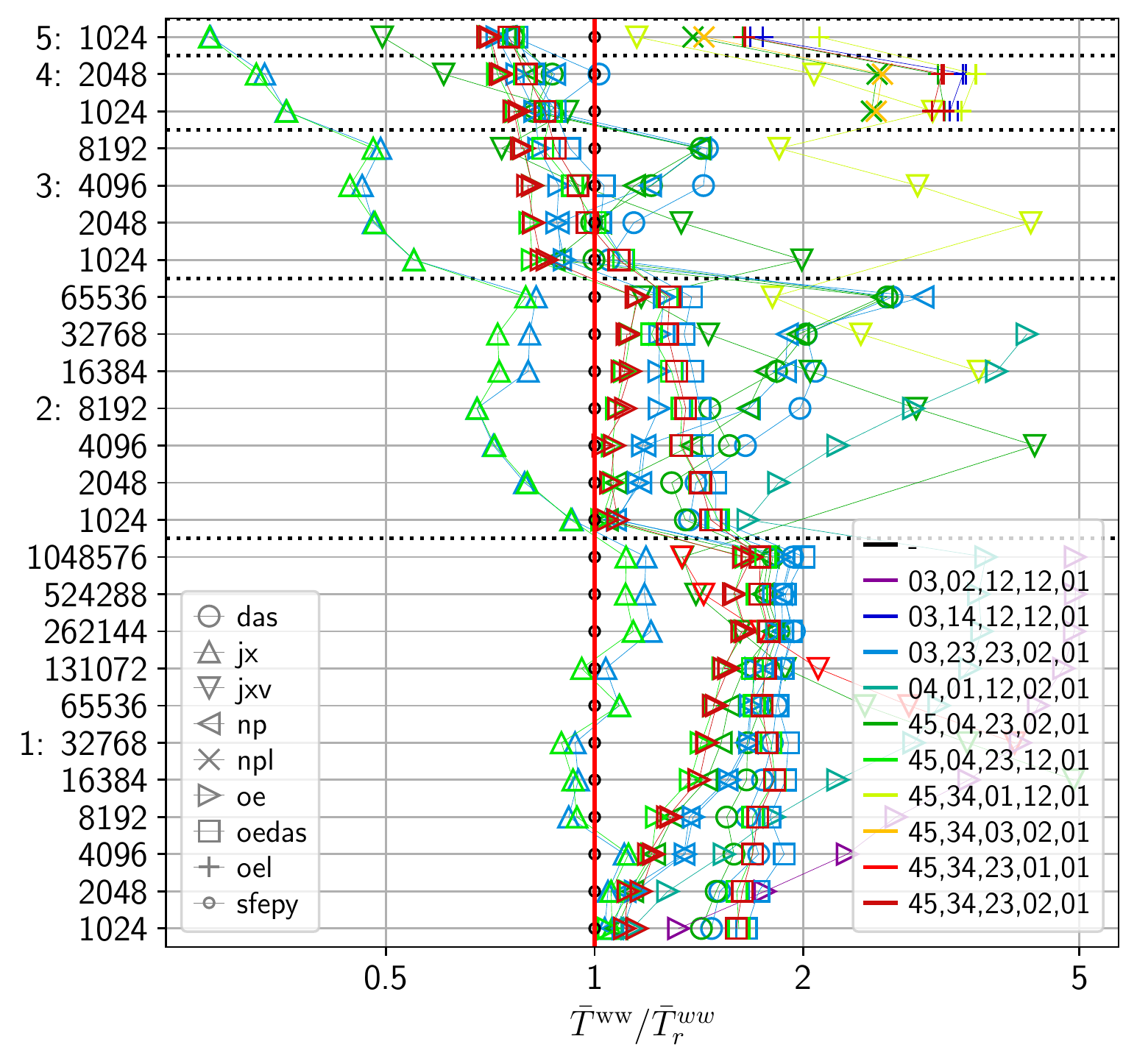}
  \includegraphics[width=\myrelwidth]{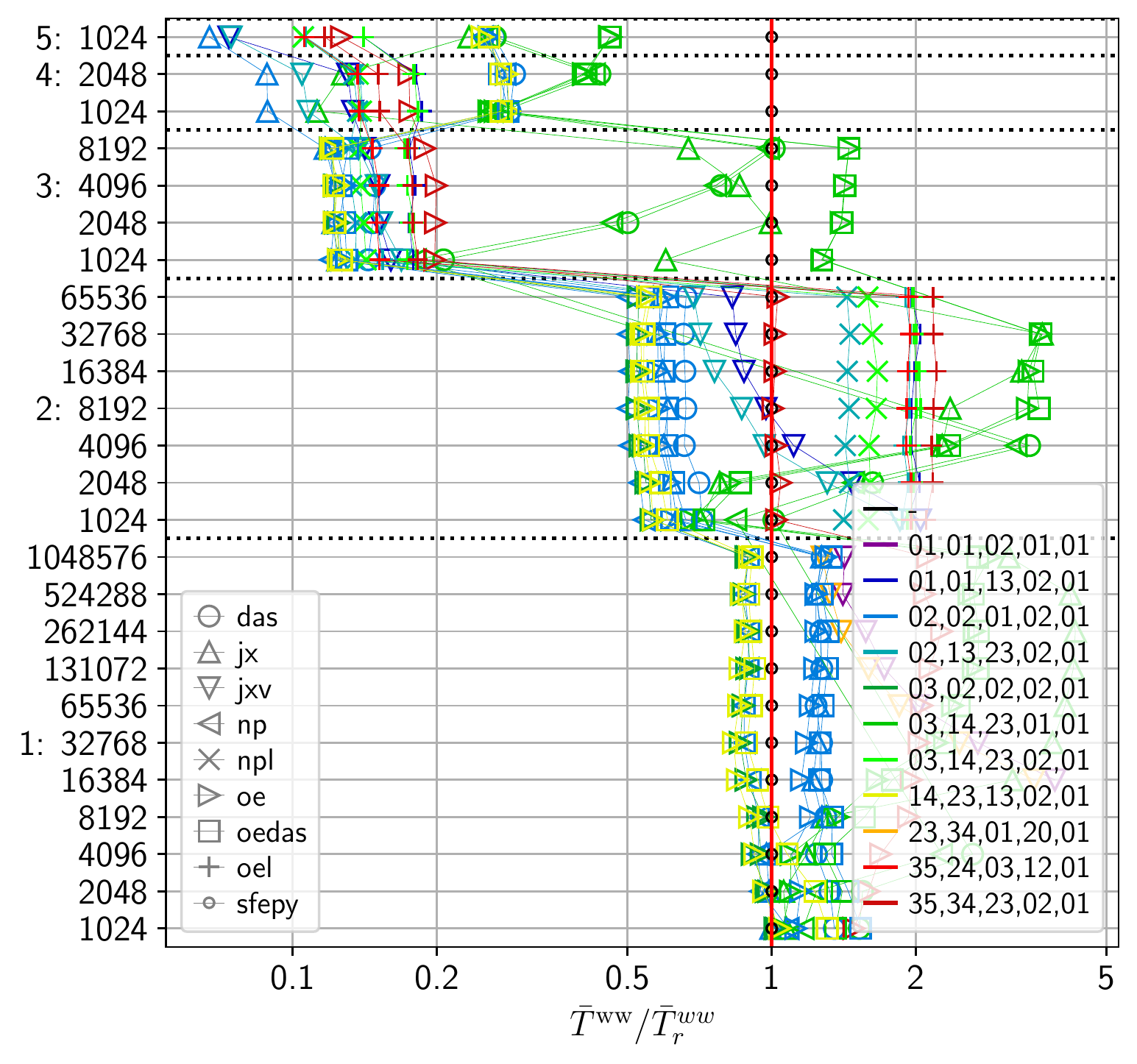}
  \includegraphics[width=\myrelwidth]{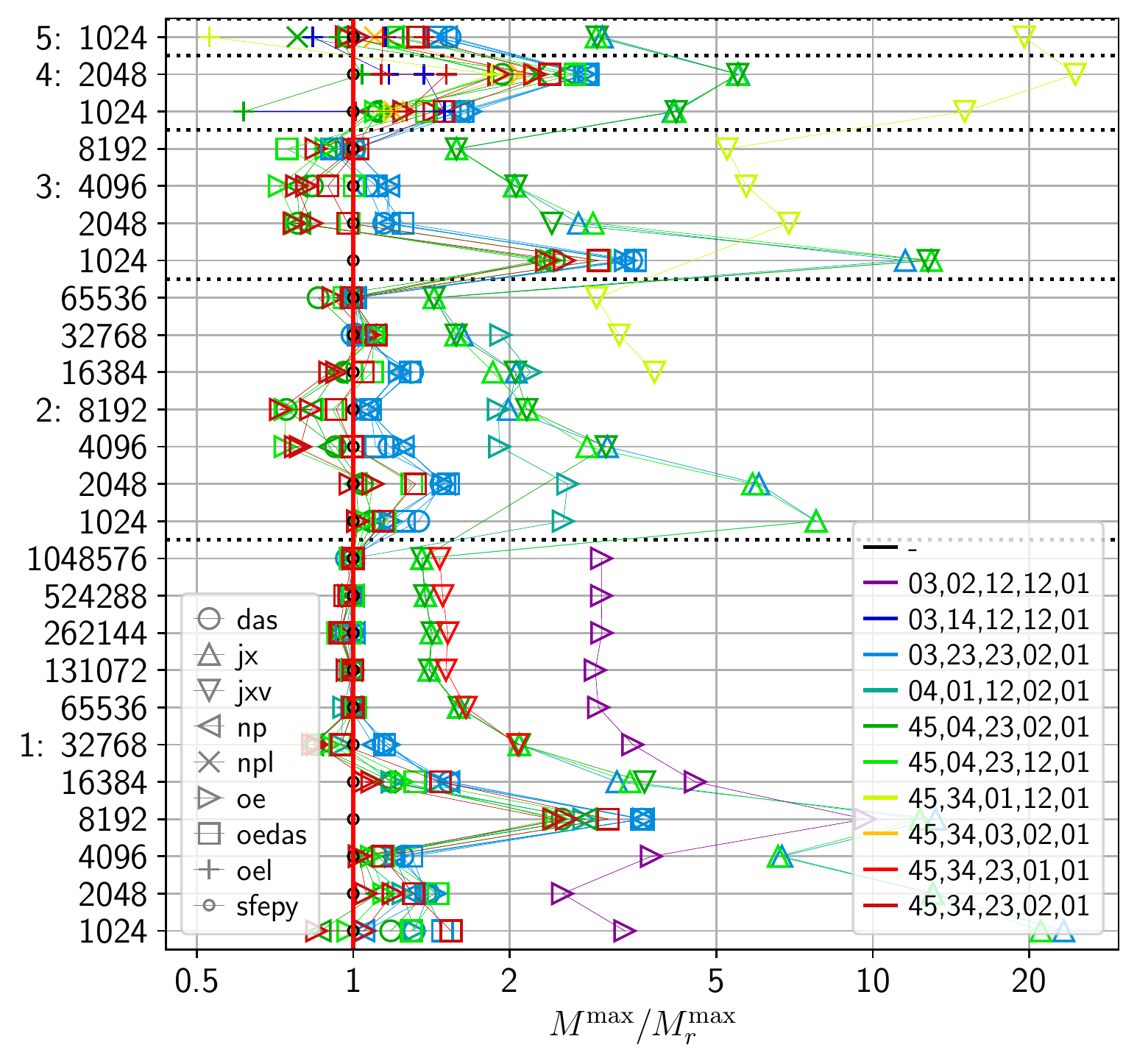}
  \includegraphics[width=\myrelwidth]{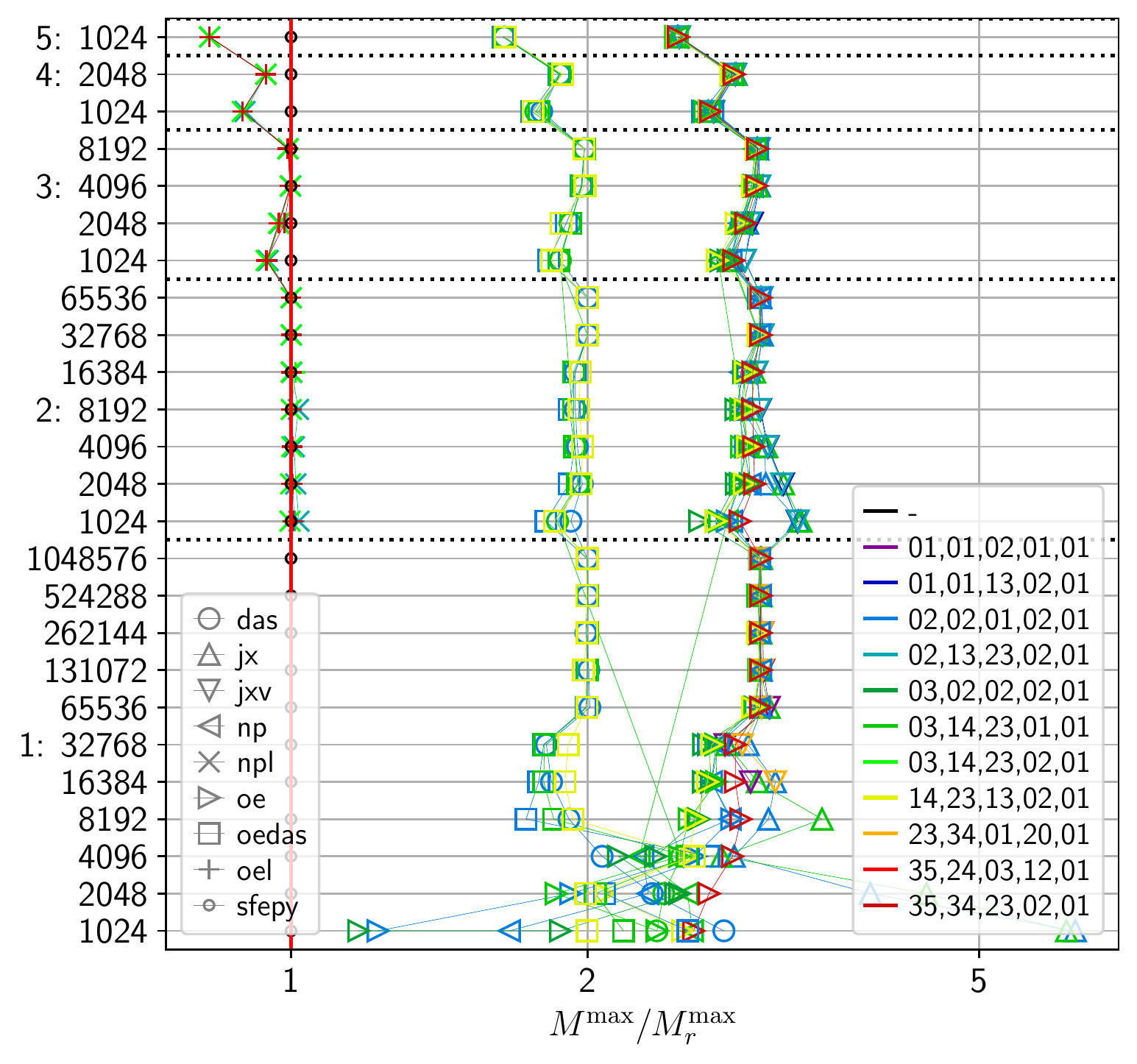}
  \caption{Relative elapsed time and memory requirements of the weighted vector
    dot product einsum expression w.r.t. the reference implementation. Left:
    residual mode, right: matrix mode, top: $\rtwwmean$, bottom: $\rmmax$. The
    backends are indicated by markers, the einsum contraction paths by color.
    The cell counts ($y$ axis) are grouped by the approximation orders.}
  \label{fig:at-4}
\end{figure}

The performance of the einsum backends applied to the evaluation of the
weighted vector dot product einsum expression is shown in Figs.~\ref{fig:atr-4}
(evaluation throughput) and \ref{fig:at-4} (relative performance).

In the residual mode, the jx backend was the fastest, followed by oe and np.
From the order 2 it consistently out-performed the reference implementation, from
the order 3 also oe, np, oedas and jxv and from the order 4 also das. The memory
consumption stayed mostly less than $5\times$ reference.

In the matrix mode, most backends performed well. The oe and np backends were
the fastest in most cases and out-performed the reference implementation. From
the order 2 also jx, oedas, das and jxv, from the order 3 npl and oel were
faster than the reference code. The jx and jxv backends were the fastest from
the order 4. The memory consumption stayed mostly less than $3\times$
reference, the npl and oel backends had smaller footprint from the order 3.


\subsubsection{Linear Elasticity}
\label{sec:le}

\begin{figure}[p!]
  \centering
  \includegraphics[width=\myrelwidth]{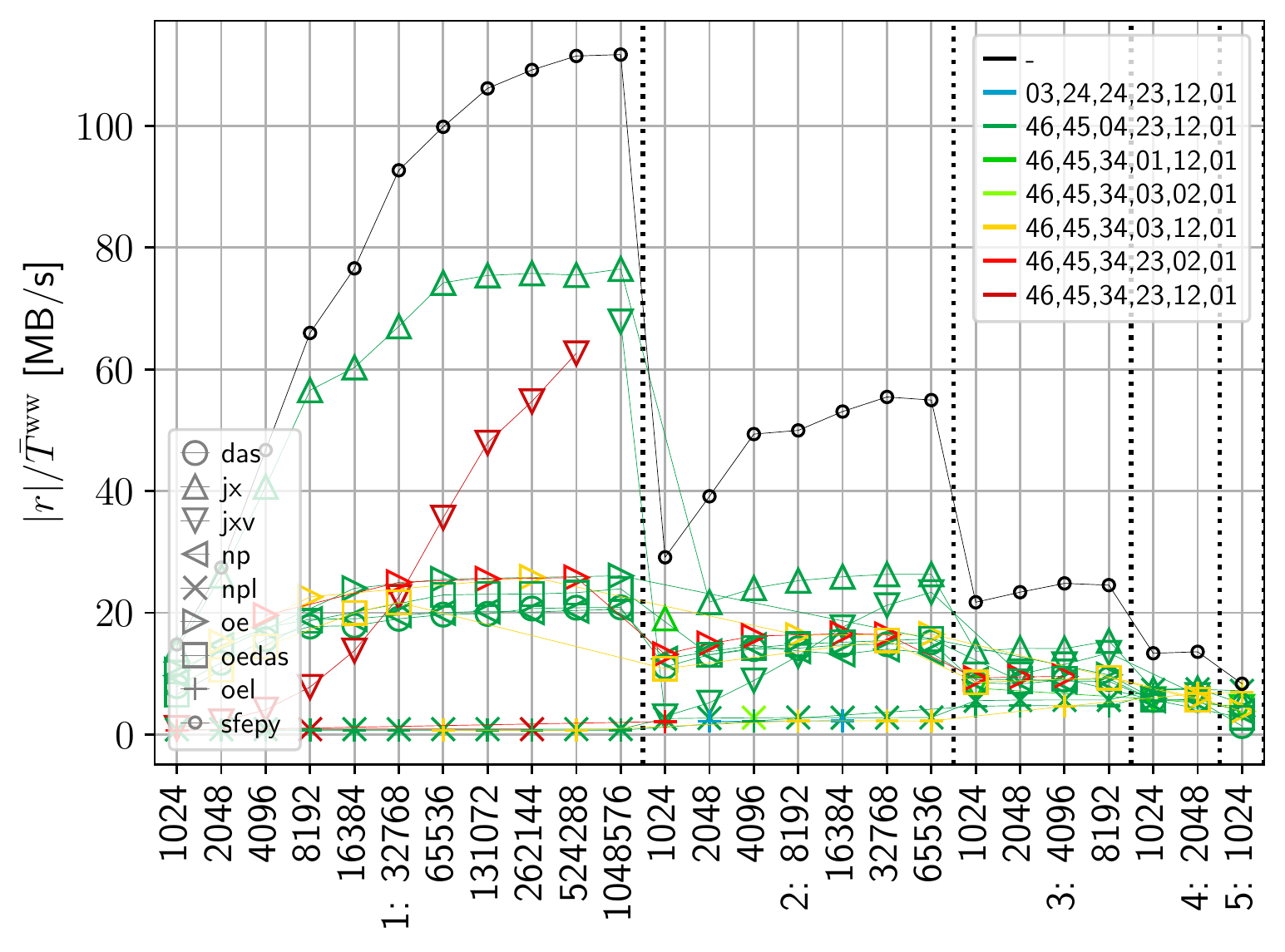}
  \includegraphics[width=\myrelwidth]{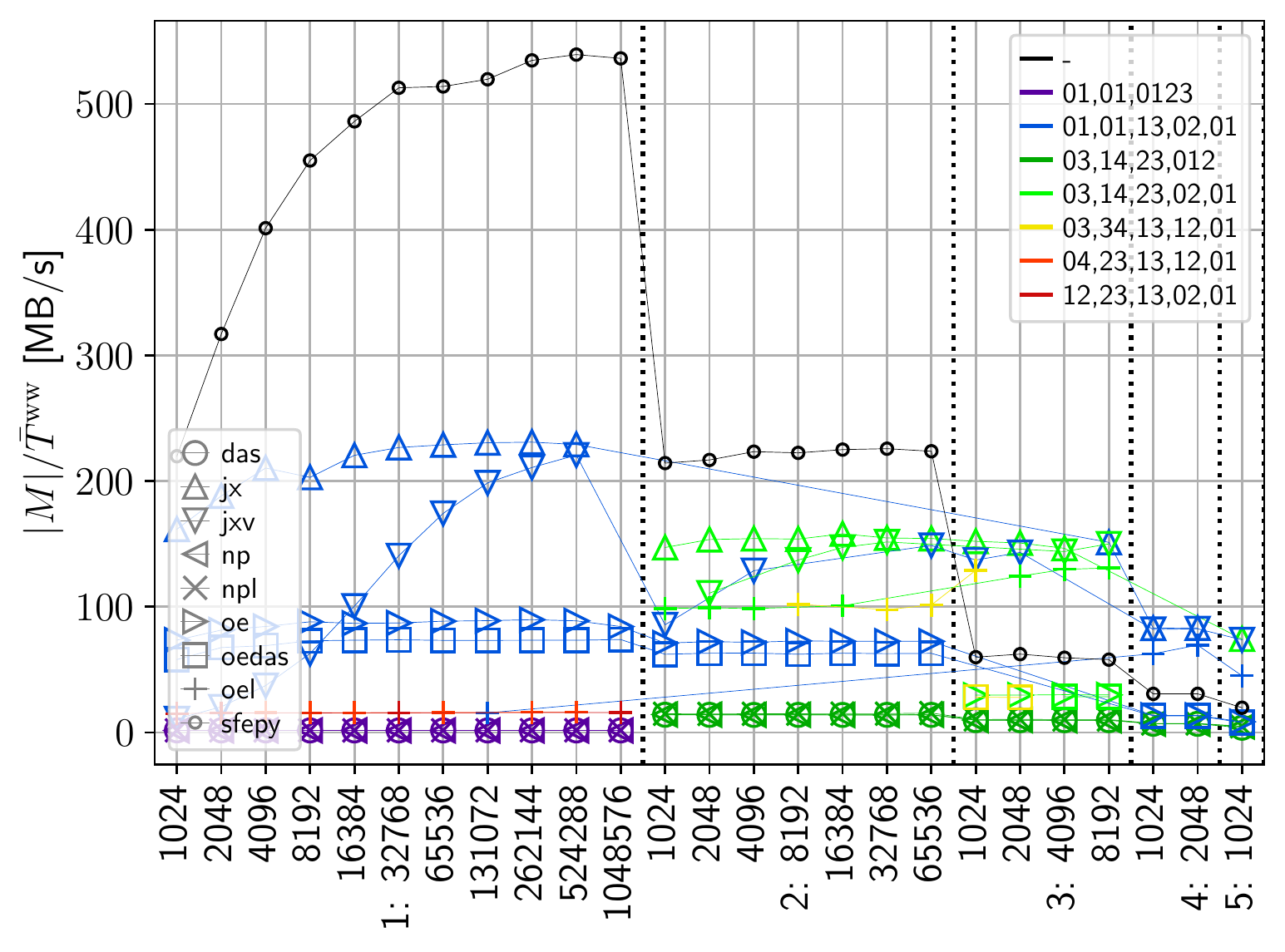}
  \caption{The highest throughput of each backend (number of megabytes
    evaluated per second) for the weak linear elasticity term einsum
    expression. Left: residual mode evaluations $|r| / \twwmean$, right: matrix
    mode evaluations $|M| / \twwmean$. The backends are indicated by markers,
    the einsum contraction paths by color. The cell counts ($x$ axis) are
    grouped by the approximation orders.}
  \label{fig:atr-5}
\end{figure}

\begin{figure}[p!]
  \centering
  \includegraphics[width=\myrelwidth]{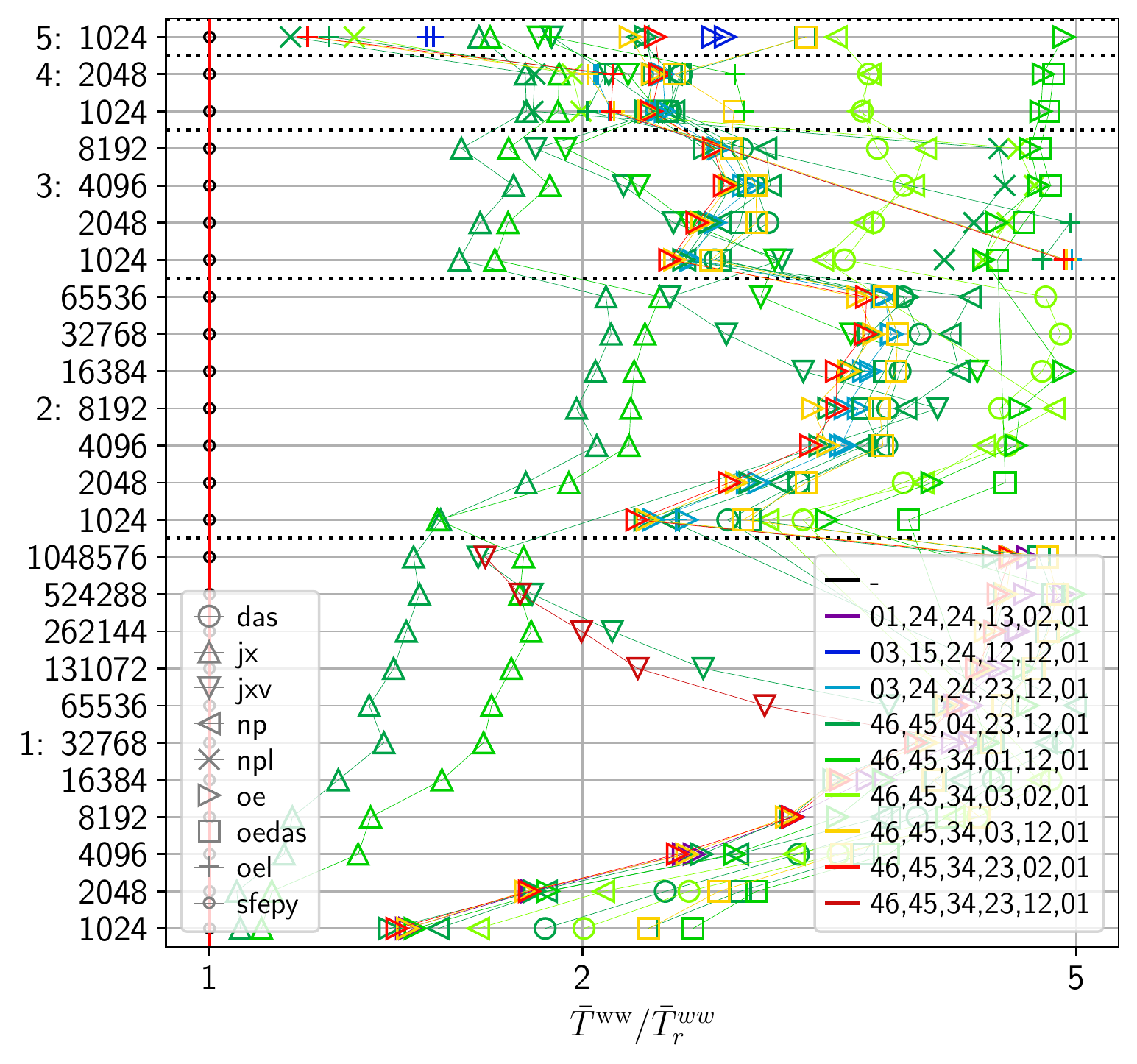}
  \includegraphics[width=\myrelwidth]{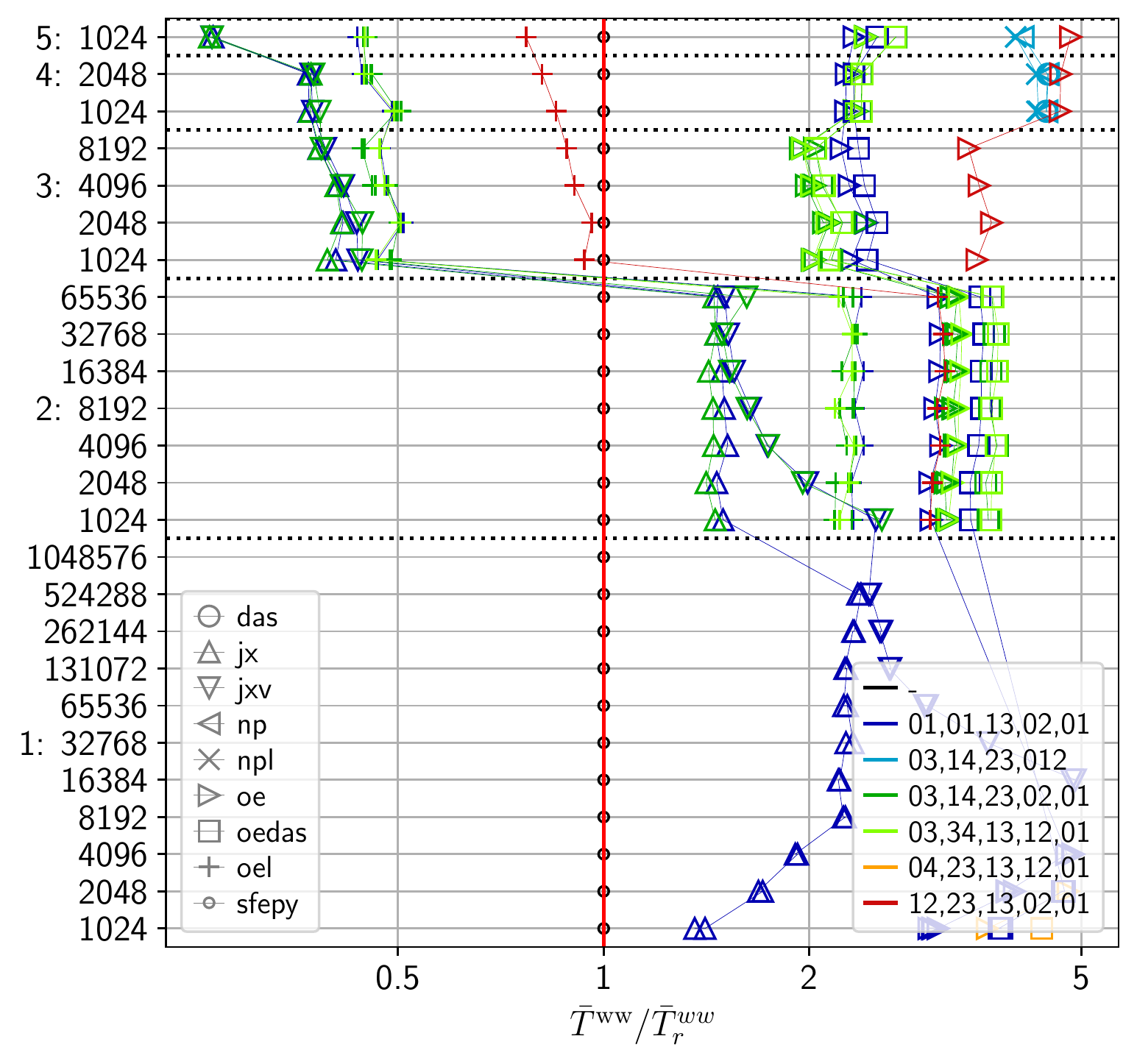}
  \includegraphics[width=\myrelwidth]{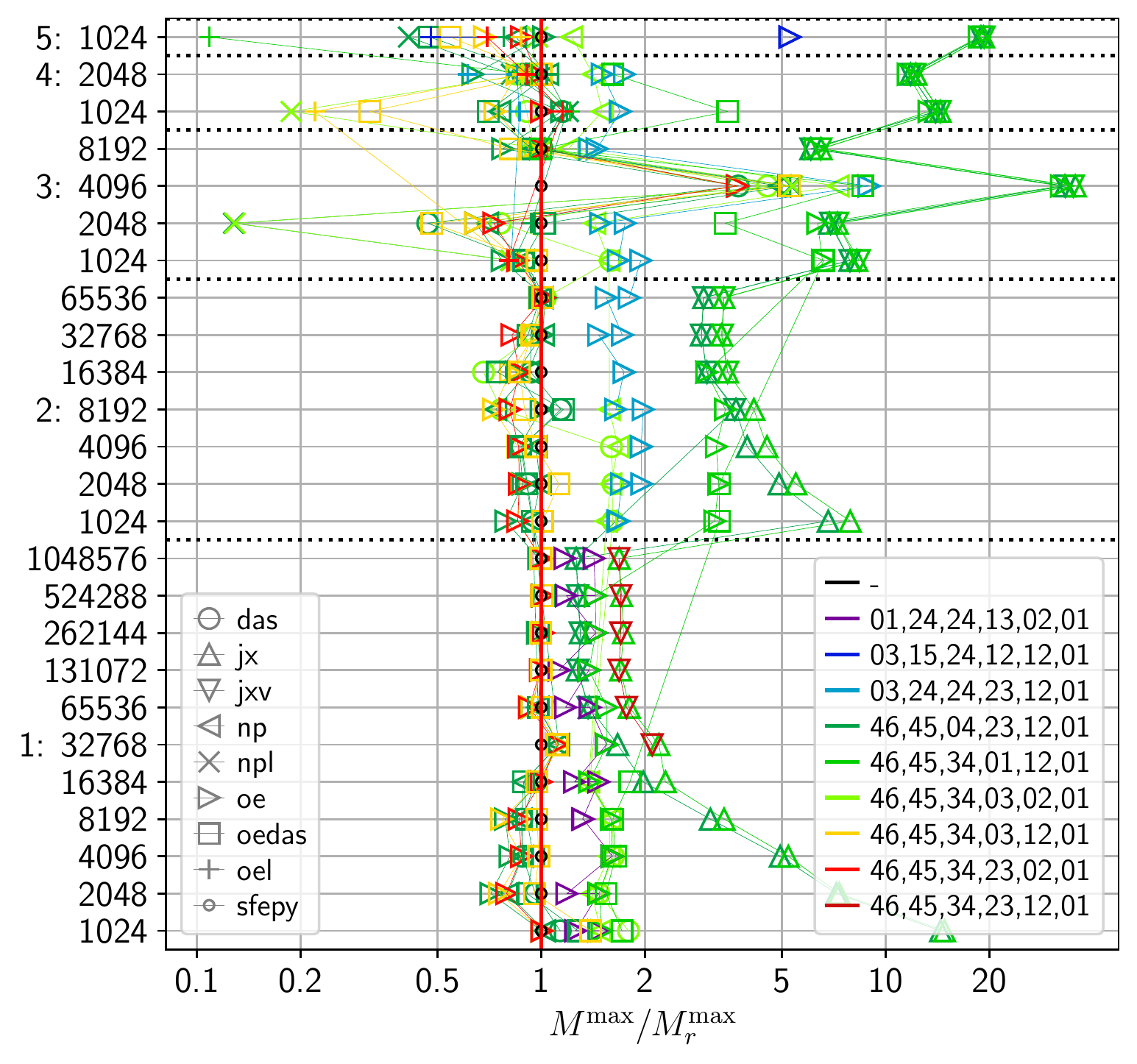}
  \includegraphics[width=\myrelwidth]{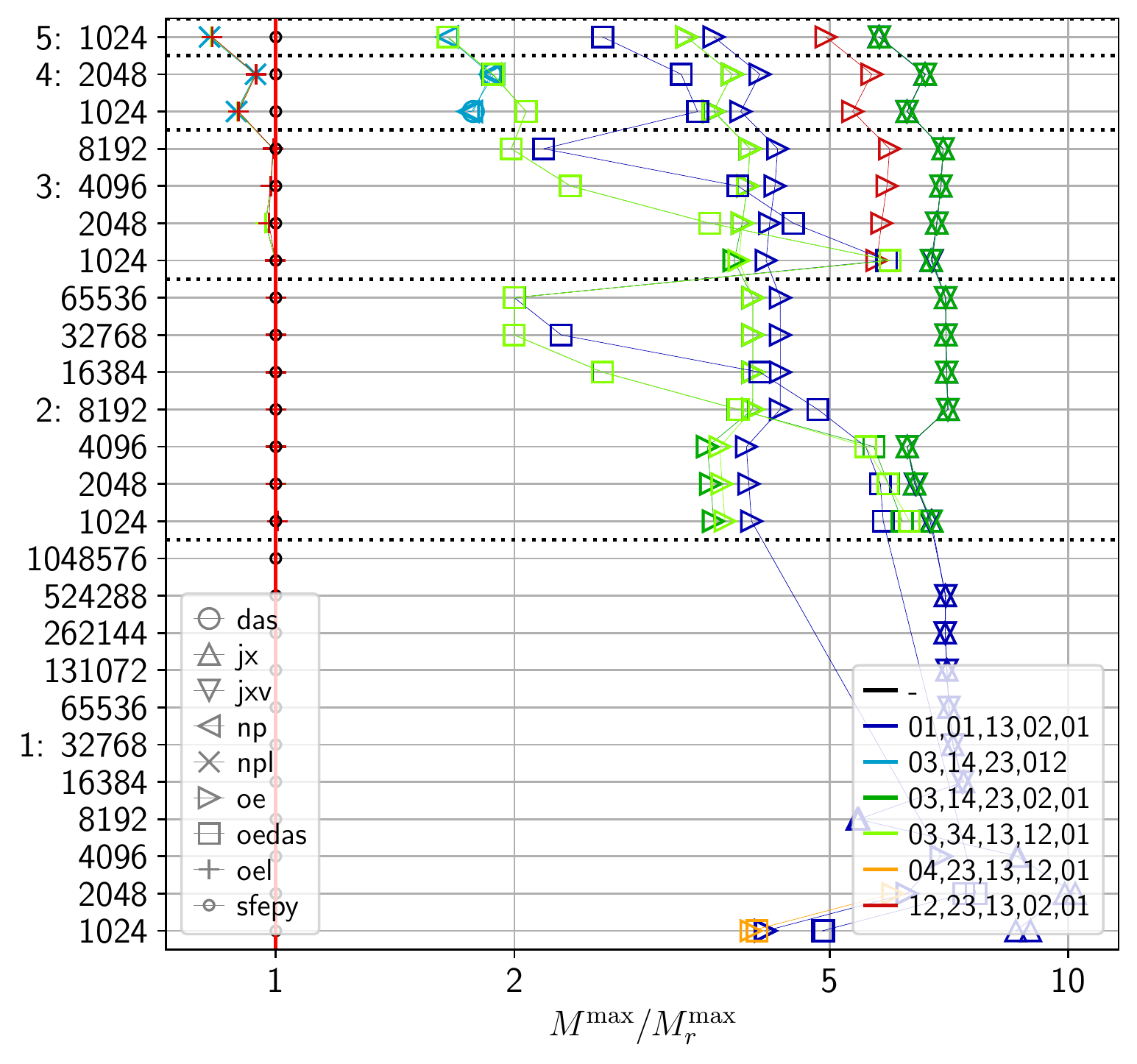}
  \caption{Relative elapsed time and memory requirements of the weak linear
    elasticity term einsum expression w.r.t. the reference implementation. Left:
    residual mode, right: matrix mode, top: $\rtwwmean$, bottom: $\rmmax$. The
    backends are indicated by markers, the einsum contraction paths by color.
    The cell counts ($y$ axis) are grouped by the approximation orders.}
  \label{fig:at-5}
\end{figure}

The performance of the einsum backends applied to the evaluation of the weak
linear elasticity term einsum expression is shown in Figs.~\ref{fig:atr-5}
(evaluation throughput) and \ref{fig:at-5} (relative performance).

In the residual mode, the jx backend was the fastest but mostly almost
$2\times$ slower than the reference implementation. For the order 5, the npl
and oel backends performed better than jx, but they were still slower than the
reference code. The memory consumption of oel, npl and das was smaller and of
oedas, das smaller or comparable to the reference implementation. Overall it
mostly stayed less than $5\times$ reference, except for jx and jxv from the
order 3, where it reached almost $30\times$ reference values.

In the matrix mode, the jx, jxv backends performed the best and were faster
than reference from the order 3, together with the oel backend.
The memory consumption was mostly less than $7\times$ reference values.

The relatively low performance in this case, especially in the residual mode,
was caused by not exploiting the sparsity of the \Verb|Psg| tensor, see the
last paragraph of Section \ref{sec:lee}.


\section{Conclusion}
\label{sec:con}

In this paper, we proposed and described a simple transpiler from generalized
einsum-like expressions, suitable for describing multi-linear finite element weak
forms, to regular tensor contraction expressions. We applied the transpiler and
presented a very simple implementation of several standard weak forms. The
transpiler supports several Python libraries as backends implementing the
einsum function which allowed us to perform a large numerical study comparing
the backends (NumPy, opt\_einsum, Dask and JAX) mutually and with the reference
implementations of the weak forms in the finite element SfePy.

An indispensable feature of the backends was their ability to optimize the
tensor contraction paths. Our study confirmed that using contraction paths
optimizations is crucial for good performance. However, the theoretical
speedup, as e.g. calculated by opt\_einsum, was found to be not a good metric
for estimating performance when applied to simple einsum expressions occurring
in the FE calculations --- here the additional memory consumed by temporary
storage was much more important. This was also one of the motivations for
realizing the presented numerical study.

Concerning the study results, we first studied the influence of the memory
layout of the expression operands on the evaluation performance, to establish
the position of the default, SfePy compatible layout, with other possible
layouts. We found that marked improvements could be achieved by using
alternative memory layouts.

Then five selected weak forms were evaluated for various mesh sizes and FE
approximation orders using the default layout. The evaluations were repeated
several times for each parameter set and statistics such as means without the
worst case were computed.

Significant differences in performance were found between the residual
(operator/matrix application to a DOF vector without the global sparse matrix
assembly) and matrix (calculating local element matrices) evaluation modes as
well as among the various weak forms --- no single backend/contraction path was
the best in all the cases.

Overall, JAX based backends were the fastest, but very memory demanding. The
JIT compilation warm-up was disregarded by omitting the worst case repetition
in calculating the elapsed time mean. Without this, the JAX backends were much
slower for small problems. GPU/TPU performance was not considered in this
study.

A good compromise between speed and memory overhead w.r.t. the reference
implementation was achieved by using the opt\_einsum or NumPy backends.
Especially their cell-loop based variants had very low memory requirements and
offered good performance for FE approximation orders greater or equal to three
in the matrix evaluation mode. Also the Dask-based backends were suitable for
large problems, where the automatic chunking of operands decreased the memory
consumption. Dask allows out-of-core calculations with data not fitting into
RAM, but this feature was not used in this study.

In many cases the einsum expression based evaluations were faster than the
reference implementation (a hand-crafted C code, called from Python via SfePy's
lightweight Cython wrappers) which is in our opinion a good result considering
the simplicity of the transpiler implementation and especially of the weak form
definitions. For forms with low numerical intensity (e.g. the weak Laplacian in
matrix mode) or unused sparsity (the linear elasticity term), going beyond
einsum might be required for further speed-up. A very preliminary result
included in the memory layouts study indicates that using Numba might be
promising.

The transpiler and einsum expression based weak form implementations are
available in SfePy from version 2021.1, allowing a rapid prototyping of
multi-physical finite element models and subsequent calculations in various
fields such as biomechanics \cite{Turjanicova_Rohan_Lukes_2019} or solid state
physics \cite{Novak_Vackar_Cimrman_2020}. All data used in preparation of this
paper are available online \cite{zenodo-time-tensors-data}.

\section*{Acknowledgements}

The work was supported from European Regional Development Fund --- Project
``Application of Modern Technologies in Medicine and Industry'' (No.
CZ.02.1.01/0.0/0.0/17\_048/0007280).

\appendix

\section{Performance of Selected Weak Forms Contractions with Threads}
\label{sec:pswfct}

This appendix complements Section~\ref{sec:pswfc} by using the same setup but
the single-threaded execution was not enforced --- default settings of each
backend library were in effect. Two additional backends --- dat and oedat --- were
included, i.e. Dask based backends with multiple threads enabled.

A parametric study run using soops \cite{soops2021github} lead to 8326 finished
simulations\footnote{This includes the reference implementation simulations
  shown in Fig.~\ref{fig:rip}. 31 backend/optimization pairs, 27 mesh
  size/order combinations, 5 terms and 2 evaluation modes, 8370 in total, 44
  failed due to insufficient memory.}, each evaluating a single weak form using
a single backend/optimization combination several times according to the repeat
parameter for various mesh sizes and function approximation orders, see
Tab.~\ref{tab:cdc}. The default \Verb|cqgvd0| layout of the weak form einsum
expression operands and result arrays was used.

\begin{figure}[htp!]
  \centering
  \includegraphics[width=0.48\linewidth]{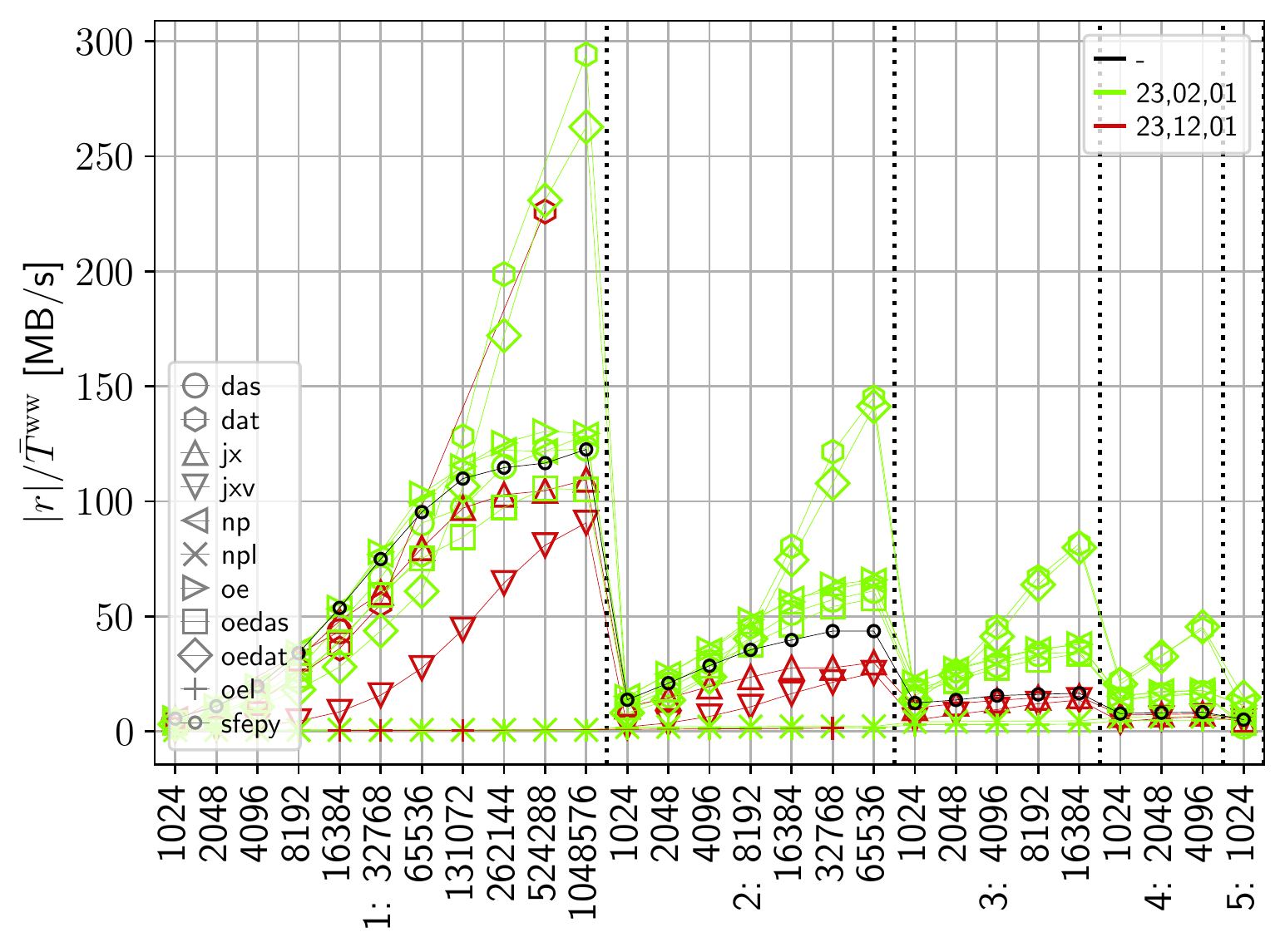}
  \includegraphics[width=0.48\linewidth]{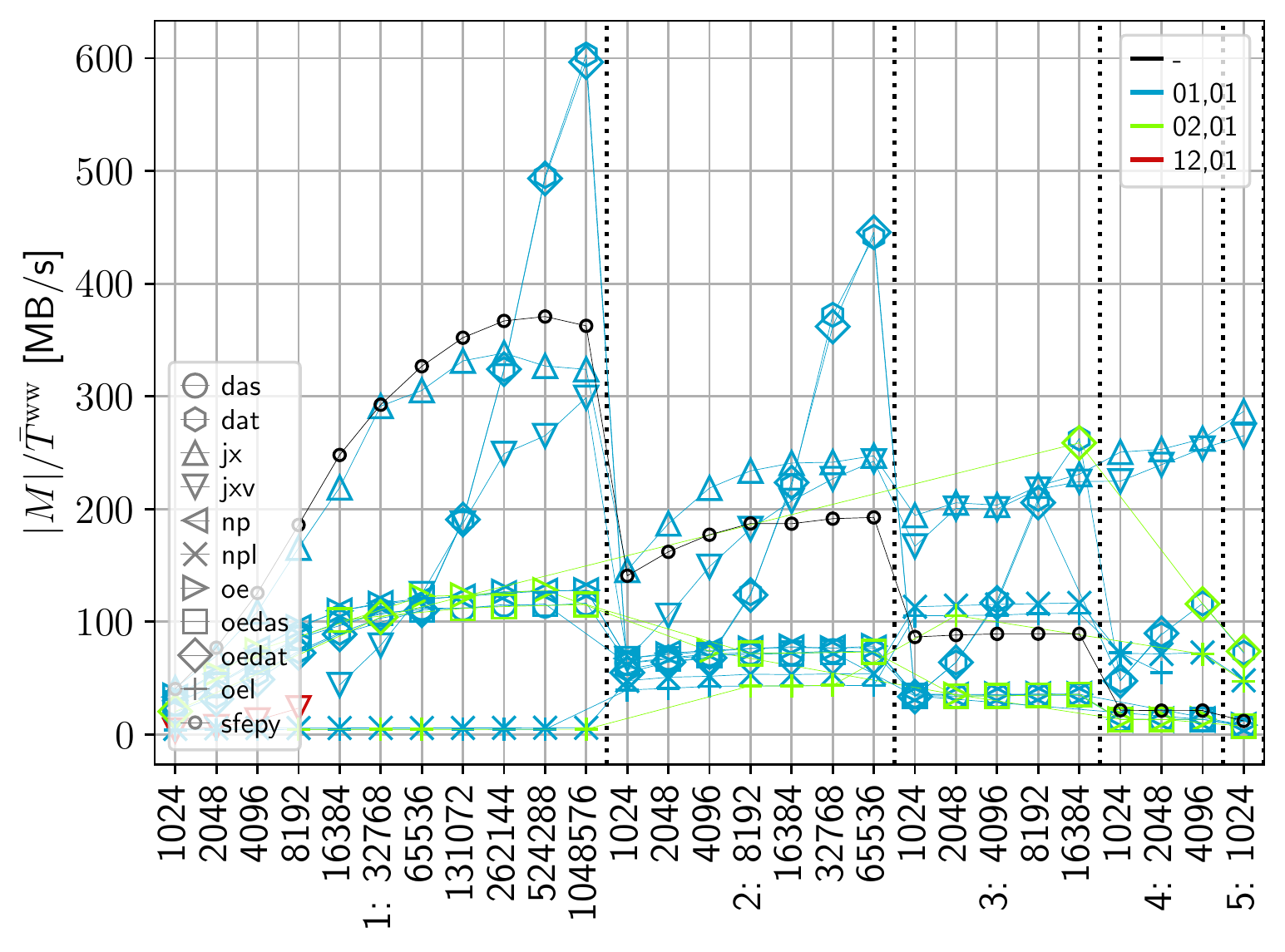}
  \caption{The highest throughput of each backend (number of megabytes
    evaluated per second) for the weak Laplacian einsum expression with allowed
    threads. Left: residual mode evaluations $|r| / \twwmean$, right: matrix
    mode evaluations $|M| / \twwmean$. The backends are indicated by markers,
    the einsum contraction paths by color. The cell counts ($x$ axis) are
    grouped by the approximation orders.}
  \label{fig:atrm-1}
\end{figure}

\begin{figure}[htp!]
  \centering
  \includegraphics[width=0.48\linewidth]{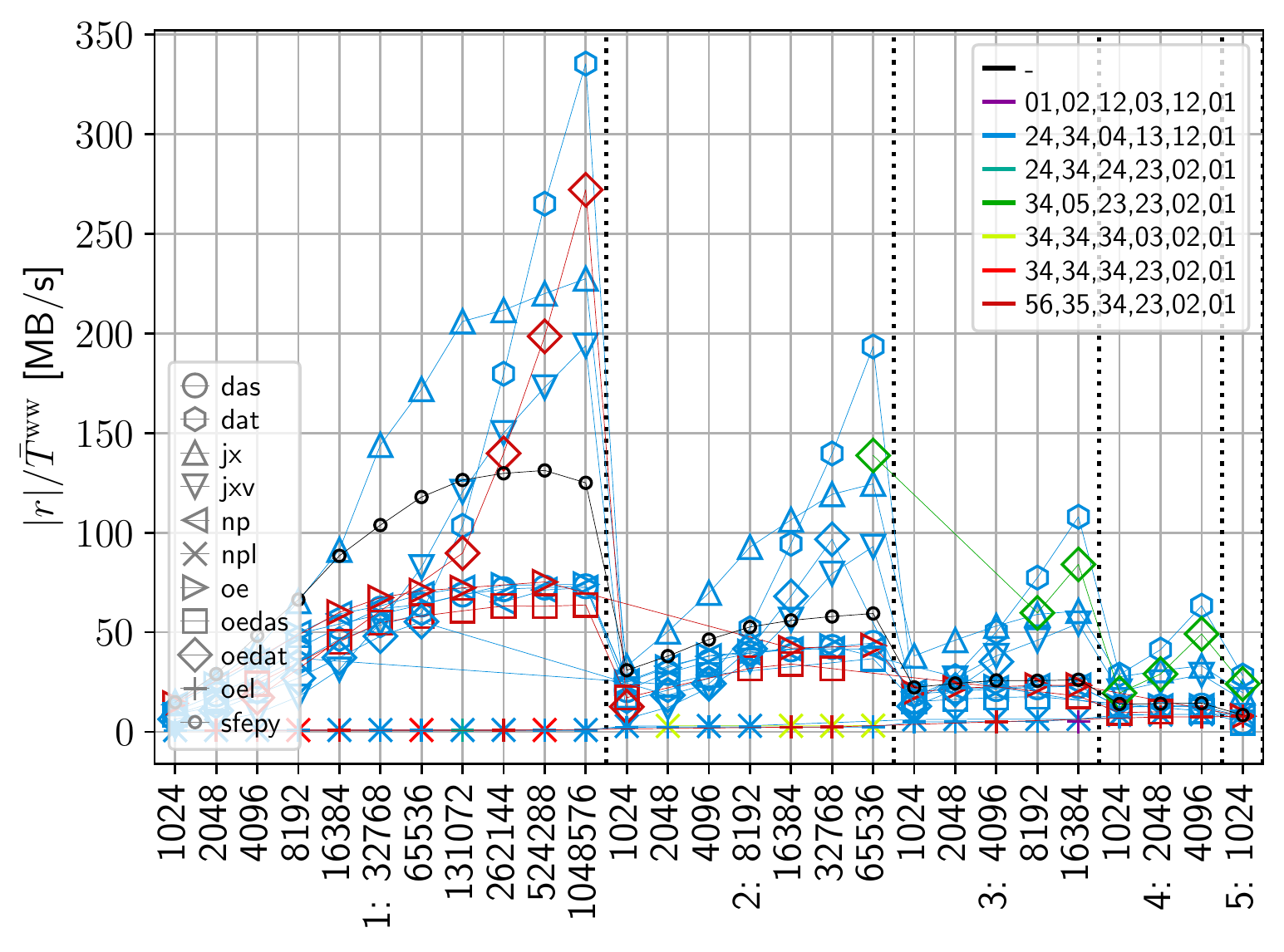}
  \includegraphics[width=0.48\linewidth]{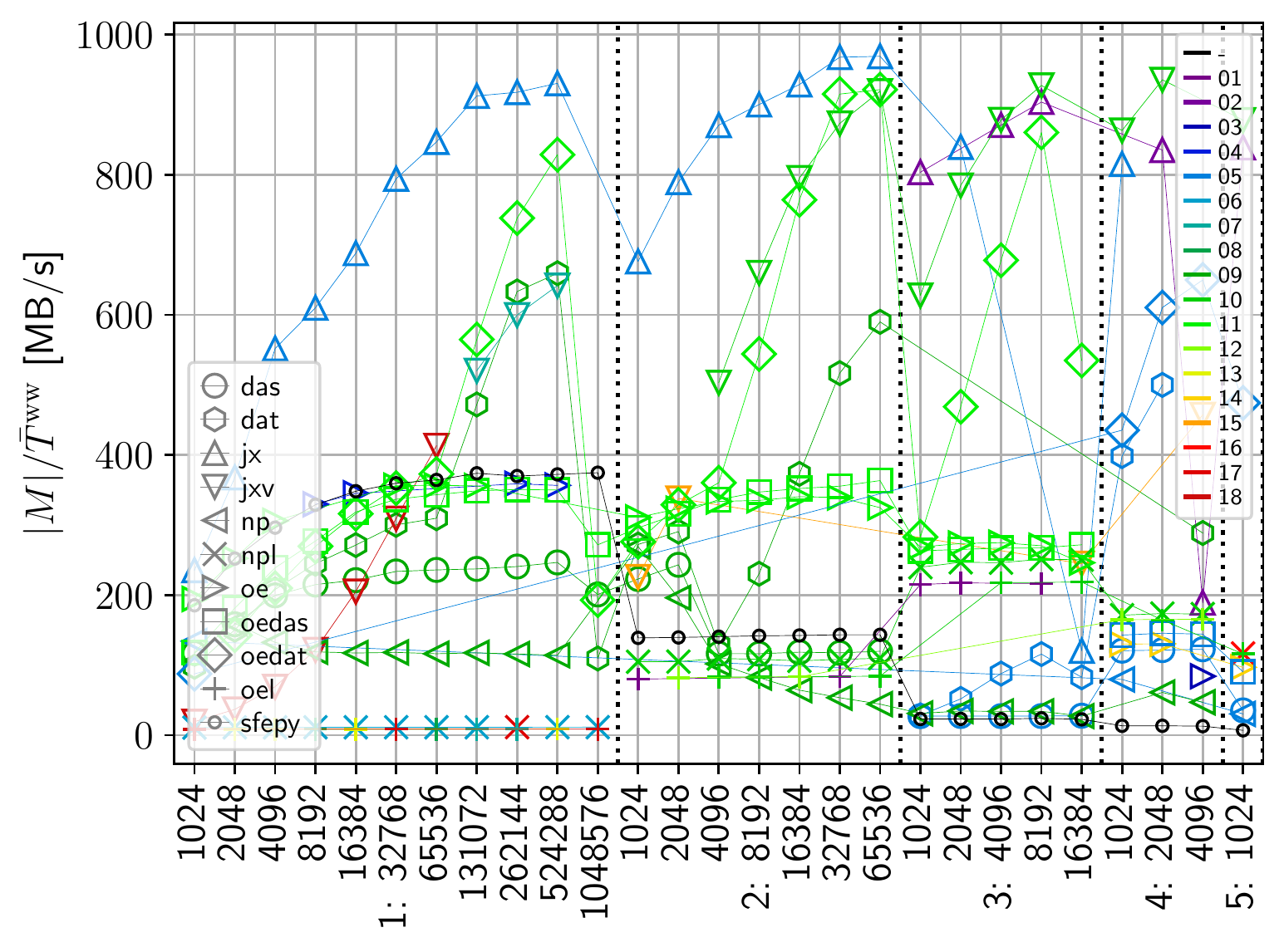}
  \caption{The highest throughput of each backend (number of megabytes
    evaluated per second) for the weak convective term einsum expression with
    allowed threads. Left: residual mode evaluations $|r| / \twwmean$, right:
    matrix mode evaluations $|M| / \twwmean$. The backends are indicated by
    markers, the einsum contraction paths by color. The cell counts ($x$ axis)
    are grouped by the approximation orders.}
  \label{fig:atrm-2}
\end{figure}

\begin{figure}[htp!]
  \centering
  \includegraphics[width=0.48\linewidth]{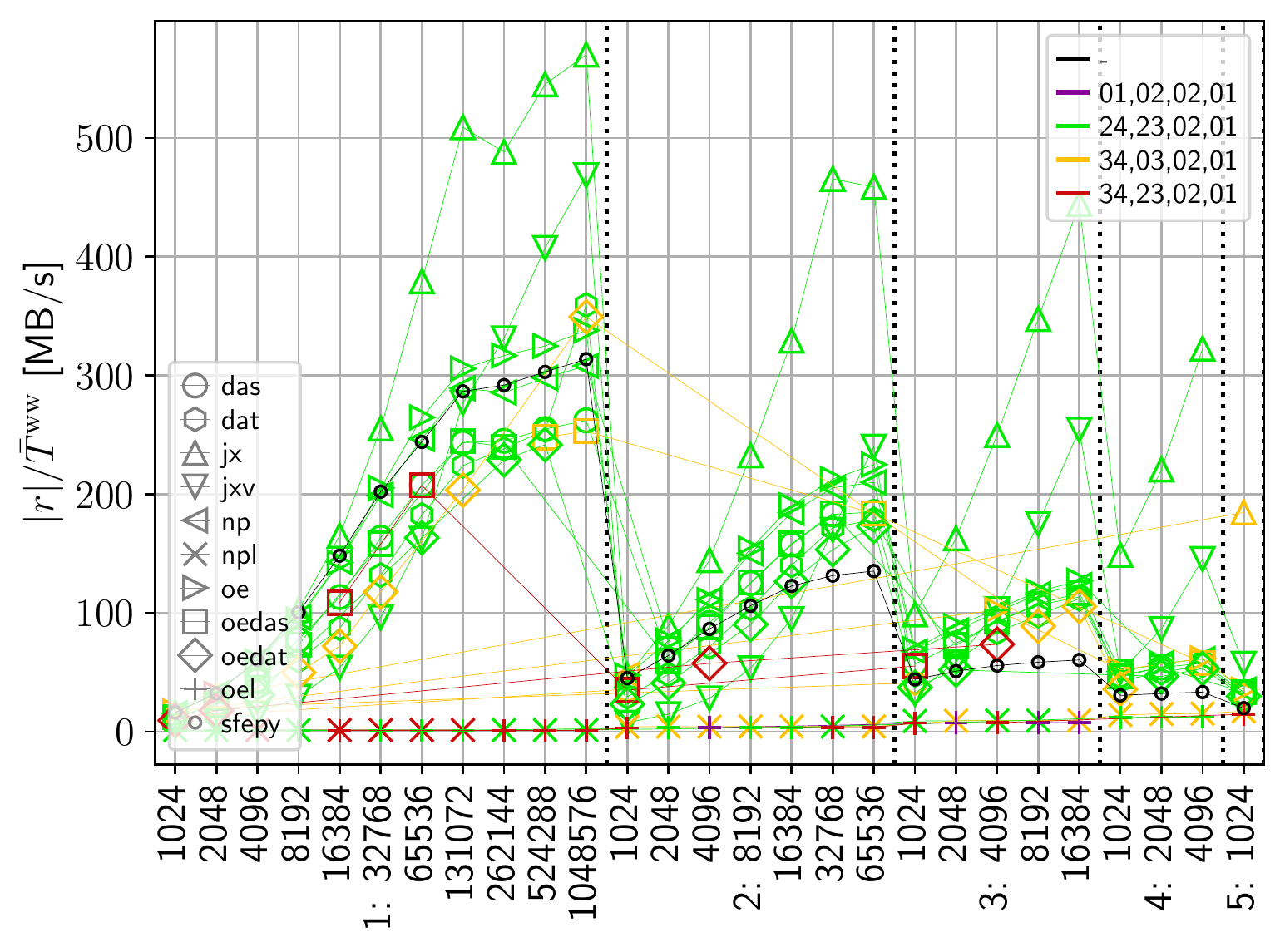}
  \includegraphics[width=0.48\linewidth]{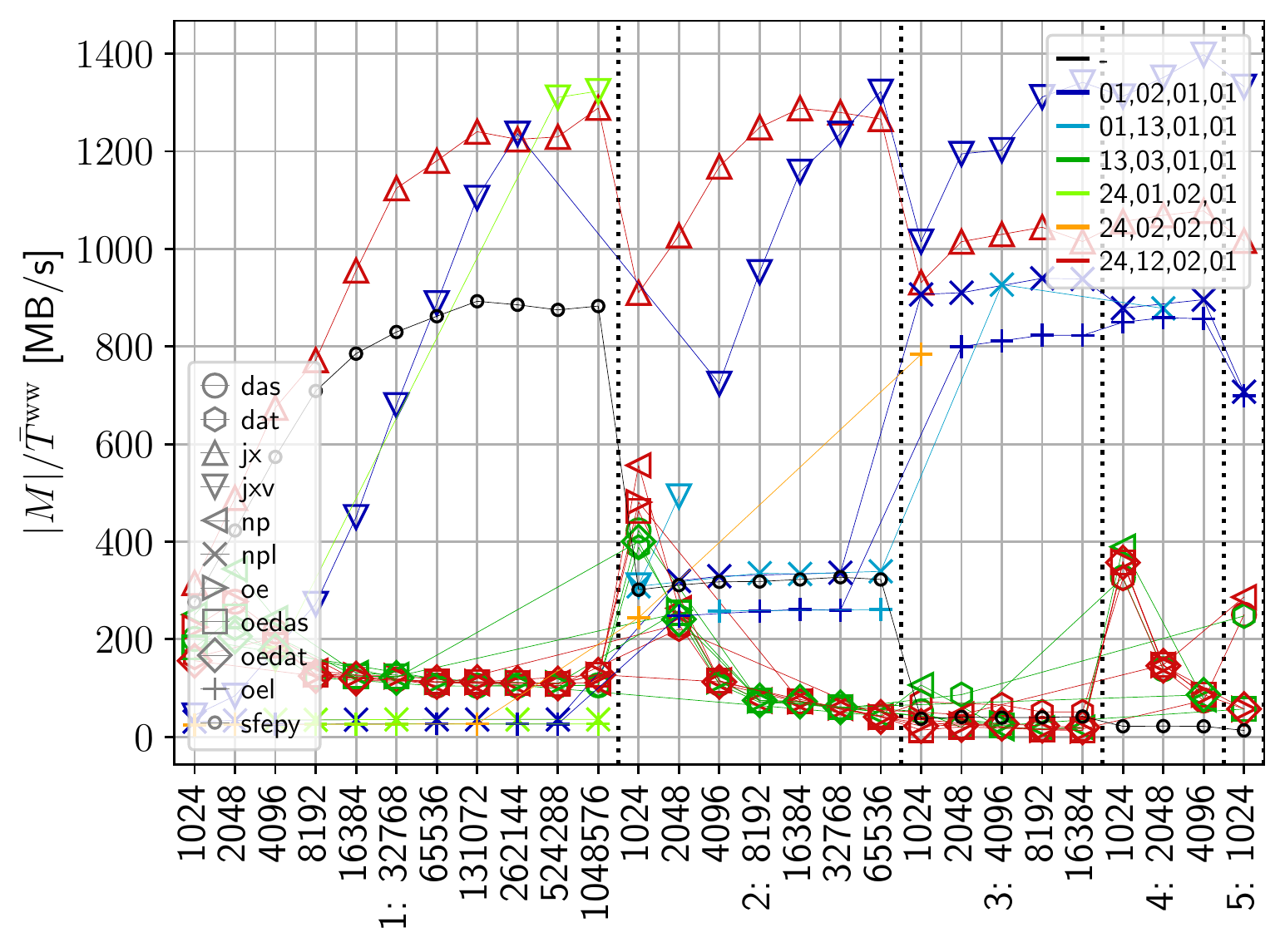}
  \caption{The highest throughput of each backend (number of megabytes
    evaluated per second) for the vector dot product einsum expression with
    allowed threads. Left: residual mode evaluations $|r| / \twwmean$, right:
    matrix mode evaluations $|M| / \twwmean$. The backends are indicated by
    markers, the einsum contraction paths by color. The cell counts ($x$ axis)
    are grouped by the approximation orders.}
  \label{fig:atrm-3}
\end{figure}

\begin{figure}[htp!]
  \centering
  \includegraphics[width=0.48\linewidth]{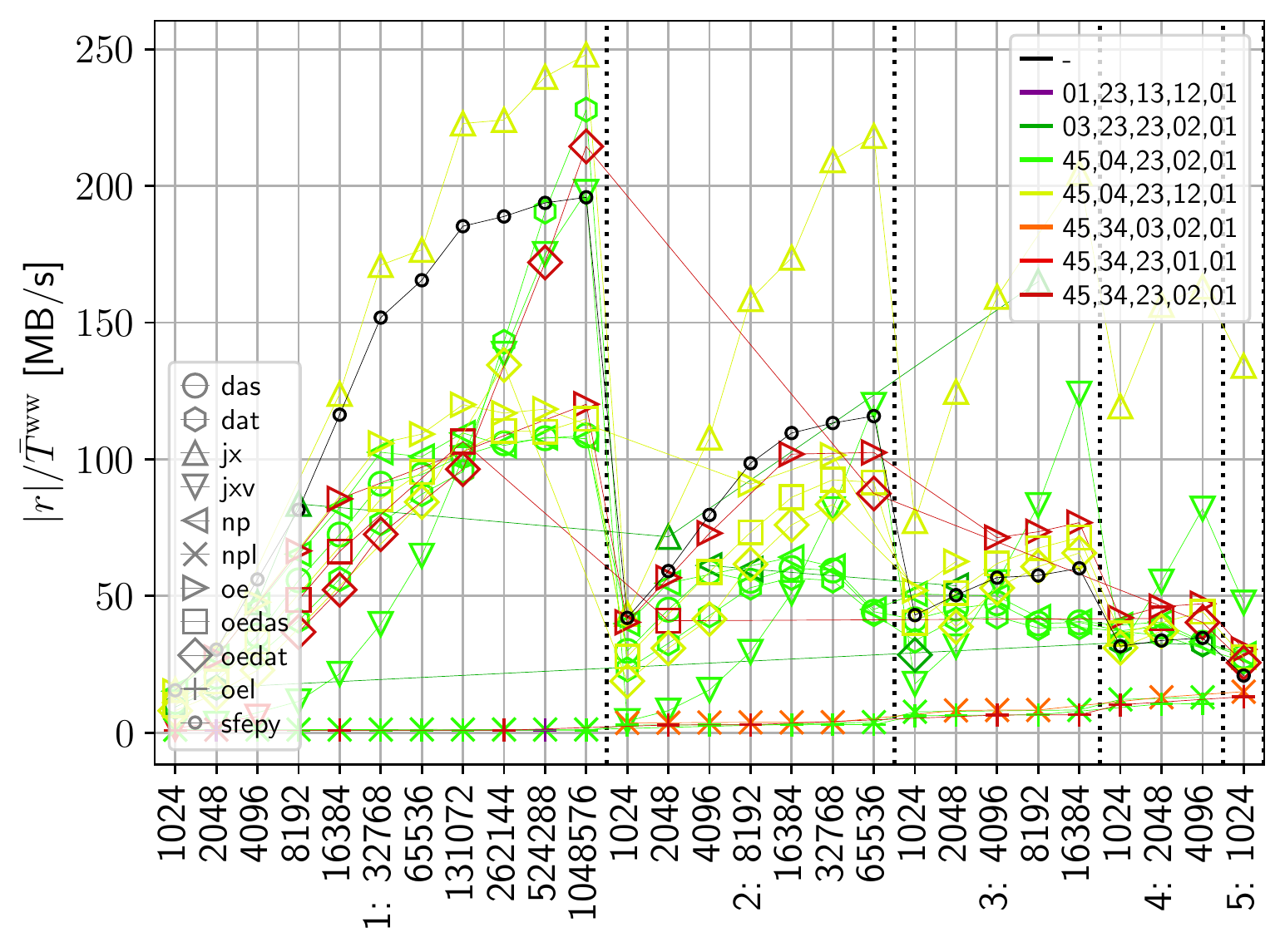}
  \includegraphics[width=0.48\linewidth]{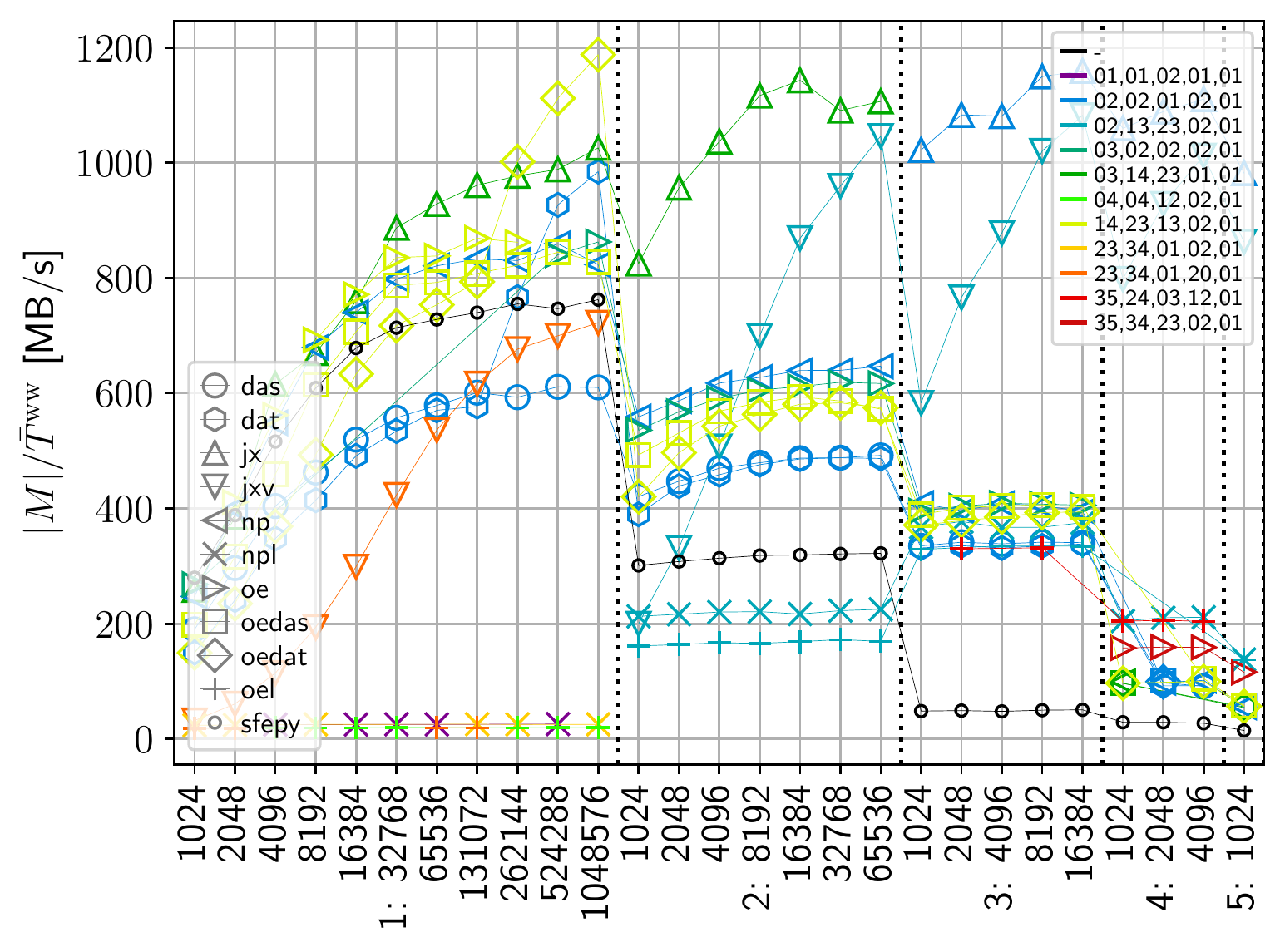}
  \caption{The highest throughput of each backend (number of megabytes
    evaluated per second) for the weighted vector dot product einsum expression
    with allowed threads. Left: residual mode evaluations $|r| / \twwmean$,
    right: matrix mode evaluations $|M| / \twwmean$. The backends are indicated
    by markers, the einsum contraction paths by color. The cell counts ($x$
    axis) are grouped by the approximation orders.}
  \label{fig:atrm-4}
\end{figure}

\begin{figure}[htp!]
  \centering
  \includegraphics[width=0.48\linewidth]{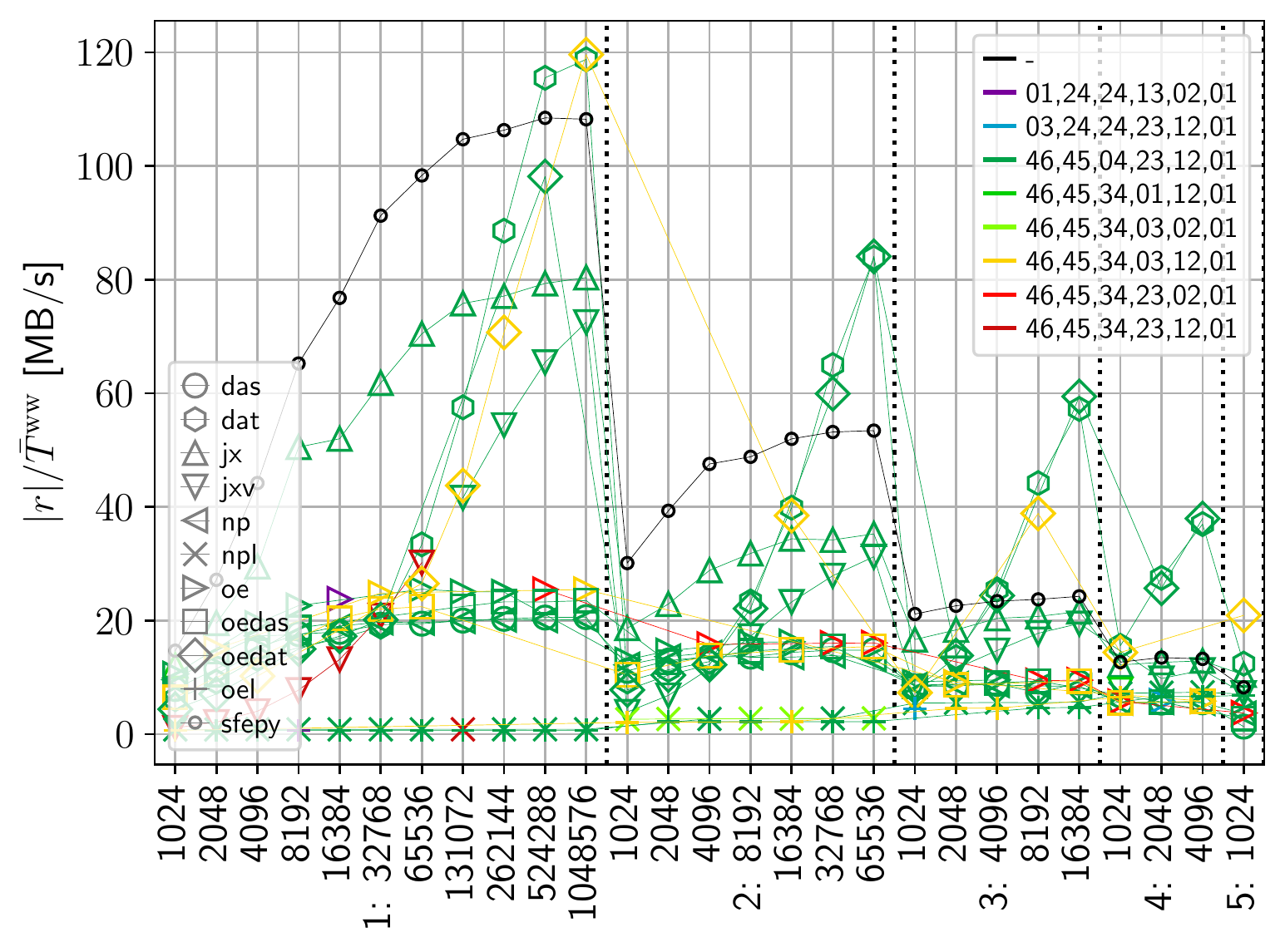}
  \includegraphics[width=0.48\linewidth]{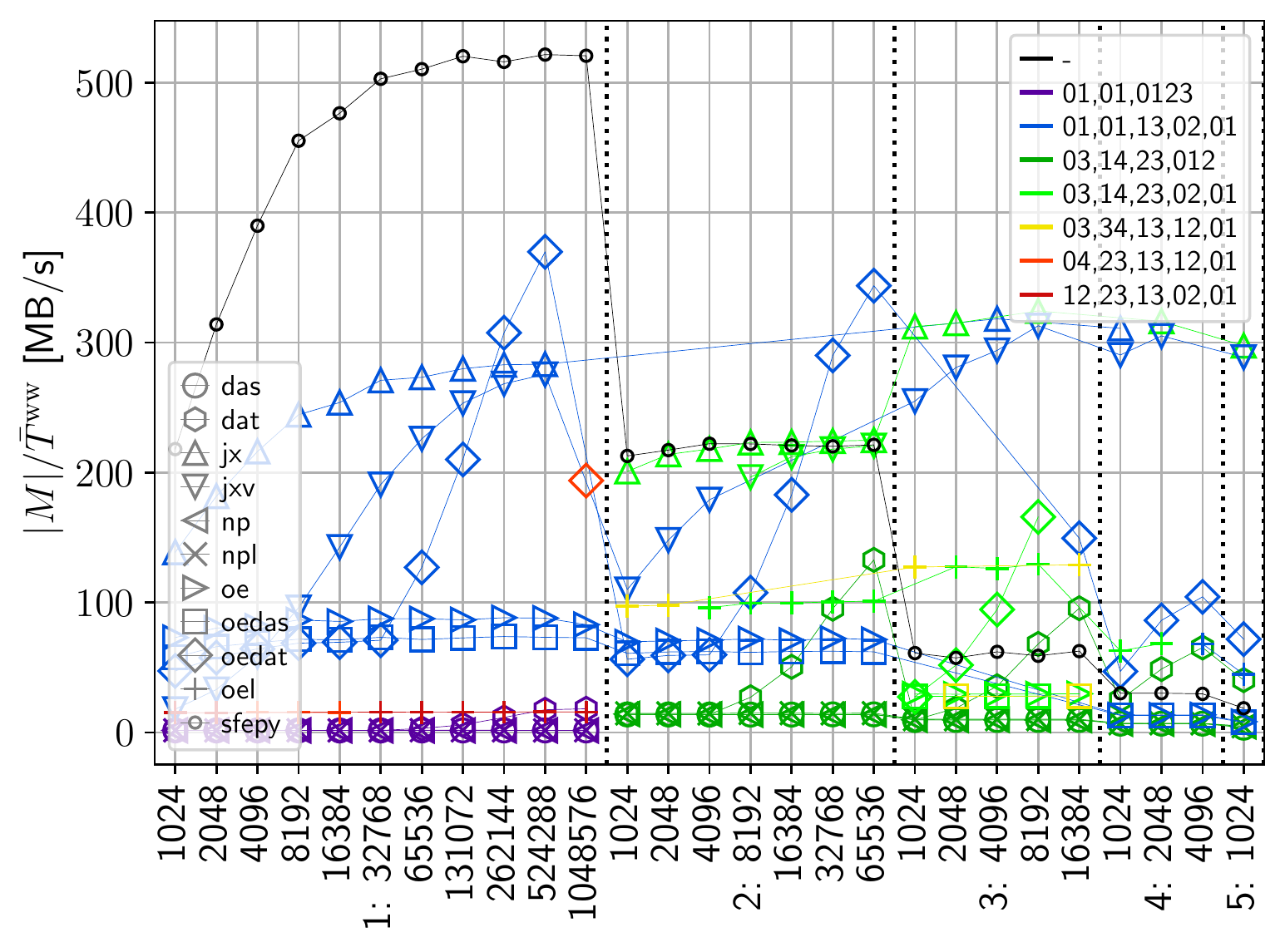}
  \caption{The highest throughput of each backend (number of megabytes
    evaluated per second) for the weak linear elasticity term einsum expression
    with allowed threads. Left: residual mode evaluations $|r| / \twwmean$,
    right: matrix mode evaluations $|M| / \twwmean$. The backends are indicated
    by markers, the einsum contraction paths by color. The cell counts ($x$
    axis) are grouped by the approximation orders.}
  \label{fig:atrm-5}
\end{figure}

The overall results are summarized in Tab.~\ref{tab:ftsm-r} for the residual
mode and in Tab.~\ref{tab:ftsm-m} for the matrix mode evaluations. The
evaluation throughput of the einsum backends applied to the evaluation of five
selected einsum expression is shown in
Figs.~\ref{fig:atrm-1}--\ref{fig:atrm-5}.

The performance of most backends was similar to the single-threaded case with
the exception of the JAX-based backends jx and jxv. Those backends had a
significantly increased performance for large problems, both in terms of the
mesh size (the number of cells) and the FE approximation order, most visible in
the matrix mode evaluations. The new backends dat and oedat also performed very
well for the largest problems considered, often replacing jx in
Tab.~\ref{tab:ftsm-r} (Laplacian, NS convective and linear elasticity weak
forms) and Tab.~\ref{tab:ftsm-m} (Laplacian, weighted vector dot and linear
elasticity weak forms). Their performance, when applied to the same contraction
paths, was very similar.

By comparing with data in Section~\ref{sec:pswfc}, we can conclude that
enabling threads resulted in about halved times for the largest problems
considered.

\begin{table}[htp!]
  \centering
  {
    \begin{tabular}{rrlllll}
\toprule
  \#cells & order &                   Laplacian &                   v. dot &          weighted v. dot &             NS convective &                  elasticity \\
\midrule
    1,024 &     1 &    $\enspace 0.01$ (oe 1.0) & $\enspace 0.01$ (oe 1.0) & $\enspace 0.01$ (jx 1.0) &  $\enspace 0.01$ (jx 1.0) &    $\enspace 0.02$ (jx 1.2) \\
          &     2 &    $\enspace 0.01$ (np 0.9) & $\enspace 0.01$ (np 0.9) & $\enspace 0.02$ (jx 1.0) &  $\enspace 0.02$ (jx 0.9) &    $\enspace 0.04$ (jx 1.6) \\
          &     3 &    $\enspace 0.02$ (oe 0.6) & $\enspace 0.02$ (jx 0.4) & $\enspace 0.02$ (jx 0.6) &  $\enspace 0.04$ (jx 0.6) &    $\enspace 0.09$ (jx 1.3) \\
          &     4 &   $\enspace 0.05$ (dat 0.3) & $\enspace 0.02$ (jx 0.2) & $\enspace 0.03$ (jx 0.3) & $\enspace 0.11$ (dat 0.5) &   $\enspace 0.20$ (dat 0.8) \\
          &     5 & $\enspace 0.12$ (oedat 0.3) & $\enspace 0.03$ (jx 0.1) & $\enspace 0.04$ (jx 0.2) & $\enspace 0.19$ (dat 0.3) & $\enspace 0.25$ (oedat 0.4) \\
    2,048 &     1 &    $\enspace 0.01$ (oe 1.0) & $\enspace 0.01$ (np 1.0) & $\enspace 0.01$ (jx 1.1) &  $\enspace 0.02$ (jx 1.1) &    $\enspace 0.02$ (jx 1.4) \\
          &     2 &    $\enspace 0.02$ (oe 0.8) & $\enspace 0.01$ (jx 0.7) & $\enspace 0.02$ (jx 0.8) &  $\enspace 0.03$ (jx 0.8) &    $\enspace 0.06$ (jx 1.7) \\
          &     3 &    $\enspace 0.04$ (oe 0.5) & $\enspace 0.02$ (jx 0.3) & $\enspace 0.03$ (jx 0.4) &  $\enspace 0.07$ (jx 0.5) &    $\enspace 0.17$ (jx 1.2) \\
          &     4 &   $\enspace 0.06$ (dat 0.2) & $\enspace 0.03$ (jx 0.1) & $\enspace 0.04$ (jx 0.2) & $\enspace 0.15$ (dat 0.3) &   $\enspace 0.22$ (dat 0.5) \\
    4,096 &     1 &    $\enspace 0.01$ (oe 1.0) & $\enspace 0.01$ (np 1.0) & $\enspace 0.02$ (oe 1.2) &  $\enspace 0.02$ (jx 1.2) &    $\enspace 0.03$ (jx 1.5) \\
          &     2 &    $\enspace 0.02$ (oe 0.8) & $\enspace 0.02$ (jx 0.6) & $\enspace 0.02$ (jx 0.7) &  $\enspace 0.04$ (jx 0.7) &    $\enspace 0.09$ (jx 1.6) \\
          &     3 &   $\enspace 0.05$ (dat 0.3) & $\enspace 0.03$ (jx 0.2) & $\enspace 0.04$ (jx 0.4) &  $\enspace 0.12$ (jx 0.5) &   $\enspace 0.25$ (dat 0.9) \\
          &     4 & $\enspace 0.09$ (oedat 0.2) & $\enspace 0.04$ (jx 0.1) & $\enspace 0.08$ (jx 0.2) & $\enspace 0.19$ (dat 0.2) & $\enspace 0.32$ (oedat 0.3) \\
    8,192 &     1 &    $\enspace 0.01$ (np 1.0) & $\enspace 0.02$ (jx 1.0) & $\enspace 0.02$ (jx 1.0) &  $\enspace 0.02$ (jx 1.0) &    $\enspace 0.03$ (jx 1.3) \\
          &     2 &    $\enspace 0.04$ (oe 0.7) & $\enspace 0.02$ (jx 0.5) & $\enspace 0.03$ (jx 0.6) &  $\enspace 0.06$ (jx 0.6) &    $\enspace 0.17$ (jx 1.5) \\
          &     3 &   $\enspace 0.06$ (dat 0.2) & $\enspace 0.04$ (jx 0.2) & $\enspace 0.08$ (jx 0.3) & $\enspace 0.16$ (dat 0.3) &   $\enspace 0.28$ (dat 0.5) \\
   16,384 &     1 &    $\enspace 0.02$ (oe 1.0) & $\enspace 0.02$ (jx 0.9) & $\enspace 0.03$ (jx 0.9) &  $\enspace 0.03$ (jx 1.0) &    $\enspace 0.06$ (jx 1.5) \\
          &     2 &   $\enspace 0.04$ (dat 0.5) & $\enspace 0.03$ (jx 0.4) & $\enspace 0.06$ (jx 0.6) &  $\enspace 0.10$ (jx 0.5) &   $\enspace 0.27$ (dat 1.3) \\
          &     3 &   $\enspace 0.10$ (dat 0.2) & $\enspace 0.06$ (jx 0.1) & $\enspace 0.12$ (jx 0.3) & $\enspace 0.23$ (dat 0.2) & $\enspace 0.42$ (oedat 0.4) \\
   32,768 &     1 &    $\enspace 0.03$ (oe 1.0) & $\enspace 0.02$ (jx 0.8) & $\enspace 0.04$ (jx 0.9) &  $\enspace 0.04$ (jx 0.7) &    $\enspace 0.10$ (jx 1.5) \\
          &     2 &   $\enspace 0.06$ (dat 0.4) & $\enspace 0.05$ (jx 0.3) & $\enspace 0.10$ (jx 0.5) & $\enspace 0.15$ (dat 0.4) &   $\enspace 0.33$ (dat 0.8) \\
   65,536 &     1 &    $\enspace 0.04$ (oe 0.9) & $\enspace 0.03$ (jx 0.6) & $\enspace 0.07$ (jx 0.9) &  $\enspace 0.07$ (jx 0.7) &    $\enspace 0.18$ (jx 1.4) \\
          &     2 &   $\enspace 0.10$ (dat 0.3) & $\enspace 0.09$ (jx 0.3) & $\enspace 0.19$ (jx 0.5) & $\enspace 0.22$ (dat 0.3) & $\enspace 0.50$ (oedat 0.6) \\
  131,072 &     1 &   $\enspace 0.07$ (dat 0.9) & $\enspace 0.05$ (jx 0.6) & $\enspace 0.11$ (jx 0.8) &  $\enspace 0.12$ (jx 0.6) &    $\enspace 0.33$ (jx 1.4) \\
  262,144 &     1 &   $\enspace 0.08$ (dat 0.6) & $\enspace 0.10$ (jx 0.6) & $\enspace 0.22$ (jx 0.8) &  $\enspace 0.24$ (jx 0.6) &   $\enspace 0.57$ (dat 1.2) \\
  524,288 &     1 & $\enspace 0.15$ (oedat 0.5) & $\enspace 0.18$ (jx 0.6) & $\enspace 0.42$ (jx 0.8) & $\enspace 0.38$ (dat 0.5) &   $\enspace 0.87$ (dat 0.9) \\
1,048,576 &     1 &   $\enspace 0.23$ (dat 0.4) & $\enspace 0.35$ (jx 0.6) & $\enspace 0.81$ (jx 0.8) & $\enspace 0.60$ (dat 0.4) & $\enspace 1.68$ (oedat 0.9) \\
\bottomrule
\end{tabular}

  }
  \caption{Summary results of residual evaluation mode, no single-thread
    limitation. For each weak form, mesh
    size and FE approximation order, the shortest elapsed time mean
    without the worst case $\twwmean$~[s] across all backends is given, as well
    as the corresponding backend and $\rtwwmean$ (in parentheses).}
  \label{tab:ftsm-r}
\end{table}
\begin{table}[htp!]
  \centering
  {
    \begin{tabular}{rrlllll}
\toprule
  \#cells & order &                   Laplacian &                    v. dot &             weighted v. dot &               NS convective &                  elasticity \\
\midrule
    1,024 &     1 &    $\enspace 0.01$ (jx 1.0) &  $\enspace 0.01$ (jx 0.9) &    $\enspace 0.02$ (oe 1.0) &    $\enspace 0.02$ (jx 0.8) &    $\enspace 0.03$ (jx 1.6) \\
          &     2 &    $\enspace 0.04$ (jx 1.0) &  $\enspace 0.06$ (jx 0.3) &    $\enspace 0.07$ (jx 0.4) &    $\enspace 0.08$ (jx 0.2) &    $\enspace 0.27$ (jx 1.1) \\
          &     3 &    $\enspace 0.17$ (jx 0.4) & $\enspace 0.30$ (jxv 0.0) &    $\enspace 0.30$ (jx 0.0) &    $\enspace 0.38$ (jx 0.0) &    $\enspace 0.97$ (jx 0.2) \\
          &     4 &    $\enspace 0.51$ (jx 0.1) & $\enspace 0.88$ (jxv 0.0) &    $\enspace 1.09$ (jx 0.0) &   $\enspace 1.33$ (jxv 0.0) &    $\enspace 3.71$ (jx 0.1) \\
          &     5 &    $\enspace 1.33$ (jx 0.0) & $\enspace 2.58$ (jxv 0.0) &    $\enspace 3.50$ (jx 0.0) &   $\enspace 3.92$ (jxv 0.0) &            $11.56$ (jx 0.1) \\
    2,048 &     1 &    $\enspace 0.02$ (jx 1.1) &  $\enspace 0.02$ (jx 0.9) &    $\enspace 0.02$ (oe 0.9) &    $\enspace 0.03$ (jx 0.7) &    $\enspace 0.05$ (jx 1.7) \\
          &     2 &    $\enspace 0.06$ (jx 0.9) &  $\enspace 0.10$ (jx 0.3) &    $\enspace 0.11$ (jx 0.3) &    $\enspace 0.14$ (jx 0.2) &    $\enspace 0.50$ (jx 1.0) \\
          &     3 &    $\enspace 0.33$ (jx 0.4) & $\enspace 0.51$ (jxv 0.0) &    $\enspace 0.56$ (jx 0.0) &    $\enspace 0.72$ (jx 0.0) &    $\enspace 1.92$ (jx 0.2) \\
          &     4 &    $\enspace 1.01$ (jx 0.1) & $\enspace 1.71$ (jxv 0.0) &    $\enspace 2.12$ (jx 0.0) &   $\enspace 2.46$ (jxv 0.0) &    $\enspace 7.28$ (jx 0.1) \\
    4,096 &     1 &    $\enspace 0.02$ (jx 1.2) &  $\enspace 0.03$ (jx 0.9) &    $\enspace 0.03$ (jx 0.8) &    $\enspace 0.03$ (jx 0.5) &    $\enspace 0.09$ (jx 1.8) \\
          &     2 &    $\enspace 0.11$ (jx 0.8) &  $\enspace 0.18$ (jx 0.3) &    $\enspace 0.21$ (jx 0.3) &    $\enspace 0.25$ (jx 0.2) &    $\enspace 0.99$ (jx 1.0) \\
          &     3 &    $\enspace 0.66$ (jx 0.4) & $\enspace 1.00$ (jxv 0.0) &    $\enspace 1.12$ (jx 0.0) &   $\enspace 1.37$ (jxv 0.0) &    $\enspace 3.80$ (jx 0.2) \\
          &     4 &    $\enspace 1.95$ (jx 0.1) & $\enspace 3.30$ (jxv 0.0) &    $\enspace 4.15$ (jx 0.0) & $\enspace 7.09$ (oedat 0.0) &         $44.14$ (oedat 0.3) \\
    8,192 &     1 &    $\enspace 0.03$ (jx 1.1) &  $\enspace 0.05$ (jx 0.9) &    $\enspace 0.05$ (oe 0.9) &    $\enspace 0.06$ (jx 0.5) &    $\enspace 0.15$ (jx 1.9) \\
          &     2 &    $\enspace 0.20$ (jx 0.8) &  $\enspace 0.34$ (jx 0.3) &    $\enspace 0.39$ (jx 0.3) &    $\enspace 0.48$ (jx 0.2) &    $\enspace 1.93$ (jx 1.0) \\
          &     3 &    $\enspace 1.22$ (jx 0.4) & $\enspace 1.84$ (jxv 0.0) &    $\enspace 2.10$ (jx 0.0) &   $\enspace 2.60$ (jxv 0.0) &    $\enspace 7.45$ (jx 0.2) \\
   16,384 &     1 &    $\enspace 0.04$ (jx 1.1) &  $\enspace 0.08$ (jx 0.8) &    $\enspace 0.10$ (oe 0.9) &    $\enspace 0.11$ (jx 0.5) &    $\enspace 0.30$ (jx 1.9) \\
          &     2 &    $\enspace 0.40$ (jx 0.8) &  $\enspace 0.67$ (jx 0.3) &    $\enspace 0.75$ (jx 0.3) &    $\enspace 0.93$ (jx 0.2) &    $\enspace 3.85$ (jx 1.0) \\
          &     3 &   $\enspace 2.05$ (dat 0.3) & $\enspace 3.60$ (jxv 0.0) &    $\enspace 4.17$ (jx 0.0) & $\enspace 9.03$ (oedat 0.0) &         $32.33$ (oedat 0.4) \\
   32,768 &     1 &    $\enspace 0.06$ (jx 1.0) &  $\enspace 0.13$ (jx 0.7) &    $\enspace 0.17$ (jx 0.8) &    $\enspace 0.19$ (jx 0.5) &    $\enspace 0.56$ (jx 1.9) \\
          &     2 &   $\enspace 0.51$ (dat 0.5) &  $\enspace 1.34$ (jx 0.3) &    $\enspace 1.58$ (jx 0.3) &    $\enspace 1.78$ (jx 0.1) & $\enspace 5.93$ (oedat 0.8) \\
   65,536 &     1 &    $\enspace 0.11$ (jx 1.1) &  $\enspace 0.26$ (jx 0.7) &    $\enspace 0.33$ (jx 0.8) &    $\enspace 0.36$ (jx 0.4) &    $\enspace 1.11$ (jx 1.9) \\
          &     2 & $\enspace 0.86$ (oedat 0.4) & $\enspace 2.60$ (jxv 0.2) &    $\enspace 3.11$ (jx 0.3) &    $\enspace 3.55$ (jx 0.1) &         $10.01$ (oedat 0.6) \\
  131,072 &     1 &    $\enspace 0.20$ (jx 1.1) &  $\enspace 0.49$ (jx 0.7) &    $\enspace 0.63$ (jx 0.8) &    $\enspace 0.66$ (jx 0.4) &    $\enspace 2.16$ (jx 1.9) \\
  262,144 &     1 &    $\enspace 0.40$ (jx 1.1) & $\enspace 0.98$ (jxv 0.7) & $\enspace 1.21$ (oedat 0.8) &    $\enspace 1.32$ (jx 0.4) & $\enspace 3.93$ (oedat 1.7) \\
  524,288 &     1 &   $\enspace 0.54$ (dat 0.7) & $\enspace 1.84$ (jxv 0.7) & $\enspace 2.17$ (oedat 0.7) &    $\enspace 2.60$ (jx 0.4) & $\enspace 6.53$ (oedat 1.4) \\
1,048,576 &     1 &   $\enspace 0.89$ (dat 0.6) & $\enspace 3.65$ (jxv 0.7) & $\enspace 4.07$ (oedat 0.6) &         $17.78$ (oedas 1.4) &         $24.93$ (oedat 2.7) \\
\bottomrule
\end{tabular}

  }
  \caption{Summary results of matrix evaluation mode, no single-thread
    limitation. For each weak form, mesh
    size and FE approximation order, the shortest elapsed time mean
    without the worst case $\twwmean$~[s] across all backends is given, as well
    as the corresponding backend and $\rtwwmean$ (in parentheses).}
  \label{tab:ftsm-m}
\end{table}


\clearpage

\bibliography{time-tensors.bib}

\end{document}